\documentclass[prb,twocolumn,a4paper]{revtex4-2}

\usepackage{amsmath,bm} 
\usepackage{amssymb}
\usepackage{braket}

\usepackage{graphicx}

\usepackage{url}
\usepackage{hyperref}
\hypersetup{
  colorlinks=true,
  linkcolor=blue,
  filecolor=blue,      
  citecolor=blue,
  urlcolor=blue,
}

\DeclareMathOperator{\Tr}{Tr}
\begin{document} 
\title{Entanglement dynamics and ergodicity breaking in a quantum cellular automaton}
\author{Kevissen Sellapillay}\email{kevissen.sellapillay@univ-amu.fr}
\affiliation{Aix-Marseille Université, Université de Toulon, CNRS, CPT, Marseille, France}
\author{Alberto D.\ Verga}\email{alberto.verga@univ-amu.fr}
\affiliation{Aix-Marseille Université, Université de Toulon, CNRS, CPT, Marseille, France}
\author{Giuseppe Di Molfetta}\email{giuseppe.di-molfetta@univ-amu.fr}
\affiliation{Aix-Marseille Université, LIS, Campus de Luminy, 13288 Marseille, France}
\date{\today}
\begin{abstract}
  Ergodicity breaking is observed in the blockade regime of Rydberg atoms arrays, in the form of low entanglement eigenstates known as scars, which fail to thermalize. The signature of these states persists in periodically driven systems, where they coexist with an extensive number of chaotic states. Here we investigate a quantum cellular automaton based on the classical rule that updates a site if its two neighbors are in the lower state. We show that the breaking of ergodicity extends to chaotic states. The dynamical breaking of ergodicity is controlled by chiral quasiparticle excitations which propagate entanglement. Evidence of nonlocal entanglement is found, showing that these nonthermal chaotic states may be useful to quantum computation.
\end{abstract}
\maketitle

\section{Introduction}
\label{S:intro}

A quantum cellular automaton can be viewed as a model of quantum computing, as well as an out of equilibrium system evolving in discrete space-time \cite{Arrighi-2019a,Farrelly-2020}. As a model of computation, an automaton belongs to the class of quantum circuits \cite{Deutsch-1985hl}, provided the additional constraints of homogeneity (a single set of local unitaries is used) and translation invariance are satisfied \cite{Schumacher-2004,Perez-Delgado-2007}. In addition, a quantum Turing machine can be build from a quantum cellular automaton, leading to its universality \cite{Watrous-1995,Arrighi-2012}. As a dynamical system, quantum automata arise naturally in periodically driven lattices. One example, relevant to the present work, is given by the arrays of atoms excited to Rydberg states \cite{Wintermantel-2020,Hillberry-2021a}. An interesting property of Rydberg arrays in the blockade regime is the existence of nonthermal states, scars, which lead to persistent oscillations \cite{Bernien-2017}. Scars are embedded into a chaotic set of states in the Hilbert space, well described by the so called PXP Hamiltonian \cite{Turner-2018}, and survive under nonintegrable quantum modifications, in the form of a quantum automaton, of the classical rule \cite{Iadecola-2020}.

Beyond simple scars, more complex quantum states can be build from \emph{classical} automaton rules, possessing nontrivial topology and error correction capabilities \cite{Yoshida-2013,Gopalakrishnan-2018}. We explore here another point of view that combines both aspects of quantum cellular automata, informational and dynamical, and extends the Clifford, integrable quantum cellular automata, to the full quantum realm.

More precisely, we investigate the ability of a simple automaton, the so called PXP based on the Toffoli gate (rule 201 \cite{Bobenko-1993}) \cite{Iadecola-2020}, to generate nonthermal, long-range entangled states potentially useful as a computing resource. Our main goal is to demonstrate that ergodicity breaking is also possible within the chaotic region of the Hilbert space.

Highly entangled states are not always useful for quantum computation \cite{Bremner-2009,Gross-2009uq}: random maximally entangled states \cite{Page-1993nr}, can be efficiently simulated classically. Moreover, from the thermodynamic point of view, ergodic thermal states do not contain information; to get information from chaotic states, ergodicity breaking is necessary. A classical paradigm of nonergodic thermal system is the spin glass, from which an associative memory can be created \cite{Hopfield-1988}. Analogously, quantum glassiness, understood as the impossibility of a system to relax to its ground state due to a multiplicity of nonergodic phases, leads to topologically protected ``logical'' states \cite{Chamon-2005}. Instead of low energy ergodicity breaking, which is related to the ground state of a complex Hamiltonian, another possibility is dynamical ergodicity breaking involving high energy states, related to a periodically driven system.

We know that the generic evolution of periodically driven system is towards an infinite temperature state satisfying the eigenstate thermalization hypothesis \cite{DAlessio-2014}. However, interactions, in particular in the form of constrained dynamics, can lead to a dynamical breaking of ergodicity, and then, to nontrivial states like scars that violate the eigenstate thermalization \cite{Mizuta-2020,Mukherjee-2020}. We ask the question whether ergodicity breaking also affects chaotic states. It would suppose that there are ``complex'' states satisfying random matrix statistics of their entanglement, but having von Neumann entropies below the Page limit \cite{Page-1993nr}. To probe the existence of a nonergodic chaotic family of states we measure the entanglement and characterize the geometry of states evolved from each state of the computational basis (initial qubits configuration), spanning the whole available Hilbert space.

In the following section we present the PXP automaton, which was first proposed by Iadecola and Vijay \cite{Iadecola-2020} to investigate the fate of PXP scars discovered by Bernien et al. \cite{Bernien-2017}, under nonintegrable deformations of the Toffoli classical cellular automaton \cite{Wilkinson-2020}. We will focus instead on the dynamics driven by quasiparticles, which are travelling excitations above the classical vacuum (domain walls and gliders). Quasiparticles in Floquet systems can carry information and create entanglement, as is well documented in integrable systems \cite{Gopalakrishnan-2018b,Friedman-2019}. In Sec.~\ref{S:qp} we first derive analytically the dispersion relation of chiral quasiparticles, and apply it to the propagation of entanglement in the weak nonintegrable perturbation case. We extend the study of quasiparticles driven entanglement to the full nonintegrable case using numerical computations of the PXP automaton. We show next, in Sec.~\ref{S:ergo}, that the evolution of initial states in the computational basis (which is preserved by the classical automaton) strongly depends on their contents in quasiparticles. Even if most states evolve into chaotic ones, their entanglement entropy and inverse participation ratio may differ significantly, leading to a nonergodic partition of the Hilbert space. In the last section we summarize the main results and discuss some perspectives.

\section{The quantum cellular automaton}
\label{S:qcm}

We consider the PXP cellular automaton depending on the parameter \(\theta\) \cite{Iadecola-2020}, whose generator is the unitary,
\begin{equation}
  \label{e:Ux}
  U_x(\theta) = \begin{pmatrix}
    \cos\theta & 0 & -i \sin\theta \\
    0 & 1 & 0 \\
    -i \sin\theta & 0 & \cos\theta 
  \end{pmatrix} \oplus I_{5} \,,
\end{equation}
acting on three qubits at sites \(\{x-1,x,x+1\} \in [0,N-1]\) of a one dimensional lattice of even \(N\) sites (\(\oplus\) denotes the direct sum of matrices and \(I_n\) is the $n$-dimensional identity matrix; periodic boundary conditions are used). It can also be written in terms of the Pauli matrices \(\bm \sigma_x = (X_x, Y_x, Z_x)\) (defined at each site \(x\))
\begin{equation}
  \label{e:ePXP}
  U_x(\theta) = e^{-i\theta (PXP)_x}, \quad
  (PXP)_x = P_{x-1} X_x P_{x+1}
\end{equation}
where \(P_x = (I_2 + Z_x)/2\) is the projector on the qubit state \(\ket{0}\). One step of the automaton is defined by the staggered product of \(U_x\), successively applied to even \((e)\) and odd \((o)\) sites:
\begin{equation}
  \label{e:U}
  U(\theta) = \prod_{x \in o} e^{-i\theta (PXP)_x}
              \prod_{x \in e} e^{-i\theta (PXP)_x} \,.
\end{equation}
In the limit \(\theta \rightarrow 0\), \eqref{e:U} approximates the evolution operator of the PXP hamiltonian, 
\begin{equation}
  \label{e:HPXP}
   H_{PXP} = \sum_x (PXP)_x \,.
\end{equation}

Equation \eqref{e:U} qualitatively describes the Floquet dynamics of a periodically driven PXP array \eqref{e:HPXP}. In the limit \(\theta \rightarrow \pi/2\) it approximates the rule 201 classical automaton, which is completely integrable \cite{Wilkinson-2020}. We focus on this limit, where \(\theta\) introduces a nonintegrable modification of the classical automaton.

The system dynamics is restricted to the subspace of the $N$-qubits Hilbert space compatible with the constraint that two neighboring \(\ket{1}\) are forbidden; it has dimension \(\mathrm{dim}(N) = F_{N-1} + F_{N+1}\), where \(F_N\) is a Fibonacci number. For large \(N\) the Fibonacci subspace dimension is \(\phi^N\), with \(\phi\) the golden ratio. The system is invariant under translation of an even number of sites, generated by
\begin{equation}
  \label{e:T2}
  T^2 \ket{s_0 \ldots s_{N-2} s_{N-1}} \rightarrow \ket{s_{N-2} s_{N-1} s_0 \ldots s_{N-3}}
\end{equation}
where the set \(\mathcal{C}\) of states \(\ket{s} = \ket{s_0 s_1 \ldots s_{N-1}}\), with \(s_x=0,1\), is the computational basis. Inversion \(x \rightarrow N-x-2\) also lets the system invariant and commutes with the translation operator (in the zero momentum sector). We note that \(U(\pi/2) \ket{s} \in \mathcal{C}\), exchanges states of the computational basis. As a consequence, the cycles \(\mathcal{C}_\ell \subset \mathcal{C}\) of length \(\ell\), of the classical automaton correspond to eigenstates of \(U_\ell = U^\ell(\pi/2)\),
\begin{equation}
  \label{e:ell}
  \ket{\ell p s} = \sum_{l=0}^{\ell-1} e^{i 2\pi p l/\ell}
                   U^l(\pi/2) \ket{s},
\end{equation}
where \(\ket{s} \in \mathcal{C}_\ell\), and \(2\pi p/\ell\) the corresponding eigenvalue (\(p = 0,\ldots,\ell-1\)).

A remarkable property of \(U(\pi/2)\) is the existence of a \(\ell = 3\) cycle:
\begin{equation}
  \label{e:vac}
  \ket{A} = \ket{0000\ldots} \rightarrow \ket{B} = \ket{1010\ldots} \rightarrow \ket{C} = \ket{0101\ldots},
\end{equation}
called the ``vacuum orbit'' \cite{Wilkinson-2020,Iadecola-2020}, because chiral quasiparticles appear as domain walls separating these states. The corresponding eigenstate with \(p=0\) is
\begin{equation}
  \label{e:ell3}
  \ket{30A} = \frac{1}{\sqrt{3}} (\ket{A} + \ket{B} + \ket{C}) \,,
\end{equation}
where \(\ell = 3\) and \(\ket{s}\) is one of the states in the vacuum orbit. In particular, the three vacuum states are eigenvectors of \(U_3 = U^3(\pi/2)\):
\begin{equation}
\label{e:U3}
U_3 \ket{s} = \ket{s}, \quad \ket{s} = \{\ket{A}, \ket{B}, \ket{C} \}\,.
\end{equation}
This cycle corresponds to the \(\mathbb{Z}_2\) scars of the PXP model \cite{Bernien-2017,Turner-2018a}. The scarring phenomenon was thoroughly investigated because its potential to create high temperature ordered states beyond eigenstate thermalization \cite{Shiraishi-2018,Papic-2021}.

\begin{figure}
  \centering%
  \includegraphics[width=1.0\linewidth]{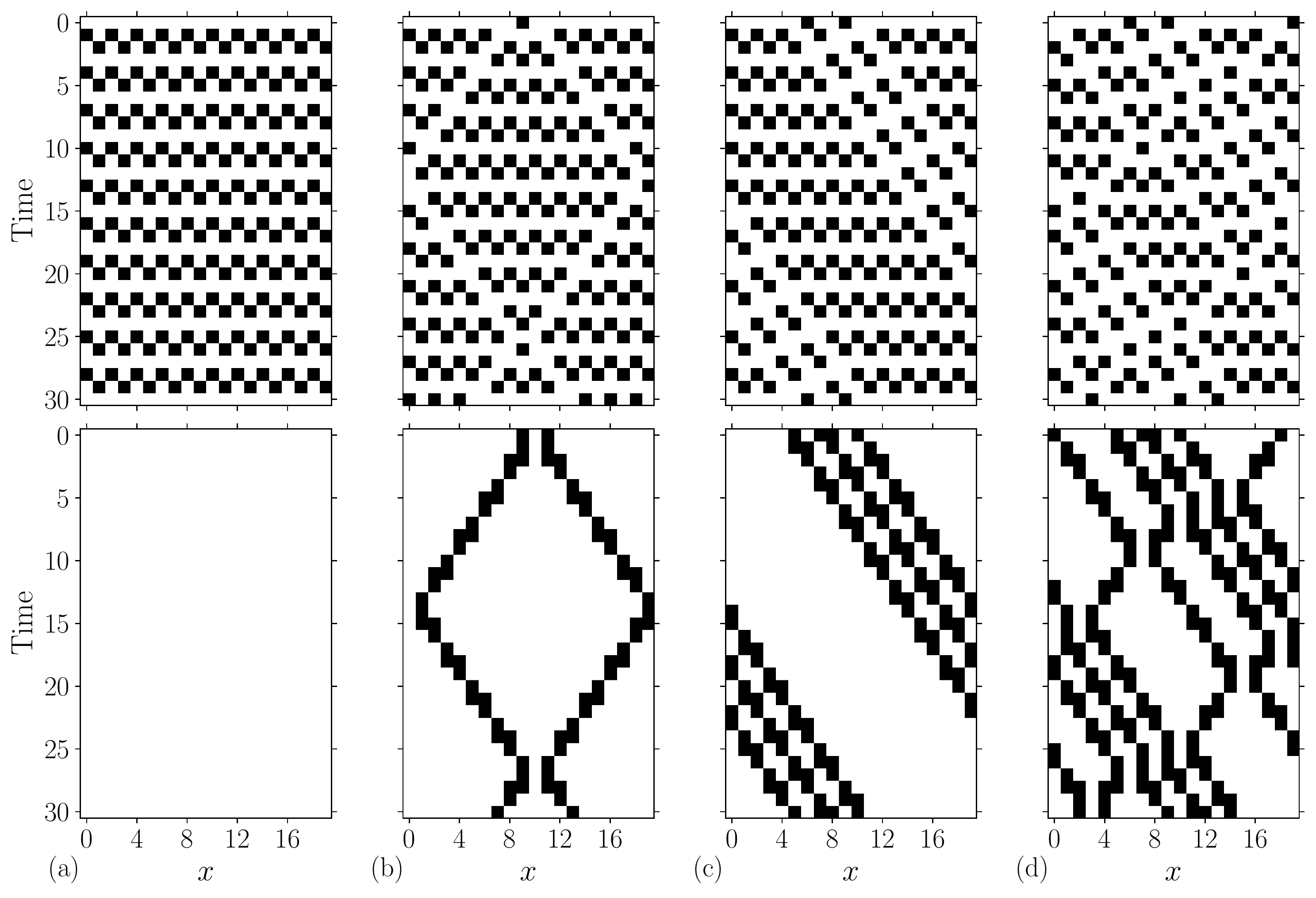}
  \caption{Classical automaton \(\ell = 3\) cycle states (\(N=20\)). The initial state \(\ket{A}\) belongs to a three steps cycle (a). Two walls, generated by a single flip, alternatively separate the three vacuum states (b). Other chiral quasiparticles can be generated by two flips (c). Interactions of quasiparticles result in shifted trajectories (d). The second row is obtained from the first one by applying the XOR operation (see text). We refer to `A' the initial vacuum state in (a), `C' the even double wall (b), and `BC' the right moving glider (c).
  \label{f:qp}}
\end{figure}

The cycle, we call `A', is shown in Fig.~\ref{f:qp}a. Simple excitations of the vacuum \(\ket{A}\) are shown in the other panels of Fig.~\ref{f:qp}. A single spin flip in the \(\ket{A}\) state, excites a chiral quasiparticle consisting in a pair of domain walls propagating at speeds \(\pm2/3\) (in full time steps). Two flips separated by two sites leads to chiral quasiparticles moving to the left or to the right depending on their parity (at speed \(1/3\)). We denote the symmetric domain walls quasiparticle by `B' or `C', depending on their parity (they separate a \(B\), or equivalently a \(C\), vacuum region to a \(A\) background), and the left and right moving gliders by `BC' or `CB' the left moving one (the vacuum pattern is \(ABCA\), for the right and \(ACBA\) for the left one).

Figure~\ref{f:qp} shows another important property of the classical automaton, it conserves the number of quasiparticles; quasiparticles weakly interact by shifting their trajectories, but fusion or annihilation of quasiparticles is forbidden \cite{Wilkinson-2020}. This is best viewed using the XOR representation of the automaton \cite{Duranthon-2021}, consisting in transforming a string \((\ldots s_x \ldots)\) in a string \(\ldots s_{x-1} \oplus s_{x+1} \ldots\) (sum modulo 2), as shown in the second row of Fig.~\ref{f:qp}.

It is worth noting that our gliders and double walls, we call generically quasiparticles, are more complex than the topological chiral ones encountered in other Floquet systems, as for example the Fredrickson-Anderson (rule 54) model \cite{Lindner-2017,Friedman-2019a}. Both types, gliders and walls, possess an internal structure: these quasiparticles separate different kinds of vacuum states (3 in the case of the shortest orbit), while the ``molecules'' of rule 54 are localized essentially in one site. Their internal structure confers to our quasiparticles richer topological properties. For instance, to destroy the double wall one should act globally on all sites between the two walls. This long-range correlations linking the two walls may impact on the way the entanglement is generated in the system.

The analogy of quasiperiodic trajectories immersed in a chaotic billard with quantum scars \cite{Serbyn-2021}, can be extended, in the case of discrete space-time dynamics, to the cycles of the automaton that are robust under nonintegrable perturbations leading to weak ergodicity breaking quantum states \cite{Iadecola-2020}. In addition to these simple orbits, the quasiparticles present in the classical automaton (\(\theta = \pi/2\)) may also be source of ergodicity breaking, in the nonintegrable (\(\theta \ne \pi/2\)) parameter region. We show in the next sections that this is effectively the case.

\section{Chiral quasiparticles}
\label{S:qp}

In order to investigate the properties of the excited states, we start with the case of the BC quasiparticle in the perturbation regime, where the parameter \(\theta\) is close to \(\pi/2\), \(\epsilon = \pi/2 - \theta\). The unperturbed operator \(U_3\) satisfies
\begin{equation}
\label{e:U3k}
U_3 \ket{LRk} = e^{-ik} \ket{LRk}\,,
\end{equation}
where \(k=4\pi n/N\) (\(n = 0, \ldots, N/2-1\)) is proportional to the momentum of the BC particle, and the eigenstate
\begin{equation}
\label{e:lrk}
\ket{LRk} = \sqrt{\frac{2}{N}} \sum_{x=0}^{N/2-1} e^{ikx} \prod_{l=0}^L X_{2x-2l} \prod_{r=0}^R X_{2x+2r+3} \ket{A}\,,
\end{equation}
contains \(L=0, \ldots, N/2-4\) left \(B\) vacuum sites and \(R = 0, \ldots, N/2-4\) right \(C\) vacuum sites with \(0\le L+R\le N/2-4\), obtained by flipping (Pauli \(X\) operators) the spins inside the \(BC\) region. For the right moving particle the roles of \(B\) and \(C\) are exchanged. Note that the choice of \(k\) is determined by the fact that \(\ket{LRk}\) are also eigenvectors of \(T^2\) (c.f. \eqref{e:T2}), therefore the units of length and time associated to BC are 2 and 3, respectively.

The perturbed evolution operator in the reference frame of the BC quasiparticle is 
\begin{equation}
\label{e:UHBC}
U^\dagger_3 U_\epsilon = e^{-i\epsilon H_{BC}}, \quad U_\epsilon = U^3(\pi/2-\epsilon)
\end{equation}
where the effective Hamiltonian \(H_{BC}\) does not depend on \(\epsilon\) to the first order in the perturbation series. An explicit computation gives the dispersion relation and eigenstates of the BC gliders (Appendix~\ref{S:glider}):
\begin{equation}
\label{e:UEk}
U_\epsilon \ket{q_1q_2k} = e^{-iE_{\bm q}(k)} \ket{q_1q_2k} 
\end{equation}
where the eigenvectors are given by the Fourier transform of \(\ket{LRk}\),
\begin{equation}
\label{e:q1q2l}
\ket{q_1q_2k} = \mathcal{N} \sum_{L,R} e^{iq_1L} e^{iq_2R} \ket{LRk}\,.
\end{equation}
with \(\bm q = (q_1, q_2)\) and \(q_1,q_2 = 4\pi (0,\ldots,N/2)/N\) and the normalization is \(\mathcal{N} \approx \sqrt{2}/N\), in the limit of large \(N\). The dispersion relation is,
\begin{equation}
\label{e:Ek}
E_{\bm q}(k) = k + 4\epsilon \big[ \cos q_1 + \cos q_2 + \sin(k + q_1 - q_2) \big] \,,
\end{equation}
which depends on the parameters \(\bm q\) and the glider wavenumber \(k\). The second term is the quasienergy due to the perturbation \(\epsilon H_{BC}\), which introduces a weak dispersion of the classical automaton glider. The wavenumbers \(\bm q\) are Fourier conjugate to the length of the BC vacuum region between the two walls delimiting the glider.

\begin{figure}
  \centering
  \includegraphics[width=0.95\linewidth]{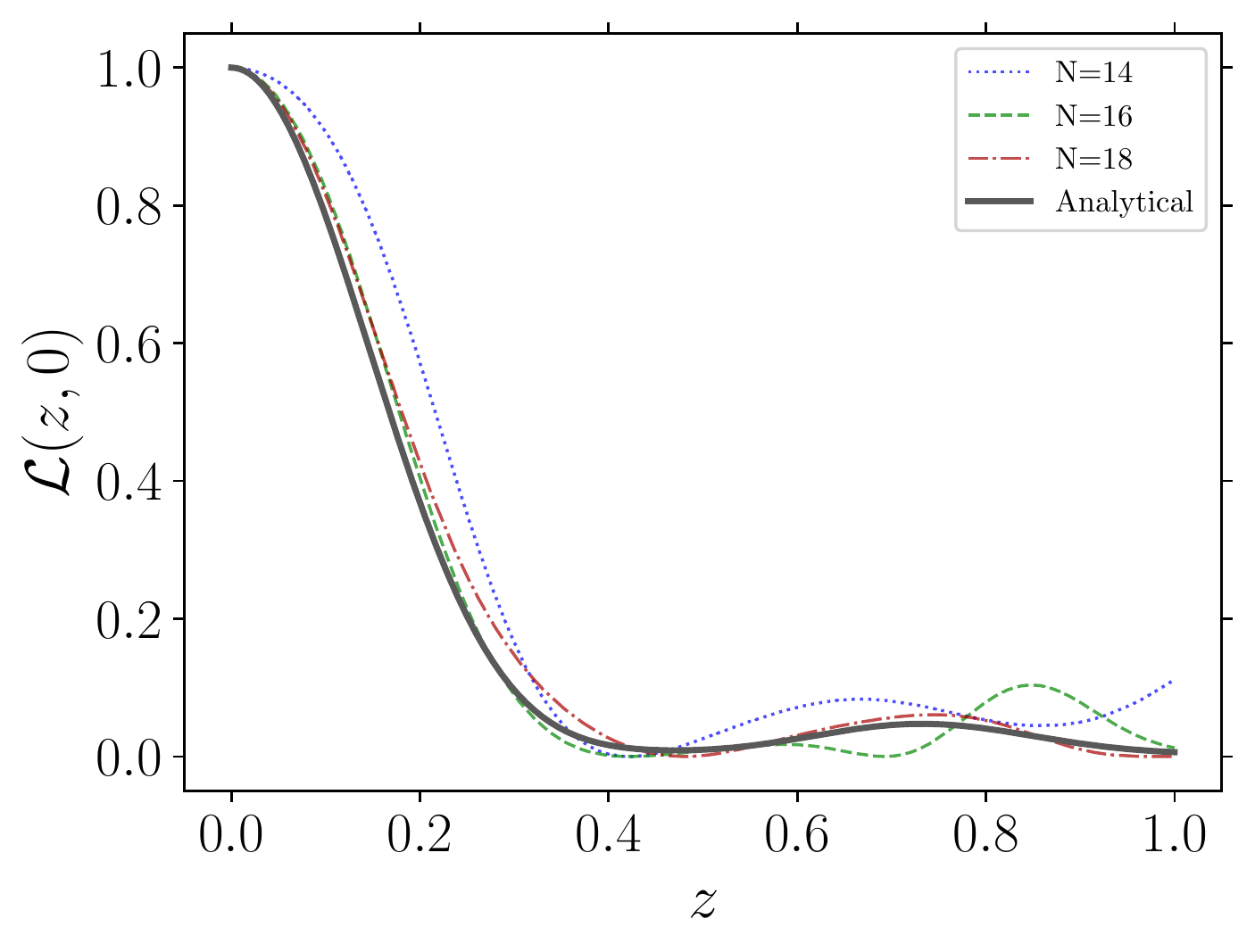} 
  \caption{Loschmidt echo computed for $N=14,16,18$, $\epsilon_{1} = 0.001$, $\epsilon_{2} = 0.002$, $t=1000$ and initial condition $L=R=k=0$ (color dashed lines), compared with the analytical result of perturbation theory, Eq.~\eqref{eq:LO_exact} (black solid line).
  \label{f:LOsimu}}
\end{figure} 

The dispersion of the quasiparticle breaks the main property of the classical automaton, which acts essentially by permuting the vectors of the computational basis \(\ket{s}\). The effect of the dispersion is to superpose different kets \(\ket{s}\) (c.f. \eqref{e:q1q2l} and \eqref{e:lrk}), thus likely creating entanglement between the qubits.

Using Eqs.~\eqref{e:lrk}-\eqref{e:q1q2l}, it is straightforward to compute the overlap of two slightly different glider states. This will inform us about the relaxation of these initial states due to the quasiparticle dispersion, at least for short times. Therefore we compute the Loschmidt echo \(\mathcal{L}(t,k)\) \cite{Peres-1984ai} of a glider in the sector of wavenumber \(k\), initially evolving from the state \(\ket{L,R,k}\) with the angle \(\theta=\pi/2-\epsilon_1\) for a time \(t\), and then backwards in time with the angle \(\theta=\pi/2-\epsilon_2\):
\begin{equation}
\label{e:losch}
\mathcal{L}(t,k) = \left| \bra{L,R,k} U^{  \dagger t}_{ \epsilon_{2}}U^{t}_{\epsilon_{1}} \ket{L,R,k} \right|^2
\end{equation}
where we used the notation of \eqref{e:U3} (note that here we have three steps of $U$ for each $t$ \eqref{e:UHBC}). In the limit of large \(N\) (see Appendix~\ref{S:glider}) we obtain a Gaussian decay \cite{Gorin-2006}:
\begin{equation}
\label{e:loscht2}
\mathcal{L}(t,k) \approx e^{-24 (\epsilon_1 - \epsilon_2)^2 t^2}\,.
\end{equation}
We compare \eqref{eq:LO_exact} with the corresponding numerical computation in Fig.~\ref{f:LOsimu}, for system of increasing size. The decay depends on the variable \(z=(\epsilon_1-\epsilon_2)t\) that, within the perturbation approximation, must be small. We observe that the dispersion relation \eqref{e:Ek} accounts for the initial evolution of the gliders, validating our hypothesis about the existence of weakly dispersive quasiparticles in the neighborhood of the classical automaton limit (see Appendix~\ref{S:num} for further numerical results).

\begin{figure}
  \centering
  \includegraphics[width=1.0\linewidth]{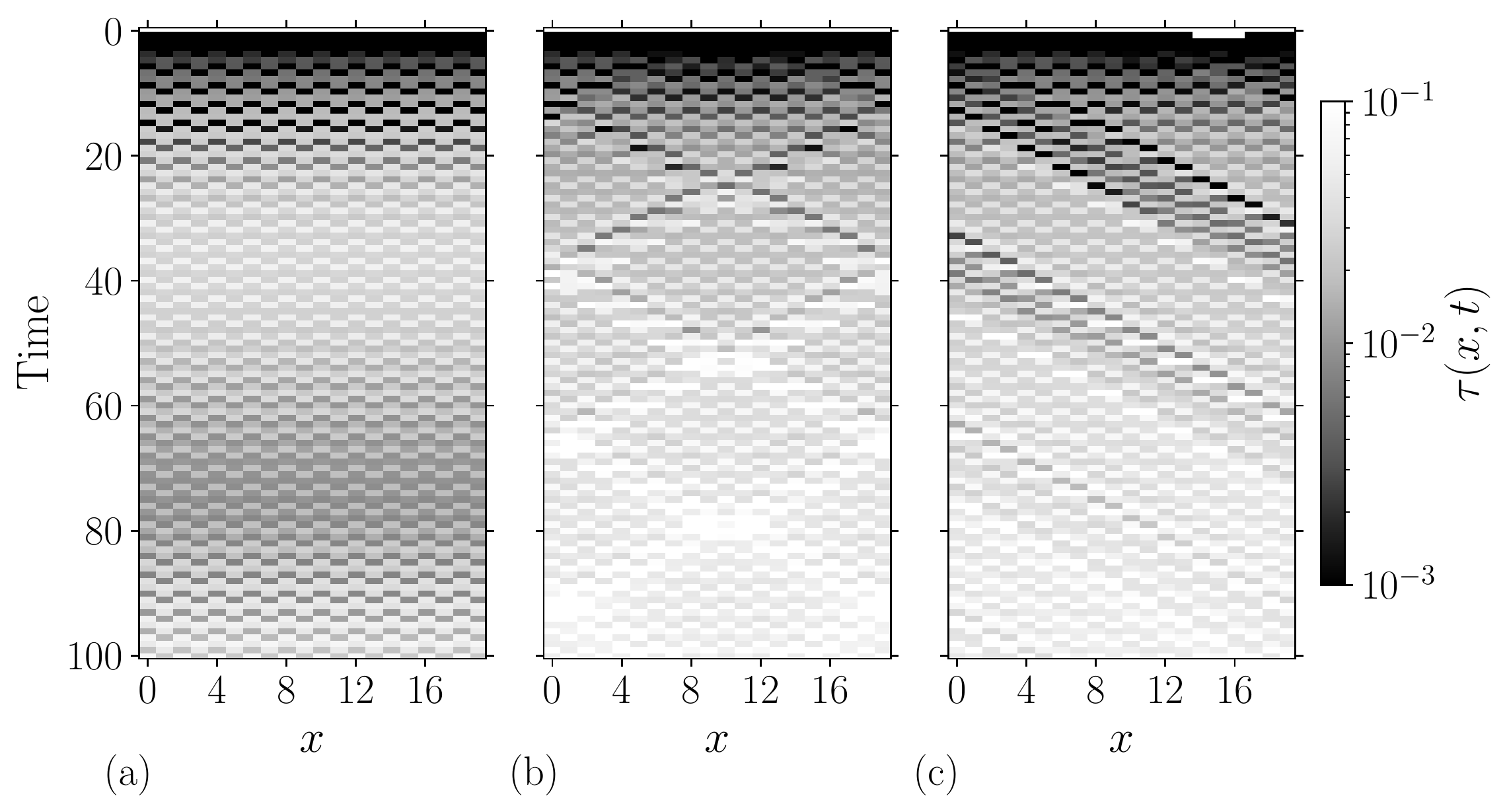}\\
  \includegraphics[width=1.0\linewidth]{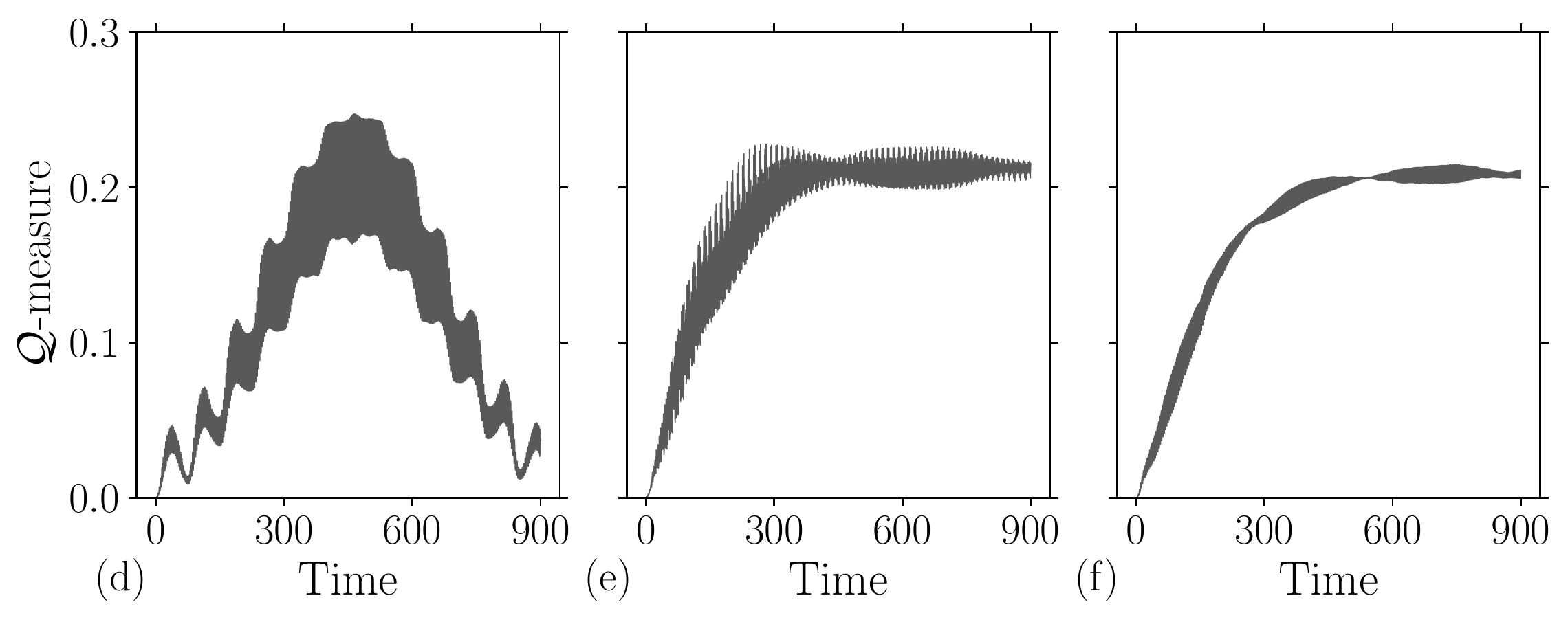}\\
  \caption{Entanglement propagation for three initial states: A-vacuum (a,d), C-quasiparticle (b,e), and BC-quasiparticle in the co-moving frame (c,f). The first row (a,b,c) shows the spatio-temporal distribution of the tangle; the second row (d,e,f) the corresponding $\mathcal{Q}$ entanglement measure. The vacuum initial state produces a recurrent entanglement (a,d); in the presence of quasiparticles the entanglement increases to saturation. Parameters: \(\epsilon = 0.01\), \(N=20\). Note the logarithmic scale of grays, allowing to enhance the weak tangle amplitudes.
  \label{f:prop}}
\end{figure}

Because the perturbation is weak, the entanglement should follow the motion of the quasiparticles. This is reinforced by the fact that, even in the presence of dispersion, the number of quasiparticles is conserved: to destroy a quasiparticle it is necessary to change the vacuum states it separates, which would need a nonlocal action. The numerical computation of a system with \(N=20\) and \(\epsilon=0.01\) confirms this scenario (c.f.\ Fig.~\ref{f:prop}). We evolve for $t$ steps the system from an initial state in the computational basis,
\begin{equation}
\label{e:psit}
\ket{\psi(t)} = U(\theta)^t \ket{s}.
\end{equation} The state of a spin at position \(x\) and time \(t\) is given by the density matrix
\begin{equation}
\label{e:rx}
\rho_x(t) = \Tr_{\overline{x}} \rho(t), \quad \rho(t) = \ket{\psi(t)} \bra{\psi(t)}\,,
\end{equation}
where we trace out the rest of the sites \(\overline{x}\). We measure the entanglement of individual spins using the tangle \(\tau(x,t)\) \cite{Coffman-2000}:
\begin{equation}
\label{e:tangle}
\tau(x,t) = 4 \det[\rho_x(t)]\,,
\end{equation}
which is proportional to the purity \(\tau_x = 2 - 2\Tr\rho_x^2\). We also define the global entanglement \cite{Meyer-2002,Brennen-2003a}:
\begin{equation}
\label{e:Q}
\mathcal{Q}(t) = \frac{1}{N} \sum_{x=0}^{N-1} \tau(x,t)\,,
\end{equation}
which is the spatial average of the tangles, and depends essentially on the distribution of the expected value of the spins \cite{Lakshminarayan-2005}:
\begin{equation}
\label{e:spin}
\braket{\bm \sigma(x,t)} = \Tr \rho_x(t) \bm \sigma \,.
\end{equation}

Figure~\ref{f:prop} represents the tangle evolution of the A (vacuum), C (double wall), and BC (glider) initial states, and their corresponding general entanglement \(\mathcal{Q}(t)\). The \(\ell=3\) cycle leads to a recurring weak entanglement (the maximum value of \(\tau\) and \(\mathcal{Q}\) is 1). The oscillations of the entanglement are here reminiscent to the scars associated to the \(\mathbb{Z}_2\)-cycle of the PXP model \cite{Bluvstein-2021,Sugiura-2021}. When quasiparticles are present, the entanglement increases and relaxes to a saturation value in a statistically stationary state (we discuss in the following section the nature of the saturated states). In Fig.~\ref{f:prop}bc one clearly identifies the trace of the quasiparticles and their weak wake, which enlarges with time. We observe that even when the entanglement measure reached its saturation level, the imprint of the domain walls is present; moreover, the self-interaction of quasiparticles due to the periodic boundary (reflective boundaries would lead to the same pattern \cite{Wilkinson-2020}), do not change their number.

\begin{figure}
  \centering
  \includegraphics[width=0.55\linewidth]{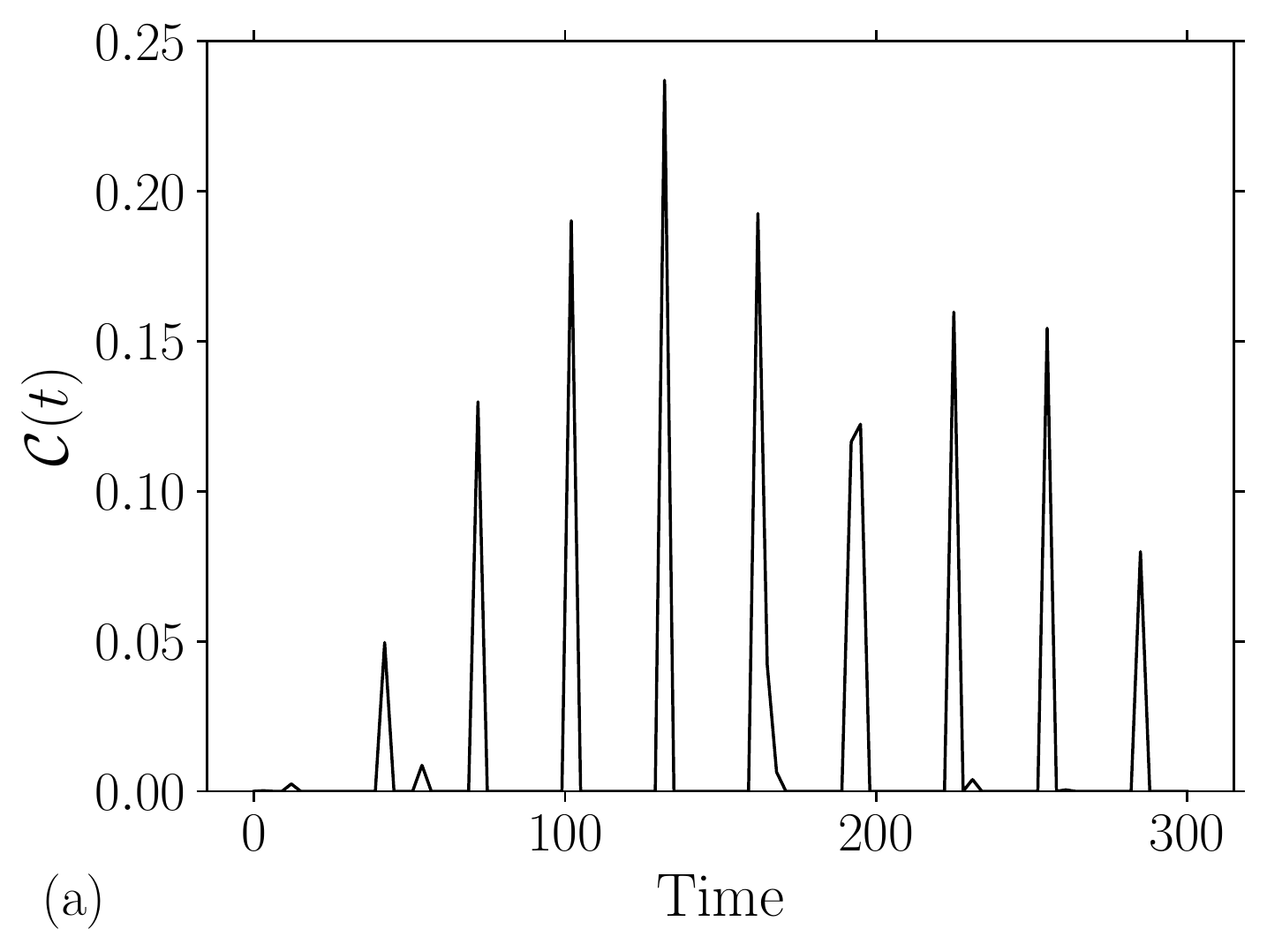}%
  \includegraphics[width=0.45\linewidth]{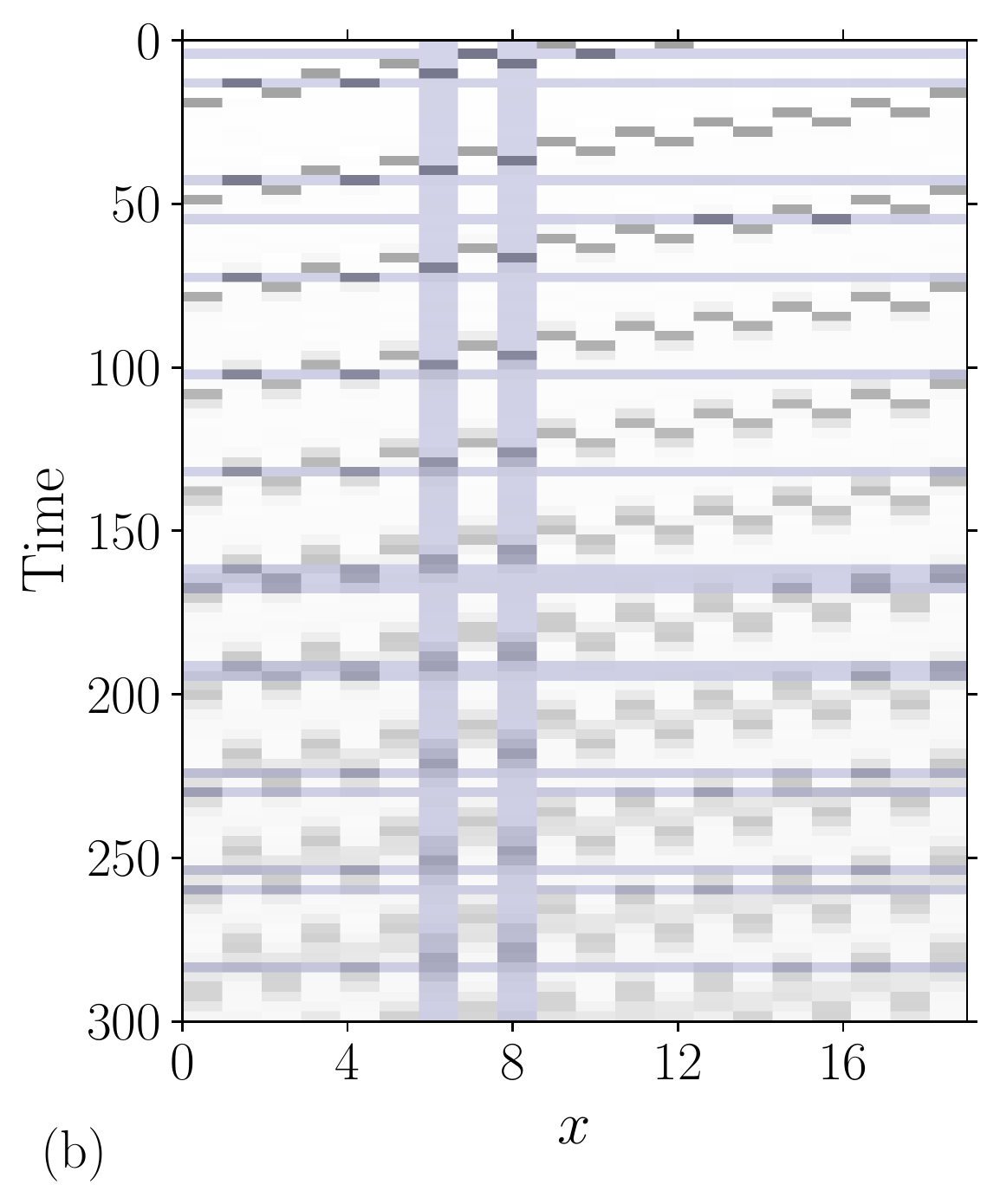}
  \caption{Concurrence between spins 5 and 7 generated by a glider. (a) The pics are separated by about \(3N/2=30\) time steps, corresponding to the passage time of the glider. (b) glider motion (note the existence of some dispersion); the 5 and 7 spins, and the passage times are underlined; each intersection correspond to a pic in (a). Parameters: \(N=20\) and \(\epsilon=0.01\); data is displayed every three time steps (to avoid the vacuum oscillations at each time step).
  \label{f:conc}}
\end{figure}

To verify that the passage of a quasiparticle creates entanglement, we can measure the concurrence of spin pairs, separated by a few sites \cite{Jurcevic-2014}. The concurrence \cite{Wootters-1998} is precisely a measure related to the entanglement of formation \cite{Plenio-2005fj}, and is computed from the eigenvalues \(\lambda_{1,\ldots,4}\) (in decreasing order) of the two qubit matrix
\[\sqrt{\rho_{xy}^{1/2} (YY)\rho^\star_{xy}(YY)\rho_{xy}^{1/2}},\]
where \(\rho_{xy}\) is the density matrix of two spins at \(x\) and \(y\):
\begin{equation}
\label{e:concurrence}
\mathcal{C}_{xy} = \max\{0,\lambda_1 - \lambda_2 - \lambda_3 - \lambda_4\}.
\end{equation}
In Fig.~\ref{f:conc} the concurrence of two qubits separated by two sites is plotted as a function of time, in the case of a BC quasiparticle; the observed peaks correspond to the successive passage of the glider; for other separations the concurrence vanishes. The locality of the concurrence is a consequence of the locality of the quasiparticle, which spans four sites, and the locality of the interaction, which correlates each site with its neighbors; a few sites away from the domain wall, the local state is near a vacuum, which is a product state.

In the next section we study the consequences of the particular behavior of quasiparticles on the thermal properties of the evolved states and their entanglement characteristics, in particular concerning the entanglement range.

\section{Ergodicity breaking by quasiparticles}
\label{S:ergo}

\begin{figure}
  \centering
  \includegraphics[width=1.0\linewidth]{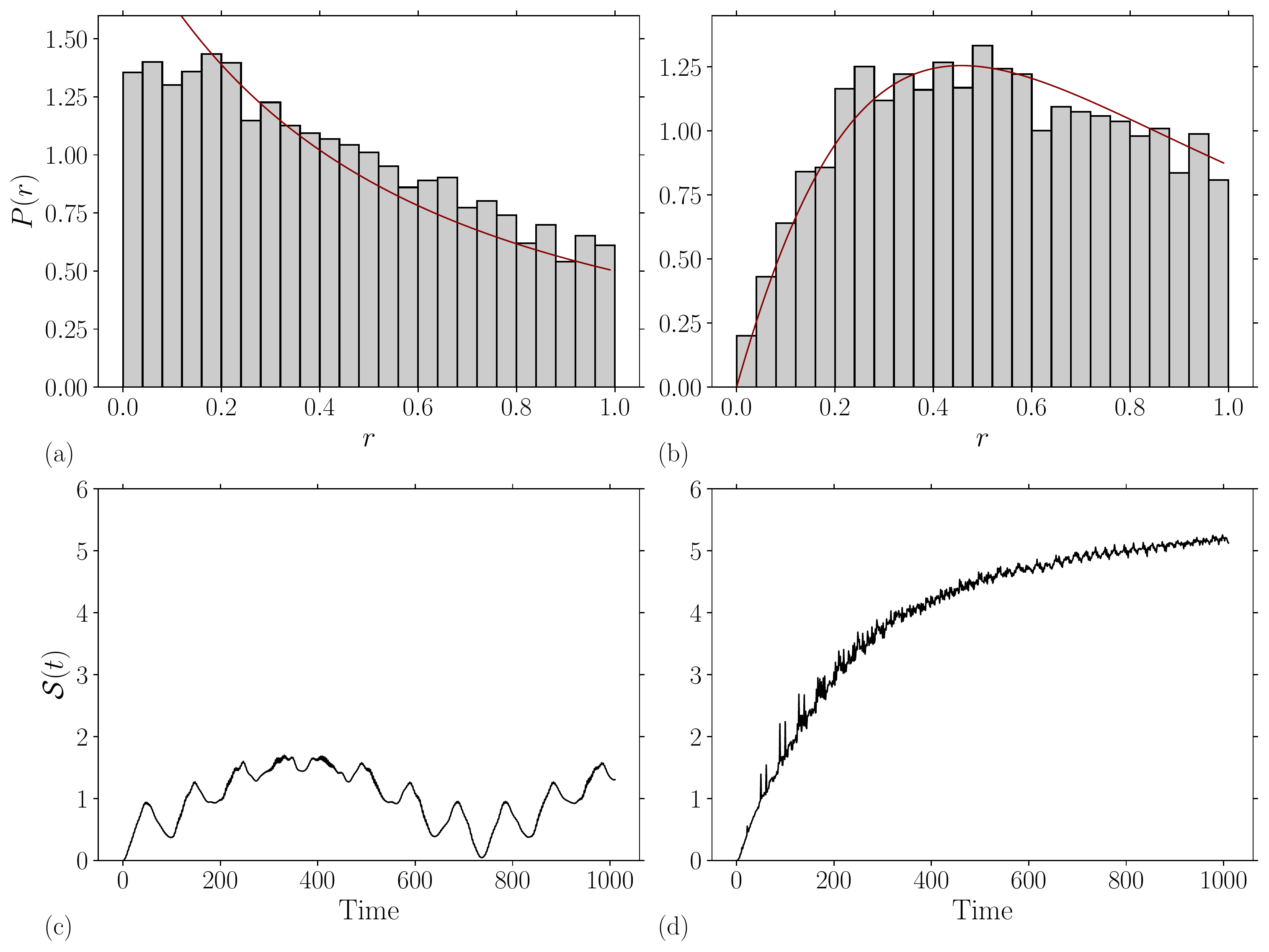}
  \caption{Entanglement spectrum level statistics. The left panels (a,c) shows the results for the A initial vacuum state, and the right ones (b,d) the BC glider. The \(r\) statistics (level spacings) match a Poissonian distribution for A (a), and an orthogonal ensemble distribution for BC (b). Their corresponding global entanglement \(\mathcal{Q}\) is recursive in the first case (a) and increasing in the second one (b). Parameters: \(N=26\), \(\epsilon=0.01\). The histograms are averaged over ten time steps (around \(t=1000\)).
  \label{f:chaos}}
\end{figure}

We just demonstrated that in the presence of quasiparticles, each qubit evolves towards a mixed entangled state, while the vacuum state shows revivals of the initial low entanglement state. The different entanglement behavior of these two kinds of initial states are reminiscent of the behavior of regular (scars-like) and chaotic (thermal-like) states. The statistics of the entanglement spectrum \cite{Li-2008fk} can distinguish between regular and chaotic states \cite{Iaconis-2021}. We then select the A and BC initial states, evolve the automaton a large number of steps \(t\) and compute the Schmidt eigenvalues \(\sqrt{p_n}\) of a partition into two halves `\(\mathrm{a}\)' and `\(\mathrm{b}\)' of the chain (where \(n=0,\ldots,N/2\)):
\begin{equation}
\label{e:schmidt}
\ket{\psi(t)} = \sum_n \sqrt{p_n(t)} \ket{n}_\mathrm{a} \ket{n}_\mathrm{b}\,.
\end{equation}
From the entanglement spectrum \(\{p_n\}\) we compute the distribution \(P(r)\) of the spacings \cite{Oganesyan-2007zr},
\begin{equation}
\label{e:pr}
r = \frac{\min\{\Delta_{n},\Delta_{n+1}\}}{\max\{\Delta_n,\Delta_{n+1}\}}
\end{equation}
where \(\Delta_n = p_{n} - p_{n-1}\). For a Poisson distribution we have
\begin{equation}
\label{e:poisson}
P(r) = \frac{2}{(1+r)^2}
\end{equation}
and for the orthogonal Gaussian ensemble \cite{Atas-2013}
\begin{equation}
\label{e:gue}
P(r) = \frac{54}{8} \frac{r(1+r)}{(1+r+r^2)^{5/2}}.
\end{equation}
The corresponding von Neumann half chain entanglement entropy is
\begin{equation}
\label{e:vN}
\mathcal{S}(t) = - \Tr_{\mathrm{b}} \rho(t) \log \rho(t) = -\sum_n p_n(t) \log p_n(t) \,,
\end{equation}
where the partial trace is over \(N/2\) consecutive sites, and \(\log=\log_2\) is the base 2 logarithm.

In Fig.~\ref{f:chaos} we represented the entanglement spectrum statistics and entropy of the A state when its entanglement is around its maximum, and of the BC state in its steady, maximum entanglement state. \(P(r)\) is averaged over 10 time steps, to avoid short time oscillations. The histograms confirm that the \(\ell = 3\) vacuum cycle leads to a regular state with localized eigenvectors satisfying the Poisson level statistics, akin to the ones of integrable dynamics. In contrast to the A dynamics, the chiral quasiparticle states tend to create a chaotic high entanglement regime well described by random matrices ensembles (Fig.~\ref{f:chaos}b). However, their entanglement entropy is far from the theoretical Page limit of a thermal state at infinite temperature \cite{Zhang-2015}. This fact suggests the question whether some structure does emerge from the quasiparticle generated chaotic states. In Appendix~\ref{S:num} we present supplemental material on the behavior of the entanglement spectrum.

\begin{figure}
  \centering
  \includegraphics[width=1.0\linewidth]{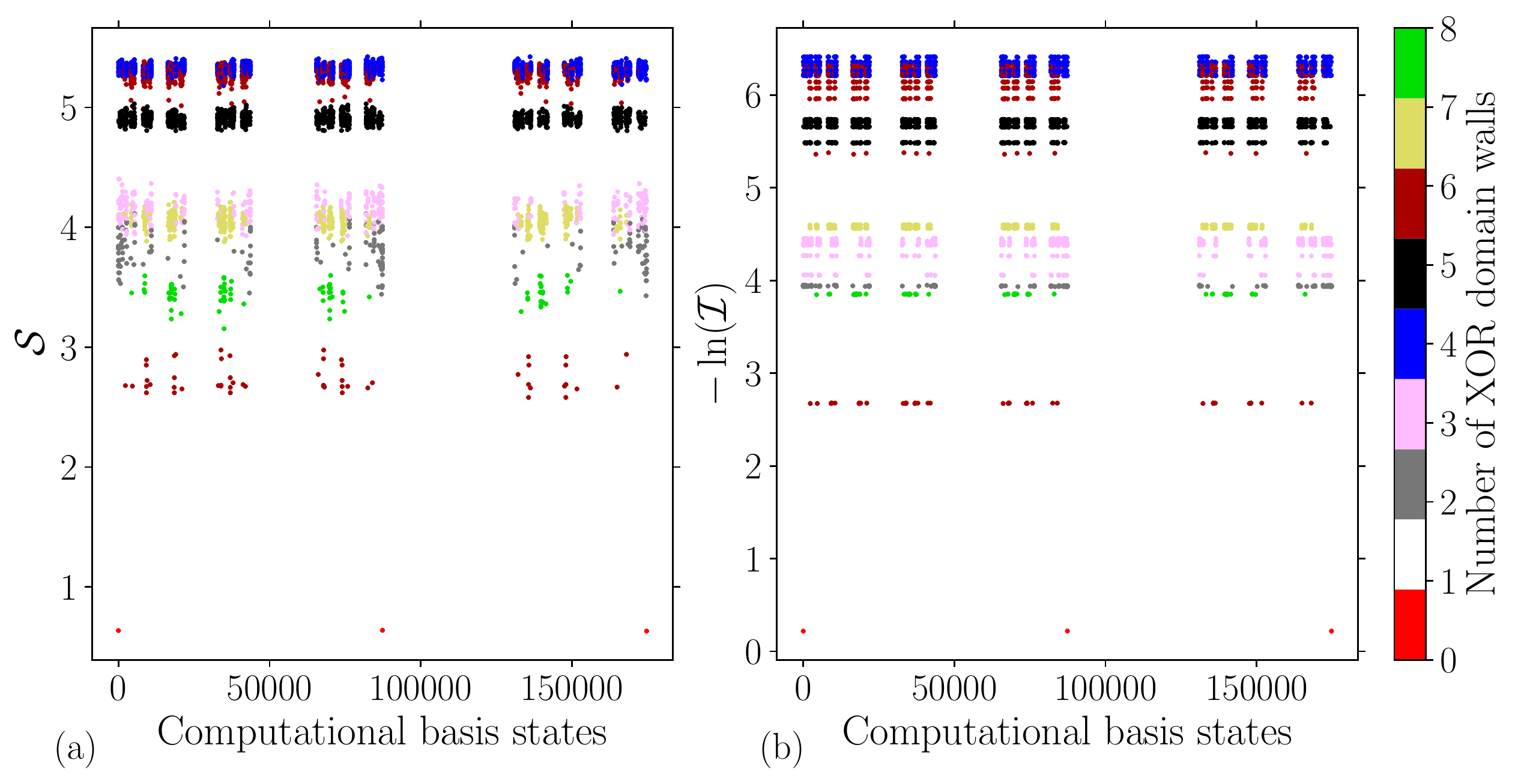}
  \caption{Entanglement entropy \(\mathcal{S}\) and logarithm of inverse participation ratio \(\mathcal{I}\). (a) The entropy is computed for a half bipartition for the set of Fibonacci states and classified by the number of domain walls. (b) Logarithm of inverse participation ratio also showing a layered structure, corresponding to the type of quasiparticles present in the states of the computational basis (qubits configurations). Note that one domain wall defects are not allowed. Parameters: \(N=18\), \(\epsilon=0.01\), measurements are made after \(t=1200\) to ensure that both indicators are in their steady state.
  \label{f:SA}}
\end{figure}

To answer this question we computed the entanglement entropy \eqref{e:vN} for the whole set of initial configurations spanned by the basis vectors in the Fibonacci subspace. We also computed the configuration inverse participation ratio \cite{Giraud-2007,Gopalakrishnan-2018},
\begin{equation}
\label{e:IPR}
\mathcal{I}(t) = \sum_s \big| \braket{s|\psi(t)} \big|^4
\end{equation}
to get information about the geometry of the chaotic states. Note that the usual definition of \(\mathcal{I}\) is in the energy basis. Using the configuration basis, it essentially counts the number of basis vector necessary to represent the state \(\ket{\psi}\); its dependence on the system size gives then information on the fractal dimension of the support \cite{Atas-2012,DeTomasi-2020}. The result is shown in Fig.~\ref{f:SA}. The state \(\ket{\psi(t)}\) was obtained numerically after \(t=1200\) steps to ensure that most configurations reach a steady state, for a system with \(N=18\) qubits, and \(\theta = 0.01\) (see also Appendix~\ref{S:num} for other parameters).

We found that the entanglement entropy organizes in a set of layers according to the number of domain walls contained in the initial state. We classify the Fibonacci states using the XOR representation of Fig.~\ref{f:qp}, into 0 for the vacuum, 2 for the B double wall, 3 for the BC or CB glider, and from 4 to 8 for any combination of these basic quasiparticles. The stratification of the entanglement entropy as a function of the number and type of quasiparticles shows that, even if the entanglement spectrum points to chaotic states satisfying the same statistics, the Hilbert space is fragmented into distinct regions. Note for instance that for 6 domain walls we observe two classes of state with low and high entanglement (c.f.\ the brown points of Fig.~\ref{f:SA}a): they correspond to two comoving gliders (low entanglement) and two countermoving (high entanglement) gliders.

The number of configurations depends on the type of quasiparticles it contains. It is then interesting to measure \(\mathcal{I}\) and correlate it with the entanglement entropy (Fig.~\ref{f:SA}b). We find that each category as defined by its entanglement correlates to a class defined by its participation ratio. In fact \(-\ln \mathcal{I}\) is an upper bound of the second Rényi entropy in the Schmidt basis \cite{Luitz-2014,Gopalakrishnan-2018}. The general trend is that the entanglement entropy increases with the delocalization of the corresponding states; as a consequence double walls dominate gliders: the largest entanglement entropy is generated by two B (or C), which have the smaller \(\mathcal{I}\) (blue dots in Fig.~\ref{f:SA}).

The remarkable stratification of the entanglement and geometry of the states evolved from different initial states, demonstrate that the dynamics of the automaton, away from its integrable limit, is nonergodic. Therefore, the chaotic nonergodic states differ to thermal ones not only because they carry different entanglement amounts but also because of their geometry: their support does not cover the entire Hilbert space but is related to the class of states with a given type of quasiparticles distribution.

In some sense the emergence of nonergodic states is not a surprise, because the quasiparticles are mostly of topological origin, they split the chain into different vacuum sectors. As a consequence, the transformation of a type of domain wall into another type needs some global operation (on an extended set of sites, as discussed in Sec.~\ref{S:qp}). This suggests us that the generated entanglement must be long range.

\begin{figure}
  \centering
  \includegraphics[width=1.0\linewidth]{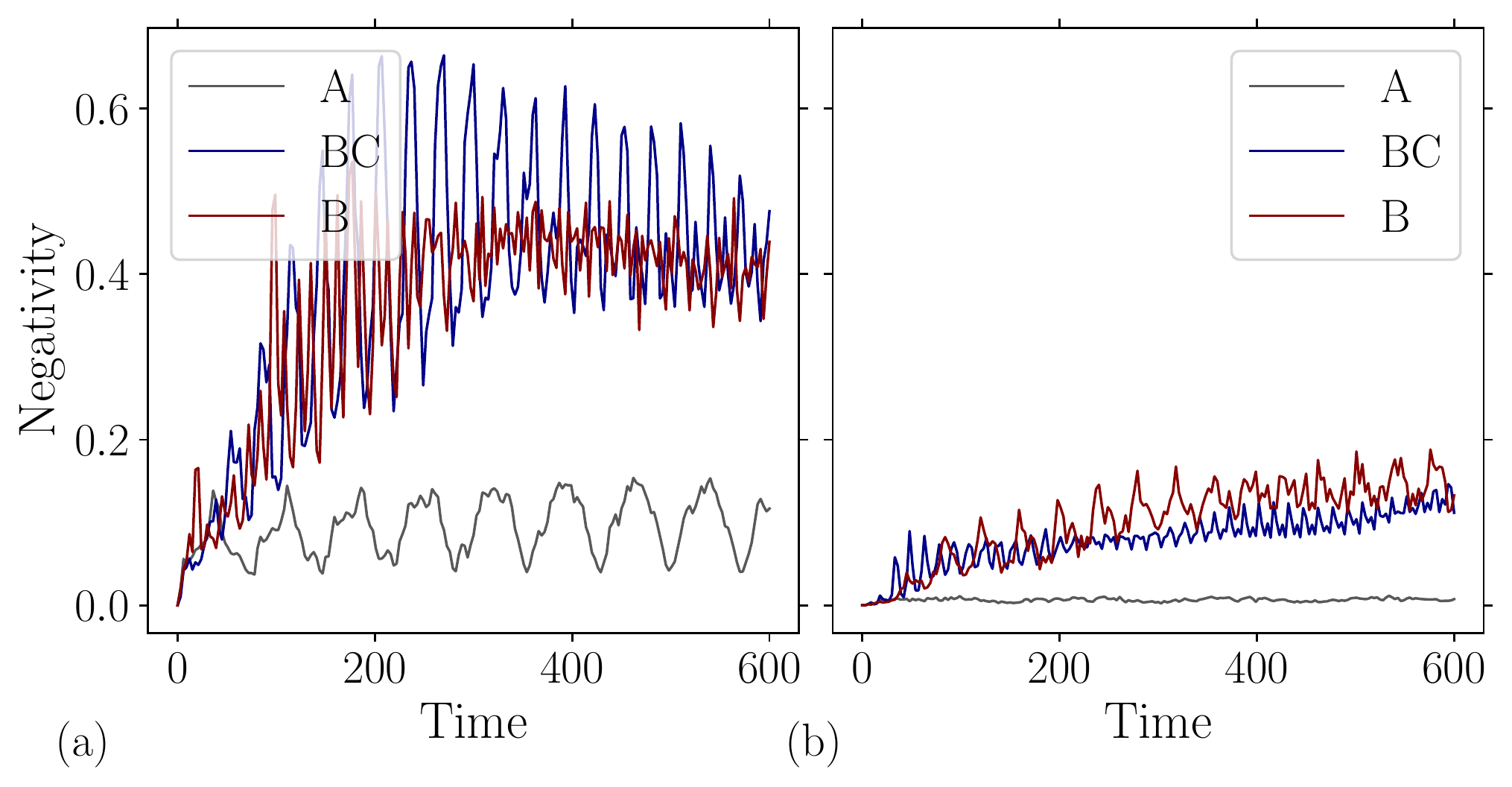}
  \caption{Negativity as a function of time for the evolution of A, BC and C. (a) adjacent subsystems; (b) disjoint subsystems. Long-range entanglement is present in the case of qausiparticles, the initial vacuum state shows revivals and short-range entanglement. Parameters: \(N=20\), \(\epsilon = 0.01\), one step out of three is displayed.
  \label{f:nega}}
\end{figure}

A convenient measure of the entanglement locality associated to a state, depending for instance on the topological order of the corresponding phase, is the negativity \cite{Lee-2013}. It extends the notion of bipartite entanglement to mixed states, allowing us to split the system into a bipartite system `ab' and the background `c'; tracing out c we get a mixed state of subsystem ab, \(\rho_\text{ab}\). The negativity \cite{Vidal-2002zr} is defined in terms of the Peres partial transpose criterion \cite{Peres-1996}
\begin{equation}
\label{e:neg}
\mathcal{N} = \frac{1}{2} \Tr \sqrt{\rho_\text{ab}^{\mathsf{T}_\text{a}} \rho_\text{ab}^{\mathsf{T}_\text{a} \dagger}} - \frac{1}{2}
\end{equation}
where \(\cdot^{\mathsf{T}_\text{a}}\) denotes the partial transposition over subspace a.

We selected states A, BC and B and compared in Fig.~\ref{f:nega} their negativity in the case where ab are adjacent Fig.~\ref{f:nega}a, and the case where the two subsystems are disjoint Fig.~\ref{f:nega}b. When the subsystems are adjacent we retrieve the behavior observed with the entanglement entropy, while, for two disjoints subsystems the entanglement of the vacuum cycle almost vanishes, but it remains finite and keeps growing in the quasiparticle case. This result is coherent with the statement about the topological properties of the quasiparticles in our case. We conclude that the evolved quasiparticle states exhibit long range entanglement, and are therefore nontrivial topological states at high temperature \cite{Choi-2020,Wildeboer-2022}.

\section{Conclusion}
\label{S:concl}

We investigated the dynamics of a nonintegrable automaton inspired by the physics of periodically driven Rydberg atoms arrays, with the goal to generate long-range entangled states. The mechanism of the creation of these states is ergodicity breaking. At variance to simple scars, we focused on chaotic states in the regime where defect-like quasiparticles possess well defined properties, like weak dispersion and weak interactions, leading to an approximate conservation of their number.

Indeed, the conservation of quasiparticles is exact only for the classical automaton; for finite but small \(\epsilon=\pi/2 - \theta\) it is only approximate. However, we observed that for \(\theta\) near the classical limit \(\pi/2\), the interactions of quasiparticles (gliders and domain walls) essentially translates into a shift of the trajectory, without change in their speed, accompanied by a slow dispersion, which implies that the number of quasiparticles is preserved by the dynamics. The effect of the dynamics is to transform an initial product state into a complex superposition of states in the same family of quasiparticle types and number. A perturbation analysis of the gliders confort this scenario.

At long times, we observe a striking fragmentation of the configuration space, the set of states which is invariant with respect to the classical automaton evolution operator and forms a basis of the whole Hilbert space. Depending on their content in quasiparticles, the entanglement and geometry of the evolved states organize into distinct subspaces. When represented as a function of the initial configurations, the von Neumann entropy of the half chain and the inverse participation ratio show a layered structure, well described in terms of the number of domain walls of the corresponding states. 

In addition, most of these nonergodic states satisfy a random matrix statistics typical of chaotic states, in the orthogonal ensemble, in contrast to the Poissonian statistics characteristic of scars-related states. The existence of nonergodic chaotic states was discussed recently. For instance, maximally entangled states, in the sense of Page, were found to have a multifractal structure in the configuration space \cite{DeTomasi-2020,Scherg-2021a}; or chaotic states in a model of quantum walk in a graph were found to have an entanglement entropy, below the Page limit, determined by the cycles structure of the graph \cite{Verga-2019b}. The interest in these nonthermal, highly entangled, and high temperature states, is that they can be useful as information resources \cite{Stephen-2019}. An extension of the present model to automata models with gauge invariance would be interesting \cite{Sellapillay-2022}.

Kitaev \cite{Kitaev-2003fk} introduced the idea that error correction could be physically implemented in the form of a symmetry-protected topological phase, the degenerated and gapped ground state of some Hamiltonian: his toric code is an explicit model now used in quantum computation \cite{Satzinger-2021a}. We have shown in this paper that it is of interest to investigate the possibility of a nonequilibrium quantum system offering quantum information features (topological excitations supporting long-range entanglement). The fragmentation of the Hilbert space into nonergodic sectors was already observed in ``fractonic'' models \cite{Sala-2020,Khemani-2020} and random circuits of unitary gates and measurements \cite{Choi-2020,Li-2021a}, among many other examples. Further work is necessary to determine if the properties of the nonthermal quantum states we obtained using the evolution of the PXP automaton, can also be viewed as a physical robust implementation of quantum information resources.

\begin{acknowledgments}
KS would like to thank Laurent Raymond for his time and for helpful discussions, especially concerning the first order perturbation calculations.
\end{acknowledgments}

\appendix
\section{Glider dispersion relation}
\label{S:glider}

\begin{figure}
  \centering
  \includegraphics[width=1.0\linewidth]{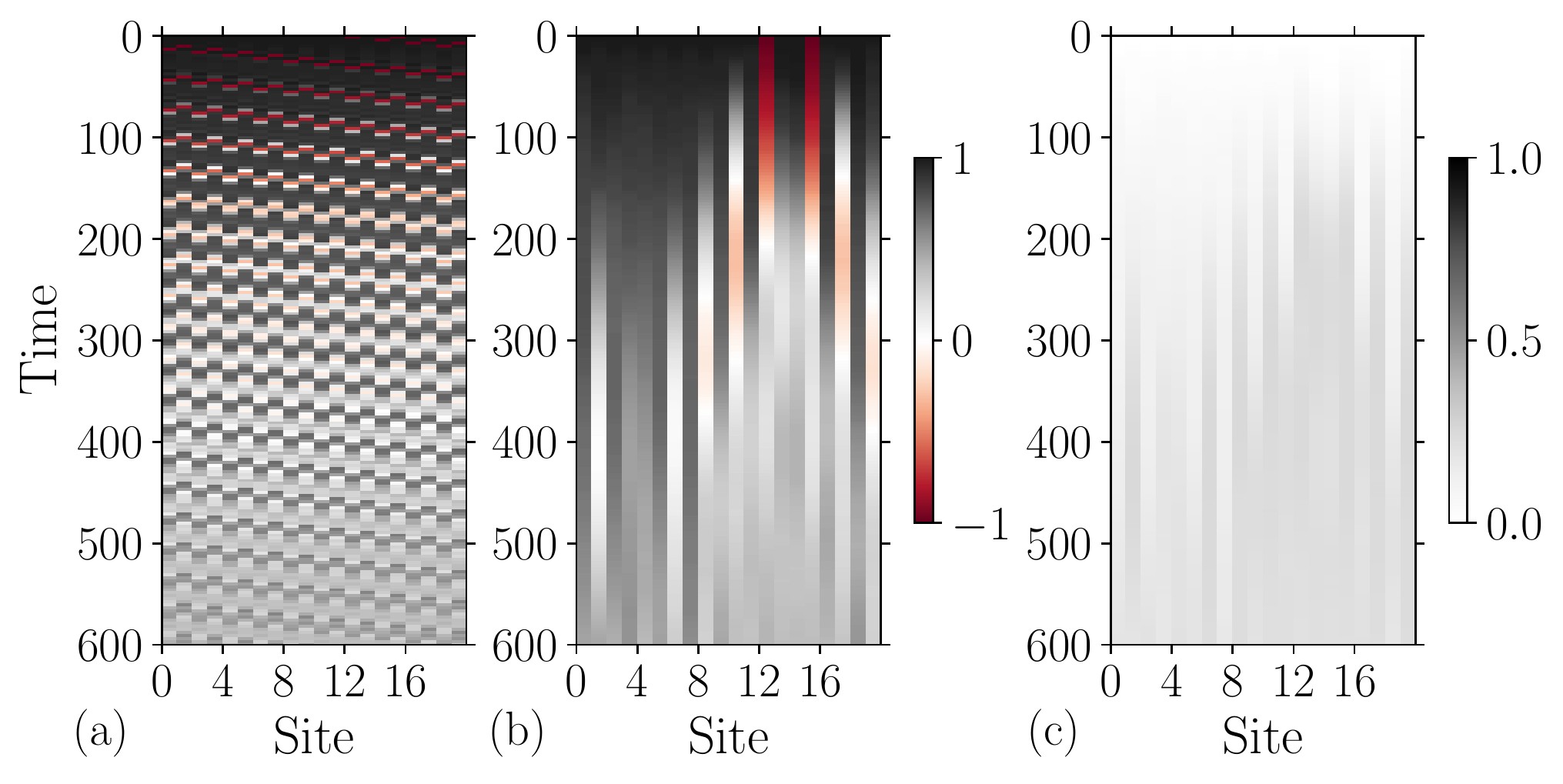}
  \caption{Expected value of the $z$-spin and the corresponding entanglement pattern. (a) Spin in fixed frame; (b) quasiparticle in its comoving frame; and (c) tangle in the quasiparticle comoving frame. Parameters: $N=20$, $\theta=\frac{\pi}{2}-0.01$ $t=600$; we show time every three steps.}
\label{f:spin}
\end{figure}

In this Appendix we compute, using an expansion in powers of \(\epsilon\) the dispersion relation of the BC quasiparticles. We write the perturbation \(U_{\epsilon}=U^{3}({ \pi/2-\epsilon})\) in the gliders subspace. Under \(U_{3}\), these gliders are shifted by two sites, therefore a glider will come back to its original position if we apply \(U^{N/2}_3\). It would be equivalent to manipulate \(U_{1}=U(\pi/2)\) and express the gliders every step instead of every three steps, however it is more convenient to write them every three steps because in this case they are also eigenvectors of \(T^{2}\) (\ref{e:T2}) as can be seen in Fig.~\ref{f:spin}a. Under \(T^{2}\) a glider will also come back to its original position after \(N/2\) applications, therefore the allowed momenta are \(k=0,4\pi/N,\ldots, 4\pi (N/2-1)/N\). 

The computation of the quasienergies and eigenvectors is equivalent for both the right moving BC and the left moving CB quasiparticles. We then define the subspace of the Hilbert space associated to the BC quasiparticles states (Eq.~(\ref{e:lrk}) of the main text),
\begin{equation}
\ket{LRk}=\sqrt{\frac{2}{N}}\sum\limits_{x=0}^{N/2-1}e^{ikx}\prod\limits_{l=0}^{L}X_{2x-2l}\prod\limits_{r=0}^{R}X_{2x+2r+3}\ket{A}.
\label{eq:lrk2}
\end{equation}
For a system size $N$, we have $L$ and $R$ defined in the domain 
\begin{equation}0\leqslant L+R \leqslant N/2-4.
\label{eq:constraint}
\end{equation}
Under $U_{3}$ and $T^{2}$, each state of the sum in (\ref{eq:lrk2}) is shifted by two, therefore $\ket{LRk}$ is an eigenvector of these operators (c.f.\ Eq.~(\ref{e:U3k}))
\begin{equation}
U_{3}\ket{LRk}=T^{2}\ket{LRk}=e^{-ik} \ket{LRk}.
\end{equation}
This subspace is protected from mixing to other states to first order in \(\epsilon\) because other states are mixtures of quasiparticles of different nature (CB, B, C, or containing more than one quasiparticle). Some of them might have the same cycle length $\ell=N/2$ under $U_{3}$, but the first order perturbation which consists in one flip as we see below, cannot connect a BC glider to them as it would have to change more than one site per state in the orbit to replicate the orbit of a quasiparticle of another nature.

To find eigenvectors of the perturbed operator we first find the perturbation to first order by going in the reference frame of the BC quasiparticle, which is equivalent to factorizing the classical automaton that moves the chiral quasiparticle to the right by two
\begin{equation}
U_{\epsilon}=U_{3}U_{3}^{\dagger}U_{\epsilon} \approx U_{3}(1-i\epsilon H_{BC}),
\end{equation}
where
\begin{equation}
\begin{split}\label{eq:1}
H_{BC}&=
U^{\dagger}_{3}V_{o}U_{3}+
U^{\dagger}_{2}V_{o}U_{2}\\&+
U^{\dagger}_{1}V_{o}U_{1}+
U^{\dagger}_{2}V_{e}U_{2}\\&+
U^{\dagger}_{1}V_{e}U_{1}+
V_{e}
,
\end{split}
\end{equation}
we obtained after an expansion to first order.

We now compute the action of $H_{BC}$ on the $\ket{LRk}$. For this end, we first compute $U_{1}\ket{LRk}$ and $U_{2}\ket{LRk}$:
\begin{multline}
U_{1}\ket{LRk}=(-i)^{N/2+L} \sum\limits_{x=0}^{N/2-1} e^{ikx} \prod\limits_{z=0}^{L}X_{2x-2z+1}\\ \prod\limits_{\substack{p \in \text{e} \backslash \\ \{ 2x-2z | z=0,\ldots,L\} \cup \\ \{ 2x+2z+2 | z=0,\ldots,R+1\}} }X_{p}\ket{A},
\end{multline}
and
\begin{multline}
U_{2}\ket{LRk}=(-i)^{3N/2-(R+4)} \sum\limits_{x=0}^{N/2-1} e^{ikx} \\
\prod\limits_{\substack{k \in \text{o} \backslash \\ \{ 2x-2z+1 | z=0,\ldots,L\} \cup \\ \{ 2x+2z+3 | z=0,\ldots,R+1\}} }X_{k}  \prod\limits_{z=1}^{R+1}X_{2x+2z+2}\ket{A}.
\end{multline}
For each term of $H_{BC}$ we compute $\bra{\tilde{L} \tilde{R} \tilde{k}} H_{BC} \ket{LRk}$, with $0\leq \tilde{L} + \tilde{R} \leq N/2-4$ and $\tilde{k} = 0, 4\pi/N,\ldots, 4\pi (N/2-1)/N$.

\begin{figure}
  \centering
  \includegraphics[width=1.0\linewidth]{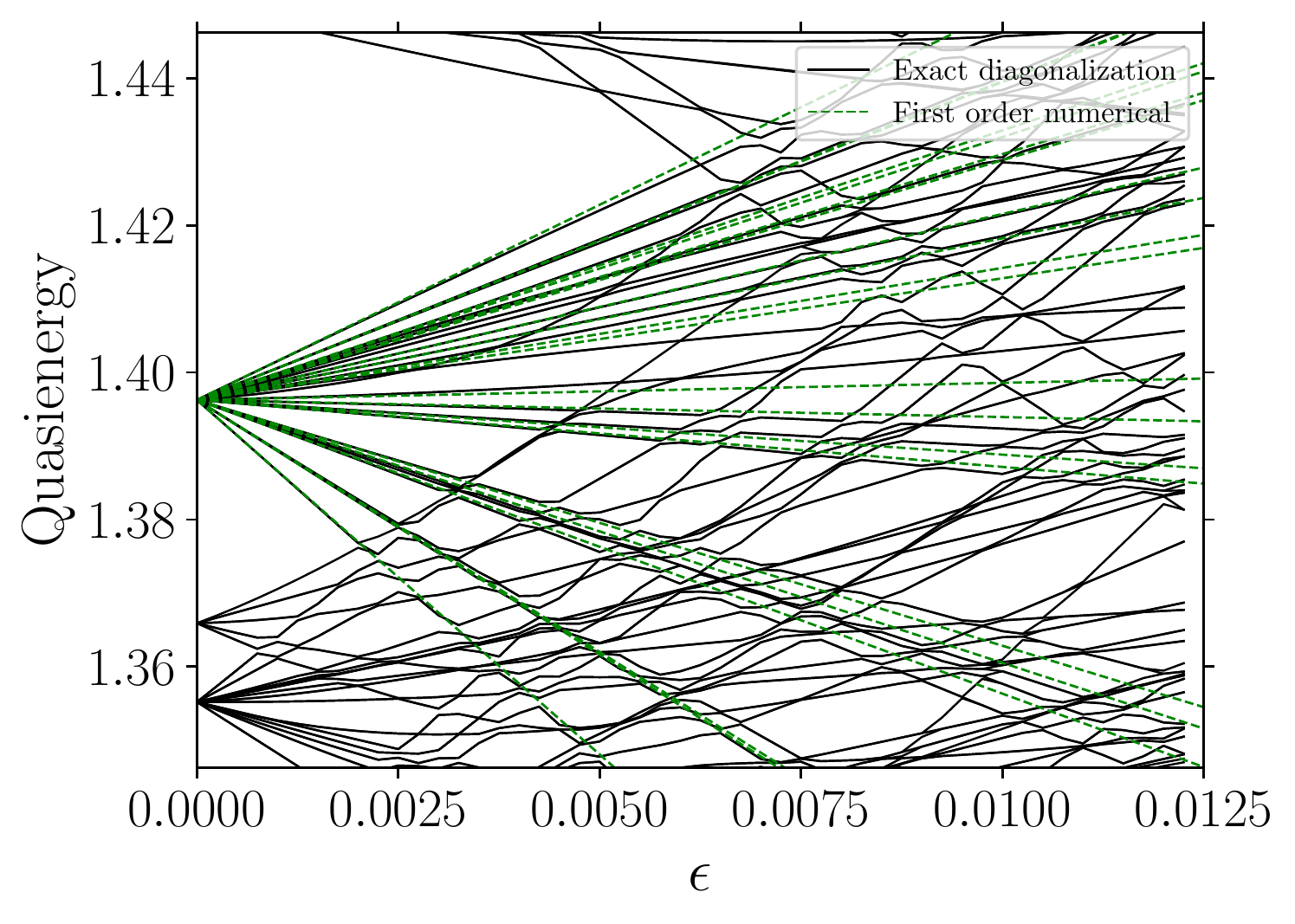}
  \caption{Quasienergy as a function of the perturbation $\epsilon$.  (Black continuous line) Exact quasienergies of $U_{\epsilon}$ using diagonalization . (Green dashed line) Numerical diagonalization of the perturbation matrix (\ref{eq:pertur}) . The parameters are $k=4\pi /9$ and $N=18$.}
\label{f:ekeps}
\end{figure}

We write the non-zero matrix elements in the $\ket{LRk}$ subspace. Observing that $[H_{BC},T^{2}] = 0$ we can do the computation with $k = \tilde{k}$.  We obtain the following expressions 
\begin{equation}
\begin{gathered}
\bra{L\pm1 R k} H_{BC}\ket{LRk}= 2\\
\bra{LR\pm1k} H_{BC}\ket{LRk}= 2\\
\bra{L-1R+1k} H_{BC}\ket{LRk}=-2 ie^{ik}\\
\bra{L+1R-1k} H_{BC}\ket{LRk}=2 ie^{-ik},
\label{eq:pertur}
\end{gathered}
\end{equation}
on the triangular domain given by \eqref{eq:constraint}.

To diagonalize this Hamiltonian in the $\ket{LRk}$ basis, we conveniently extend the original triangular domain of $(L,R)$ to all values $0\leq L, R\leq  N/2-4$, which thus becomes a square. We remark that the matrix elements of the effective Hamiltonian are independent of $L,R$. The states outside \eqref{eq:constraint} are unphysical, we temporarily use them to diagonalize the Hamiltonian. Furthermore we impose periodic boundary conditions on both $L$ and $R$ axis, and use the following ansatz
\begin{equation}
\ket{q_{1}q_{2}k} = \mathcal{N} \sum_{L=0}^{N/2-4} \sum_{R=0}^{N/2-4} e^{iq_{1}L} e^{iq_{2}R} \ket{LRk},
\end{equation}
with 
\begin{equation}
\begin{gathered}
\mathcal{N} = \frac{1}{N/2-4} \rightarrow \frac{2}{N}\\
q_{1,2} = \frac{2\pi n_{1,2}}{N/2-3}\rightarrow [0,2\pi]\\
k = \frac{4\pi n}{N}\rightarrow [0,2\pi]\\
n_{1,2} = 0,\ldots,\frac{N}{2}-4\\
n = 0,\ldots,\frac{N}{2}-1,
\end{gathered}
\end{equation}
where the large $N$ limit is taken for the first four equations and is assumed in the rest of the computation. Below, we verify numerically these approximations.

\begin{figure}
  \centering
  \includegraphics[width=1.0\linewidth]{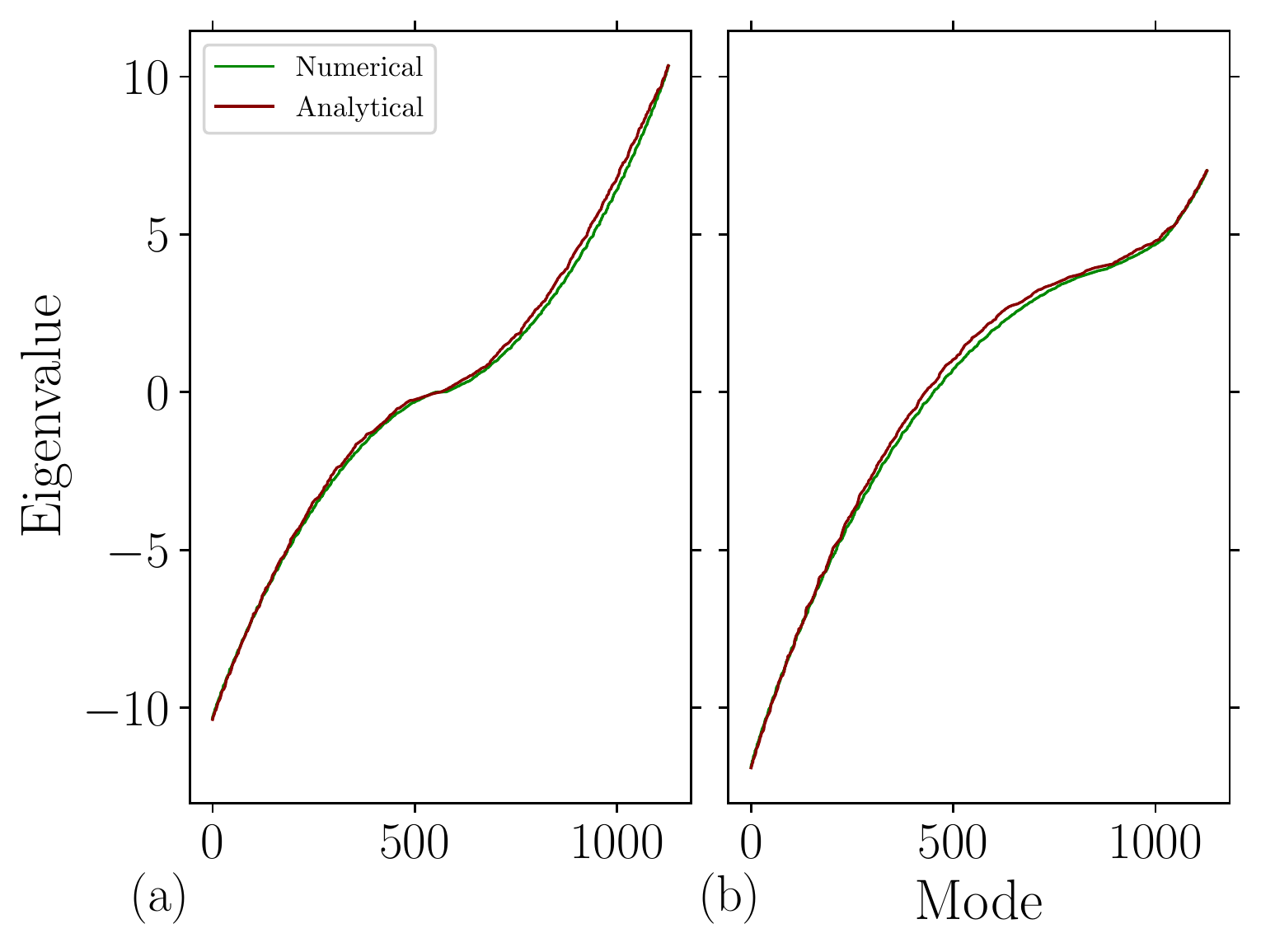}
  \caption{Sorted first order eigenvalues for $N=100$. (a) $k=0$  and (b) $k=160\pi/100$. We compare the results of the numerical diagonalization of the perturbation matrix (\ref{eq:pertur}) and the analytical ones obtained using Eq.~(\ref{eq:energies}).}
\label{f:ekN}
\end{figure}

The parameters $\bm{q} = (q_{1}, q_{2})$ also define a square domain just as the $(L,R)$ ones. At the end of the calculation we will keep only the momenta defined in the triangular domain corresponding to $0\leq n_{1}+n_{2}\leq N/2-4$ in order to have the same number of eigenvalues as initially. This can be done due to the symmetry along the diagonal of the square. In the large $N$ limit, this corresponds in a Brillouin zone given by $0 \leq q_{1}+q_{2} \leq 2\pi$. We apply $H_{BC}$ on $\ket{q_{1}q_{2}k}$, we find 
\begin{equation}
\label{eq:energies}
H_{BC}\ket{q_{1}q_{2}k} = E_{\bm{q}}^{(1)}(k) \ket{q_{1}q_{2}k},
\end{equation}
where
\begin{equation}
\label{e:Ek1}
E_{\bm{q}}^{(1)}(k) = 4[\cos q_{1} + \cos q_{2}-\sin(q_{2}-q_{1}-k)] \ket{q_{1}q_{2}k}.
\end{equation}
Therefore we have a modified definition of quasiparticles due to the perturbation with the following dispersion relation
\begin{align}
\begin{split}
U_{\epsilon}\ket{q_{1}q_{2}k} &= U_{3}U_{3}^{\dagger}U_{\epsilon} \ket{q_{1}q_{2}k}\\
&\simeq U_{3}(1-i\epsilon H_{BC}) \ket{q_{1}q_{2}k}\\
&= U_{3}(1-i\epsilon E_{\bm{q}}^{(1)}(k) )\ket{q_{1}q_{2}k}\\
&\simeq e^{-ik}e^{-i \epsilon E_{\bm{q}}^{(1)}(k)}\ket{q_{1}q_{2}k}\\
&=e^{-i(k+ \epsilon E_{\bm{q}}^{(1)}(k) )}\ket{q_{1}q_{2}k}.
  \label{eq:u_energy}
\end{split}
\end{align}
The dispersion relation for the right-moving quasiparticle is thus given by 
\begin{equation}
E_{\bm{q}}(k)=k+4\epsilon[\cos q_{1} + \cos q_{2} - \sin(q_{2}-q_{1}-k)],
\end{equation}
written in units of length equal to 2 and time equal to 3.

We now compare the eigenvalues of the perturbation matrix and the numerical eigenvalues. We show in Fig.~\ref{f:ekeps} a plot of the eigenvalues obtained form the exact matrix diagonalization of a $N=18$ system. In this case we have 21 eigenvalues given by the dimension of the triangular domain defined by $0\leq L+R\leq N/2-4$. We observe that, for this relatively small number of spins, the dependency on $\epsilon$ is indeed linear validating the use of the power series, however, already for $\epsilon < 10^{-2}$ some eigenvalue crossings appear, breaking the perturbation series. 

In addition, formula \eqref{eq:energies}, assumes that $N$ is large enough such that the boundary in the $\bm q$ plane is negligible. It is easy to estimate the size of a system for which the number of modes in the perimeter of the triangle will be smaller than the ones in its bulk, it is given by 
$$3(N/2-3) \ll  \frac{1}{2} (N/2-3)(N/2-2),$$
which leads to the criterion $N \gg 16$. We then verify that the analytical formula (\ref{eq:energies}) fits the numerical eigenvalues of the perturbation matrix \eqref{eq:pertur}, in the limit of large systems. Indeed, by extending the domain and imposing periodicity we increase the number of the degrees of freedom contributing from the bulk of the triangular domain, with respect to its boundary. In Fig.~\ref{f:ekN} we display the comparison of the analytical distribution of eigenvalues and the numerical ones for a large system ($N=100$) in two $k$ sectors, and obtain a satisfactory matching.

\begin{figure}
  \centering
  \includegraphics[width=0.8\linewidth]{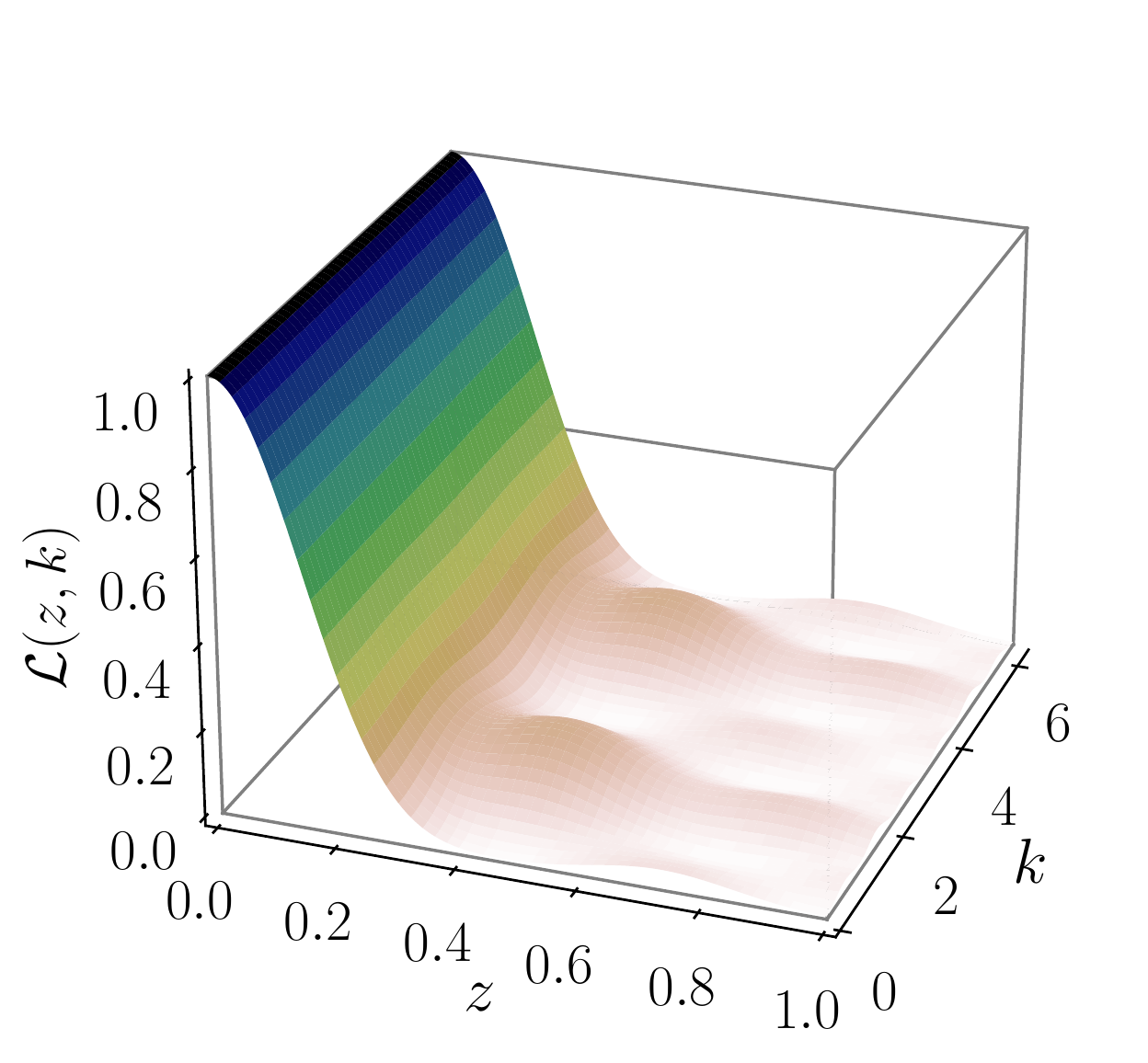} 
  \caption{Loschmidt echo numerical integration of the perturbation theory formula \eqref{eq:LO_exact} as a function of $k$.  \label{f:LO}}
\end{figure} 

Using the previous results, eigenvectors and dispersion relation, we can compute the Loschmidt echo, as defined by \eqref{e:losch}. In the limit of large \(N\) we replace the summations by integrals, 
\begin{equation}
\ket{LRk} = \int_{0}^{2\pi} \frac{dq_{1}}{2\pi} \int_{0}^{2\pi-q_{1}} \frac{dq_{2}}{\pi} e^{-iq_{1}L}e^{-iq_{2}R}\ket{q_{1}q_{2}k},
\end{equation}
and from \eqref{eq:u_energy}, we find
\begin{equation}
\mathcal{L}(z,k)=\left| \frac{1}{2\pi^{2}}\int_{0}^{2\pi}dq_{1}\int_{0}^{2\pi-q_{1}}dq_{2} 
e^{i z  E_{\bm{q}}^{(1)}(k)} \right|^{2},
\label{eq:LO_exact}
\end{equation}
where $z=t(\epsilon_{2}-\epsilon_{1})$, Fig.~\ref{f:LO} gives a representation of this function. A series expansion in powers of $z$ gives 
\begin{equation}
\mathcal{L}(z,k) \approx 1-24z^{2} \approx e^{-24 z^{2}}.
\end{equation}
We recover a short-time parabolic decay which can be approximated by a Gaussian decay \cite{Quan-2006}. The computation of the integral \eqref{eq:LO_exact} is in good agreement with the numerical computation using the full unitary evolution operator, as shown in Fig.~\ref{f:LOsimu}. This comparison also shows indirectly the usefulness of the perturbation dispersion relation as a qualitative guide of the structure of the quasiparticle spectrum, even if it includes some uncontrolled approximations. 

\section{Numerical complements}
\label{S:num}

\begin{figure}
  \centering
  \includegraphics[width=1.0\linewidth]{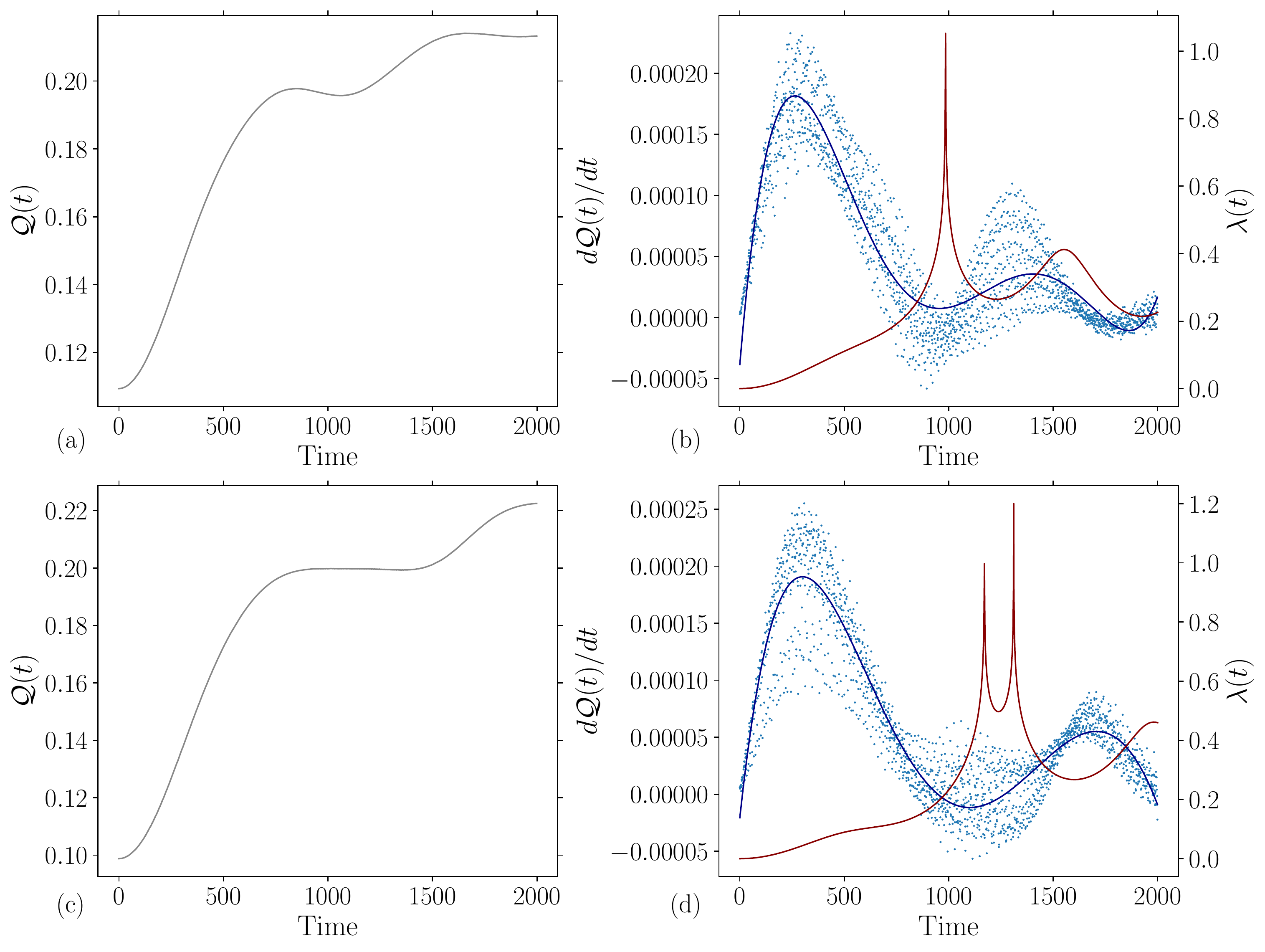}
  \caption{(a,c) $\mathcal{Q}(t)$ entanglement measure. (b,d) Coincidence of the peaks of $\lambda(t)$ (red) with the minimum of the derivative of $d\mathcal{Q}/dt$ (blue).  Parameters are (a,b) $N=16$, (c,d) $N=18$, $\epsilon_{1}=0$, $\epsilon_{2}=0.001$, $t=2000$ and initial condition $\ket{000}$.
  \label{f:logLO}}
\end{figure}

\begin{figure}
  \centering
  \includegraphics[width=0.5\linewidth]{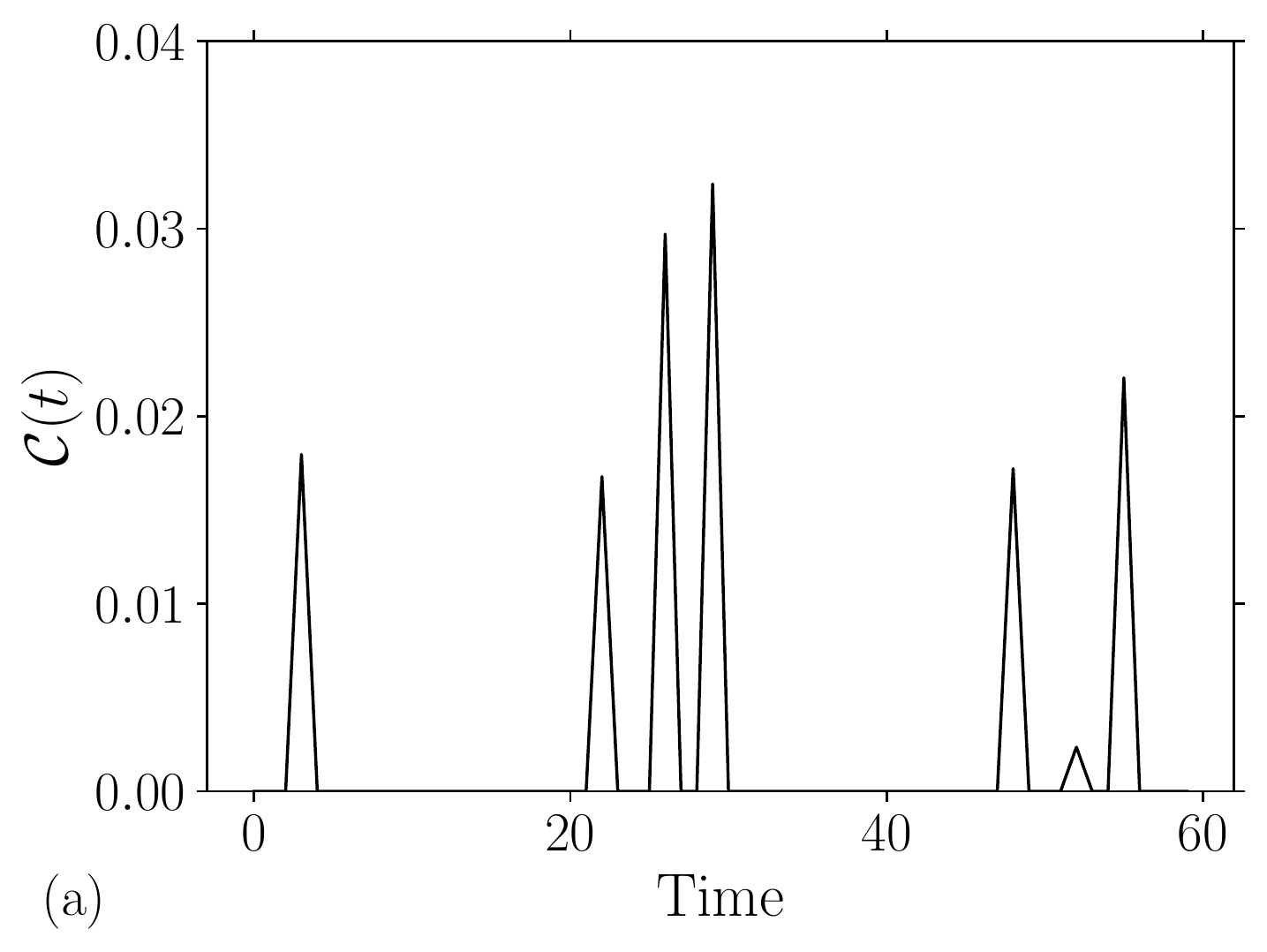}
  \includegraphics[width=0.45\linewidth]{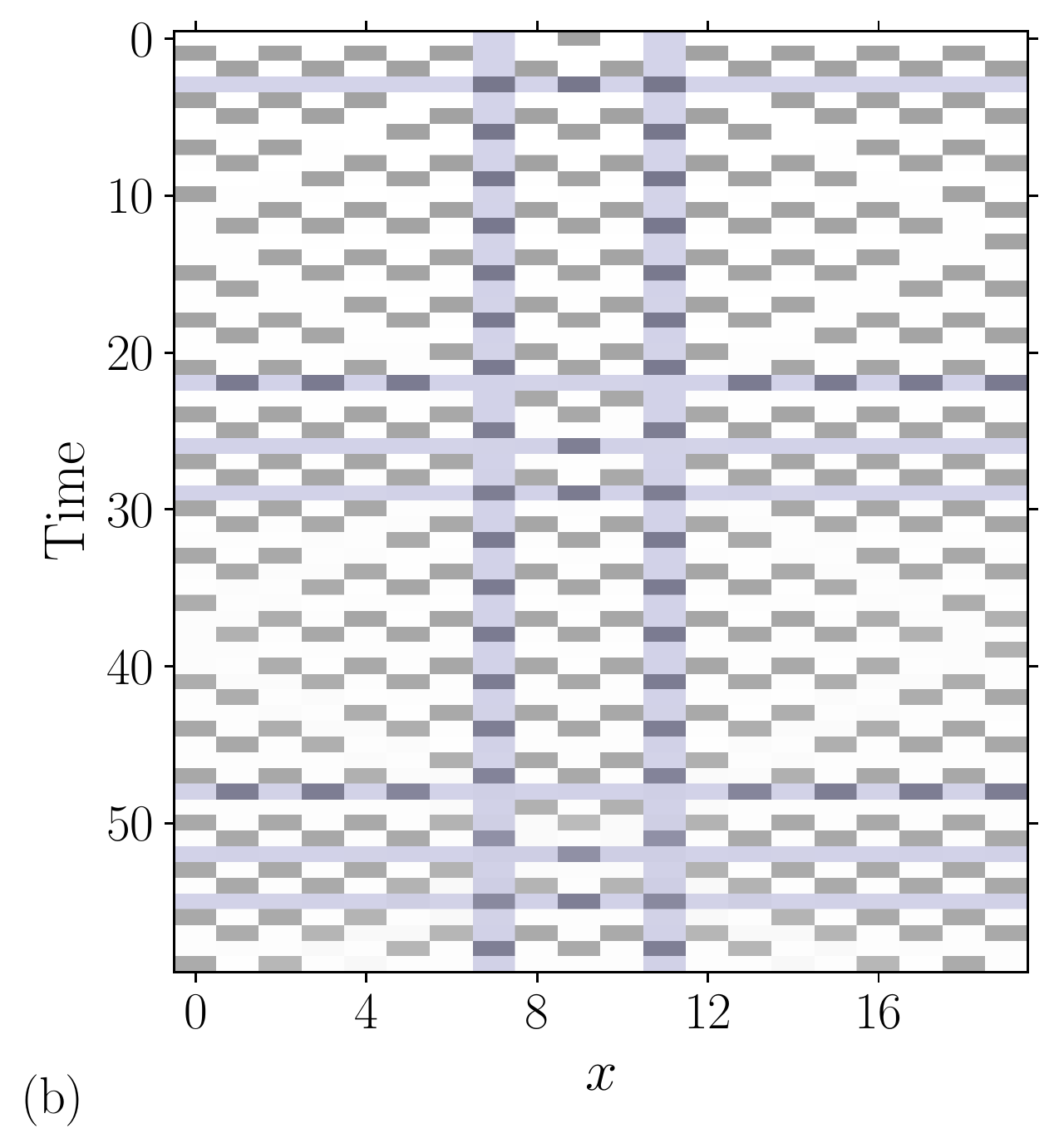}
  \caption{Concurrence between spins 7 and 11 generated by a B quasiparticle. (a) The peaks correspond to the passage time of the quasiparticle. (b) quasiparticle motion (note the existence of some dispersion); the 7 and 11 spins, and the passage times are underlined; each intersection correspond to a peak in (a). Parameters: \(N=20\) and \(\epsilon=0.01\).
  \label{f:conc2}}
\end{figure}

In Fig.~\ref{f:LOsimu}, the Loschmidt echo was numerically studied for different system sizes and showed agreement with the analytical curve. The Loschmidt echo is expected to behave as
\begin{equation}
\mathcal{L}(t)=e^{-N\lambda(t)},
\end{equation}
for large $N$. The quantity $\lambda(t)$ is the rate function \cite{Heyl-2018}, and it is intensive. Figure~\ref{f:logLO} shows the Loschmidt rate function computed for the initial state $\ket{L=0,R=0,k=0}$ and system sizes $N=16,18$. In order to be able to compare $\lambda(t)$ with the entanglement measure $\mathcal{Q}(t)$, we put $\epsilon_{1}=0$ (automaton limit), $\epsilon_{2}=\epsilon=0.001$. We also compute the global entanglement measure $\mathcal{Q}(t)$ for the same state evolution $U^{t}_{\epsilon} \ket{000}$.

We observe the appearance of singularities in $\lambda(t)$. These kinks are usually interpreted as signs of dynamical quantum phase transition \cite{Heyl-2018}. We find that they coincide with local minima of the time derivative of the entanglement measure. In the case of $N=18$ the two peaks are in a plateau of $\mathcal{Q}$, making more difficult their identification with inflexion points. Therefore, the change between the initial low entanglement state and the large time high entanglement one can be thought as the result of a dynamical transition, well captured by the Loschmidt ratio.

\begin{figure}
\centering
  \includegraphics[width=1.0\linewidth]{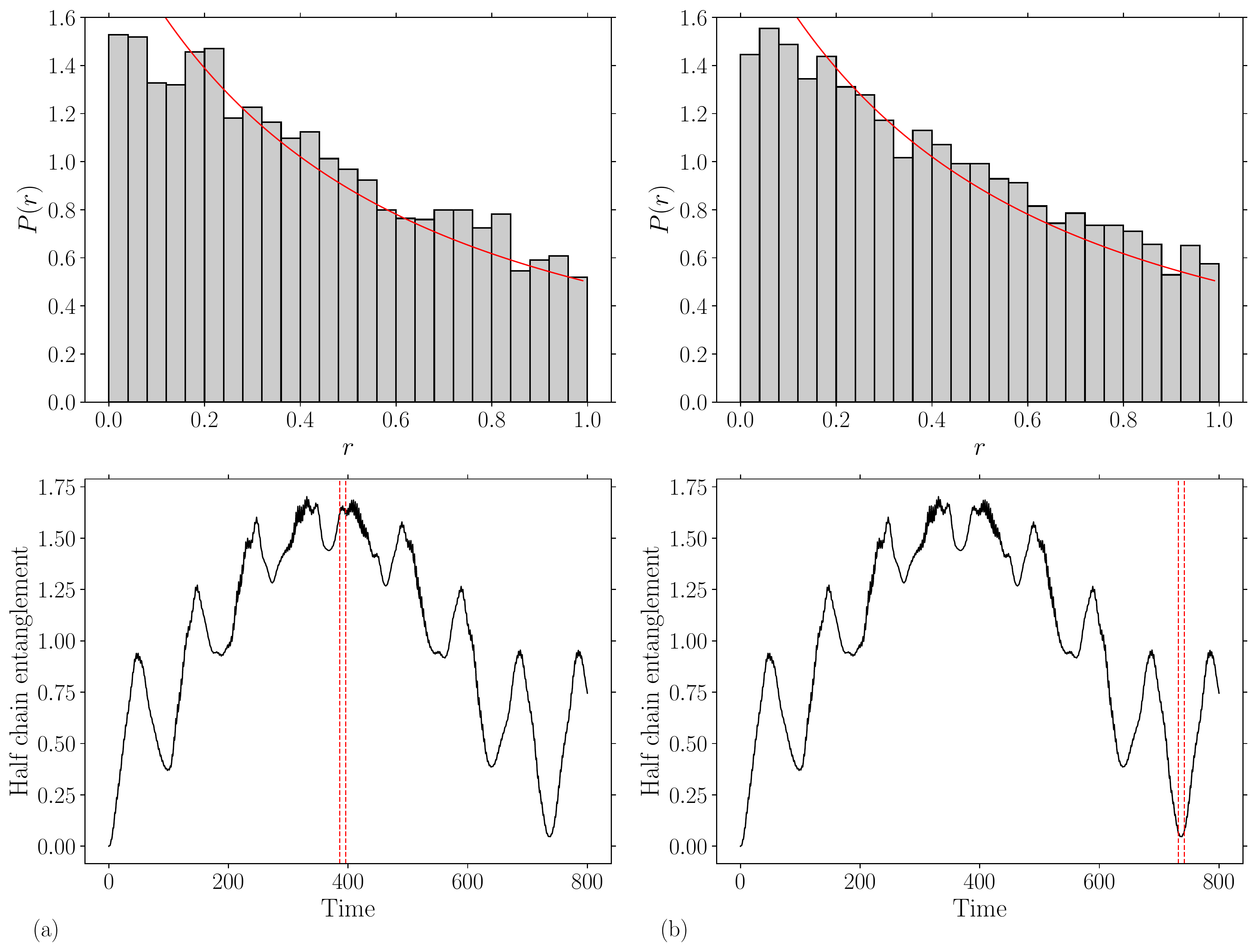}
  \caption{Half chain entanglement spectrum spacings statistic for $\ket{A}$ initial state $N=18$, $\epsilon=0.01$. The data is taken on 10 spectra (a) near the peak of half chain entanglement (b) in the local minimum of half chain entanglement.}
\label{f:HCEpeaks}
\end{figure}

\begin{figure}
\centering
  \includegraphics[width=0.7\linewidth]{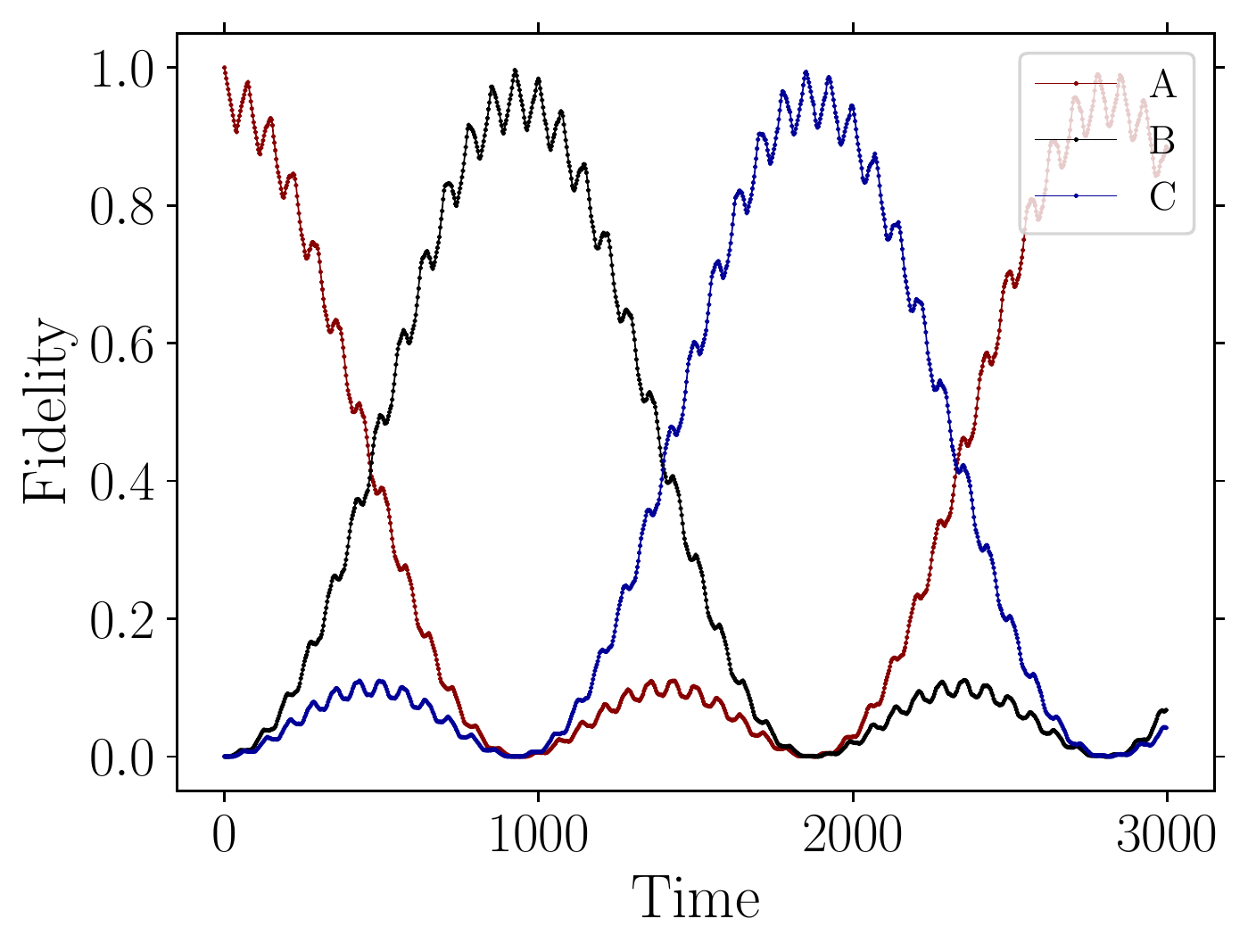}
  \caption{Fidelity of $\ket{A}$ for $N=20$, $\epsilon=0.01$ and $t=3000$. }
\label{f:fiderevivals}
\end{figure}

In Fig~\ref{f:conc} we showed the concurrence generated by a BC glider between spins separated by one site. In Fig.~\ref{f:conc2} we plot the same quantity for the B quasiparticle with spins separated by three sites. The peaks of concurrence coincide with the passage of the B quasiparticle. These quasiparticles are different in nature than the gliders because they are two waves going in opposite directions whereas the glider can be considered as one wave going in one direction. Because of the periodic boundary conditions the two waves interact at each cycle. As a consequence, we observe that entanglement power of the B (C) quasiparticles is larger that the one of the gliders. In the gliders case isolated peaks produce at a given pair of sites, whereas in the cases of B we find bunches of peaks. This mechanisms can also favor the generation of long range entanglement.

In Fig.~\ref{f:chaos}c the half chain entanglement was found to oscillate for an A initial vacuum state, and the corresponding histogram of the spacings between levels of the entanglement spectrum was close to a Poisson distribution. These oscillations are reminiscent of the scarring phenomenon in the PXP Hamiltonian. It is then natural to ask whether the entanglement spectrum level statistics depends on the interval of time in which we collected the data. In Fig.~\ref{f:HCEpeaks} we present two histograms obtained at two different phases of the entanglement oscillation, around its maximum and around its minimum, and found that in both cases the level statistics always approximately follows a Poisson distribution. Disagreement is larger at very small spacings, showing in sensibility to the emergence of level crossings (c.f.\ Fig.\ref{f:ekeps}).

\begin{figure}
\centering
  \includegraphics[width=0.7\linewidth]{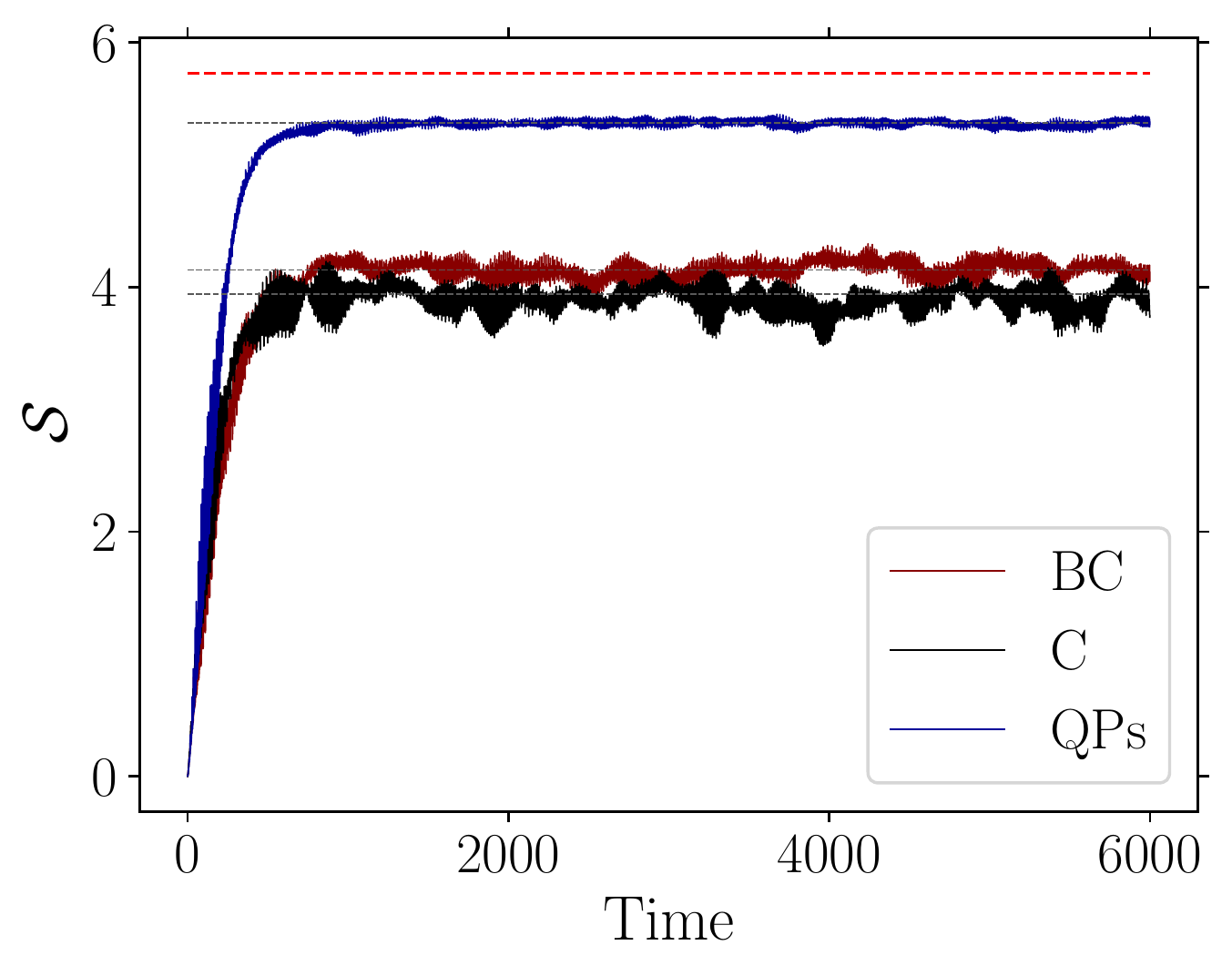}
  \caption{Long time half-chain entanglement for BC, C and interacting quasiparticles $N=18$, $\epsilon=0.01$ and $t=6000$. }
\label{f:longtimehce}
\end{figure}

\begin{figure}
\centering
  \includegraphics[width=0.7\linewidth]{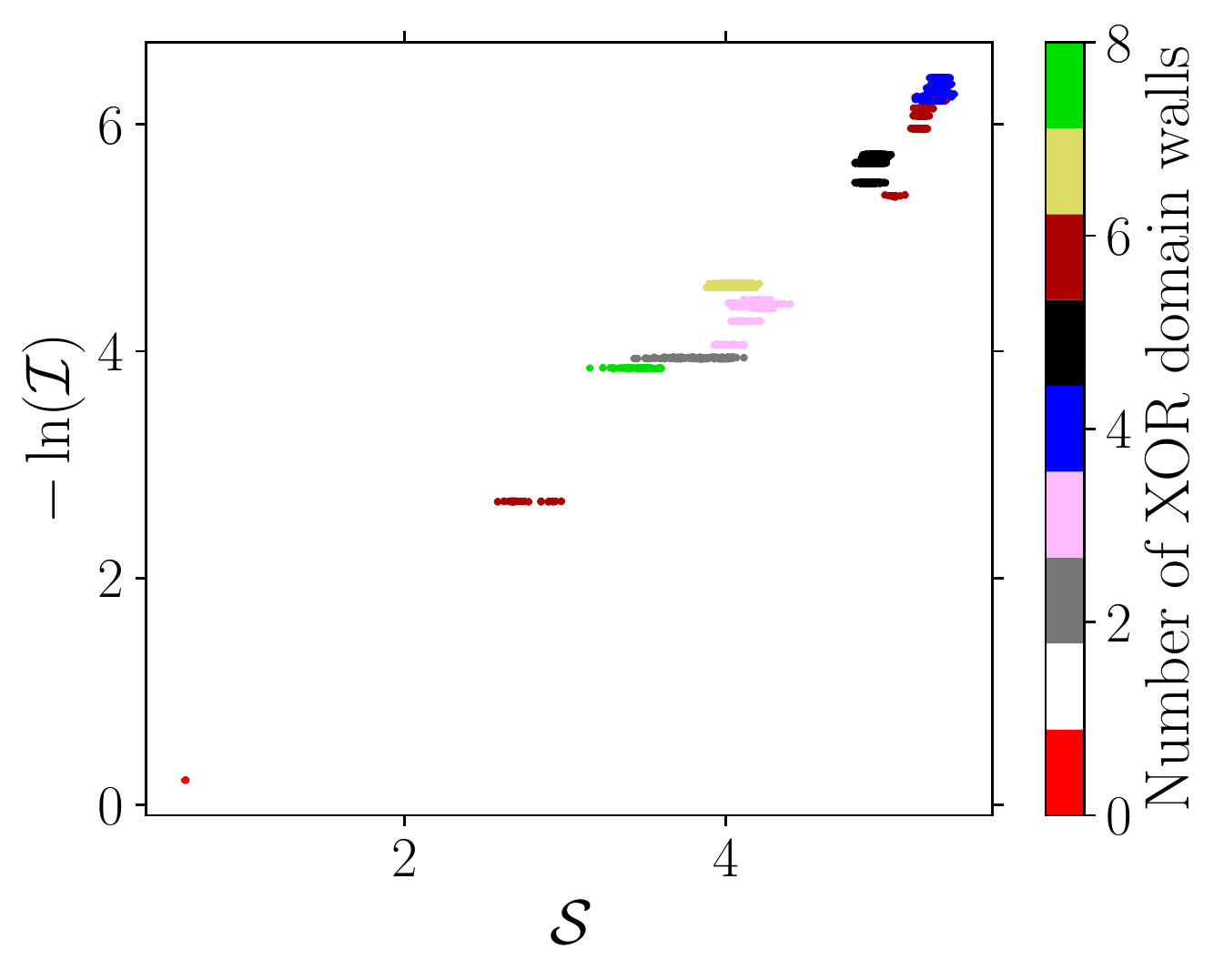}
  \caption{Logarithm of the inverse participation ratio as a function of the half-chain entanglement entropy, for all evolved spin configuration states. Parameters: $N=18$, $\epsilon=0.01$ and $t=1200$. }
\label{f:log001}
\end{figure}

To better understand this oscillation we compute the fidelity $\mathcal{F}(t)$ defined as
\begin{equation}
\mathcal{F}(t)=\lvert \braket{\psi(t)|A}\rvert^{2}
\end{equation}
Figure~\ref{f:fiderevivals} shows the revival phenomenon, typical of the scars states in the PXP limit of the model (for $\theta \rightarrow 0$). Just like the scars, the states seem to live in a small subspace and does not spread to the whole Hilbert space. We note that due to the fact that the vacuum cycle of the classical automaton (we are near $\theta = \pi/2$), the fidelity splits into three distinct oscillations, labeled A, B and C. Moreover, contrary to the PXP model where there is no perfect revivals of the state because the peaks slowly decay in time, here we find almost perfect revivals, much as of the exact scarring observed in other Floquet systems \cite{Mizuta-2020,Sugiura-2021}.

\begin{figure}
\centering
  \includegraphics[width=1.0\linewidth]{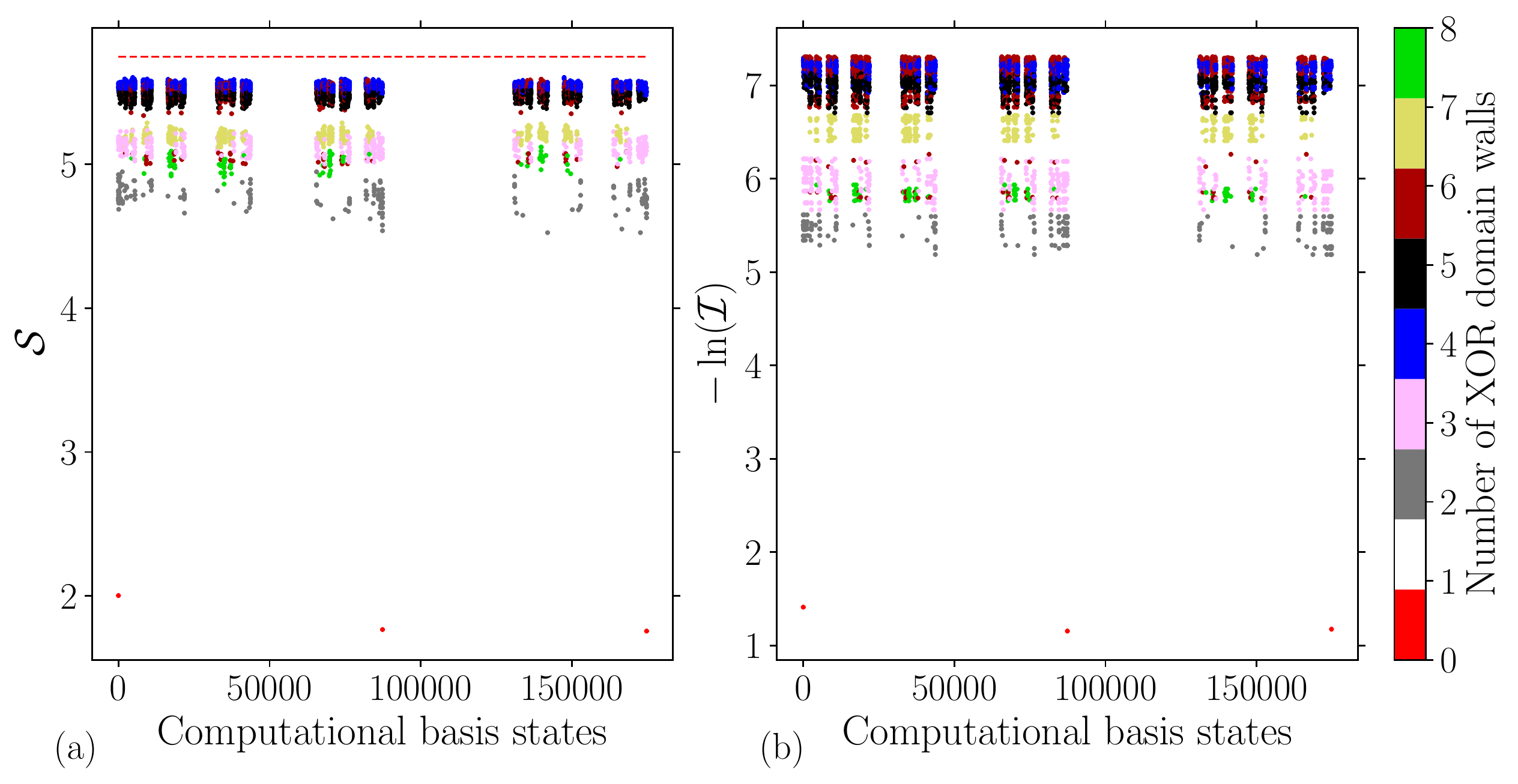}
  \caption{Half-chain entanglement and logarithm of inverse participation ratio of all evolved computational basis states for $N=18$, $\epsilon=0.1$ and $t=600$. }
\label{f:hcelog01}
\end{figure}

\begin{figure}
\centering
  \includegraphics[width=0.7\linewidth]{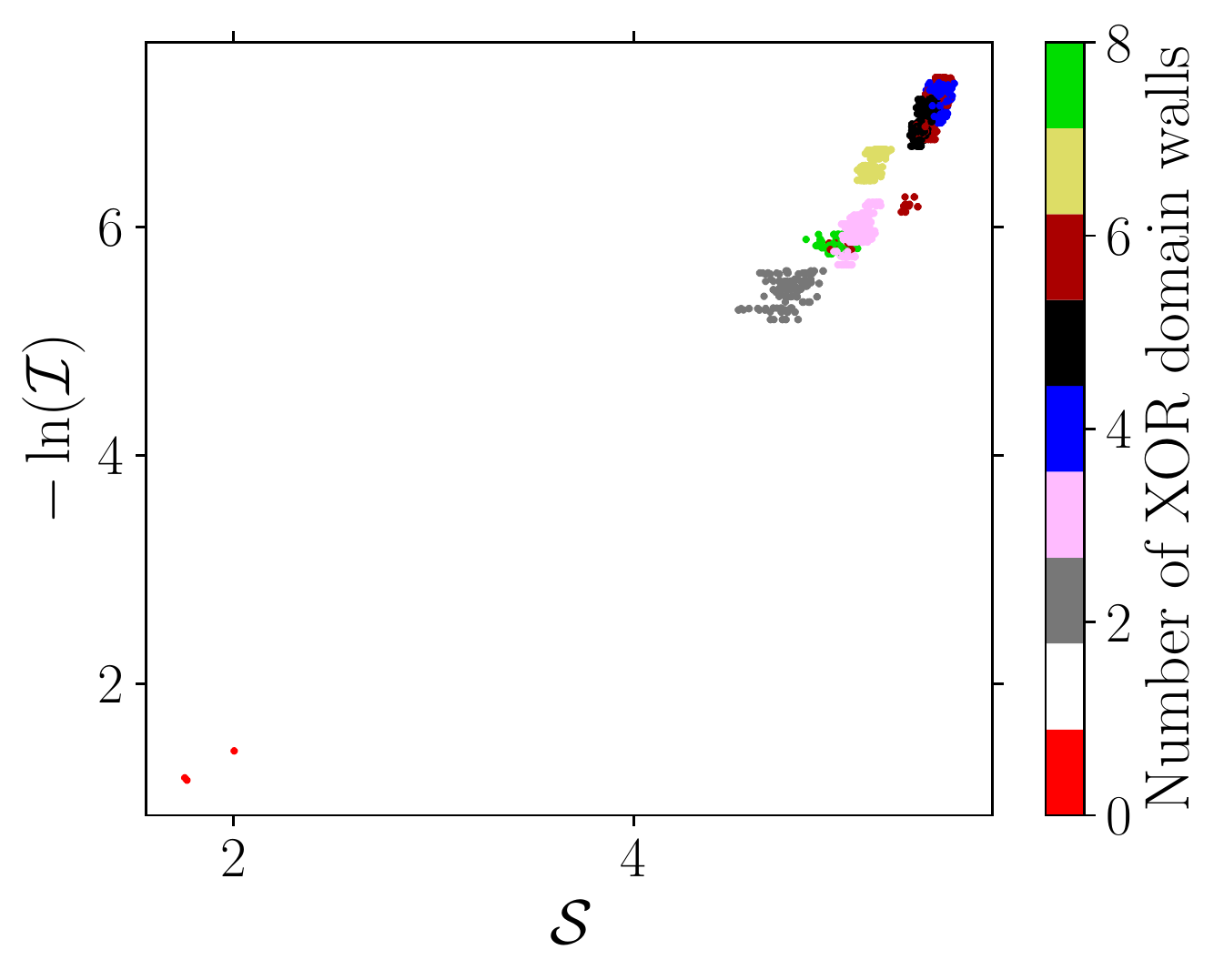}
  \caption{Logarithm of the inverse participation ratio as a function of half-chain entanglement for $N=18$, $\epsilon=0.1$ and $t=600$. }
\label{f:log01}
\end{figure}

In Fig.~\ref{f:SA}, we showed the dependency of the entanglement, as measured by the von Neumann entropy of a half chain, on the set of initial product states (configuration basis). More precisely, \(\mathcal{S}(t)\) was obtained by evolving all states of the Fibonacci subspace for a given lattice size, and fixed values of $\epsilon$ and number of time steps. However, the observed fragmentation could be an artifact stemming from the finite time used in the simulation. To verify this is not the case, we computed the half-chain entanglement for long times with different initial conditions, BC, C and interacting BC+C quasiparticles, and plotted it in Fig.~\ref{f:longtimehce}. 

We observe that for these states the entanglement $\mathcal{S}$ display two different stages, a fast initial linear growth due to the spreading due to the quasiparticles followed by a saturation stage, in which the entropy fluctuates around a stable mean value. In addition, this saturation value depends on the initial state showing that there is indeed an ergodicity breaking. The choice of $t=1200$ for Fig.~\ref{f:SA} in the main text is taken as the minimal common time that $\mathcal{S}(t)$ for each state in Fig.~\ref{f:longtimehce} reaches a well defined saturation value.

We add now some complementary numerical results to illustrate the fragmentation of the Hilbert space. In Fig.~\ref{f:log001} we give another representation of  Fig.~\ref{f:SA} by plotting the logarithm of the inverse participation ratio as a function of the half-chain entanglement entropy. This clearly shows the stratification of the states according to their content in quasiparticles, using the same set of parameters that in the main text, $\epsilon=0.01$ and $N=18$. We probe how this fragmentation depends on $\epsilon$, that is on the strength of the nonintegrable effects. For this end we plot the half-chain entanglement and the inverse participation ratio for larger values $\epsilon=0.1$ and $\epsilon=1$, beyond the validity of the perturbation theory.

We present the half-chain entanglement in Fig.~\ref{f:hcelog01}a, and the logarithm of the inverse participation ratio in Fig.~\ref{f:hcelog01}b, for all evolved states with $t=600$ and $\epsilon=0.1$. The same data is displayed in Fig.~\ref{f:log01} as a function $-\ln\mathcal{I}$ of $\mathcal{S}$. Both figures show that the different ergodic sectors are still present and well classified by the number of domain walls of the initial states in the region of $\epsilon = 0.1$. This result tends to validate the idea that, even beyond the limits of the perturbation theory, the contribution of quasiparticles to the entanglement is dominant.

\begin{figure}
\centering
  \includegraphics[width=1.0\linewidth]{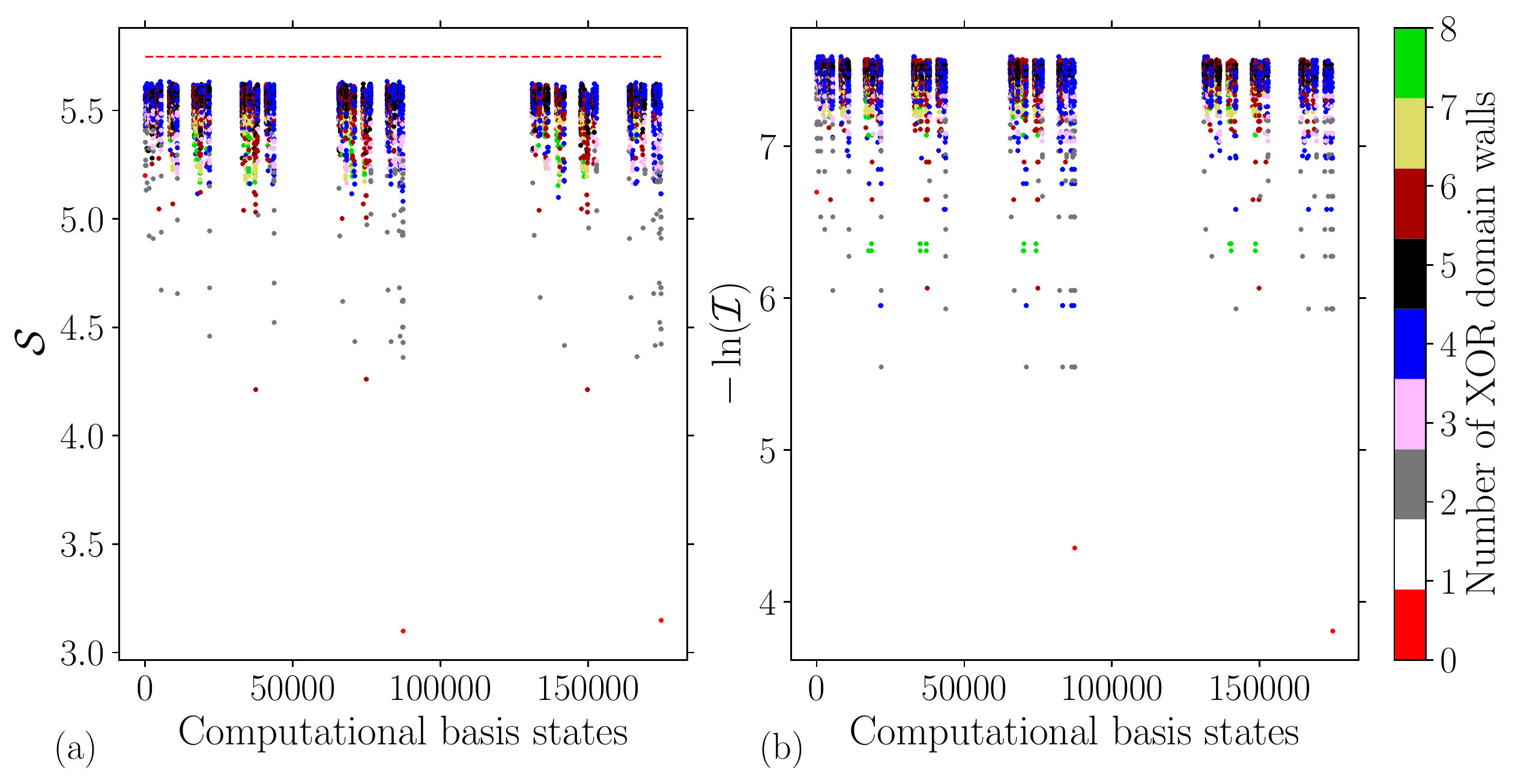}
  \caption{Half-chain entanglement and logarithm of inverse participation ratio of all evolved computational basis states for $N=18$, $\epsilon=1$ and $t=900$. }
\label{f:hcelog1}
\end{figure}

\begin{figure}
\centering
  \includegraphics[width=0.7\linewidth]{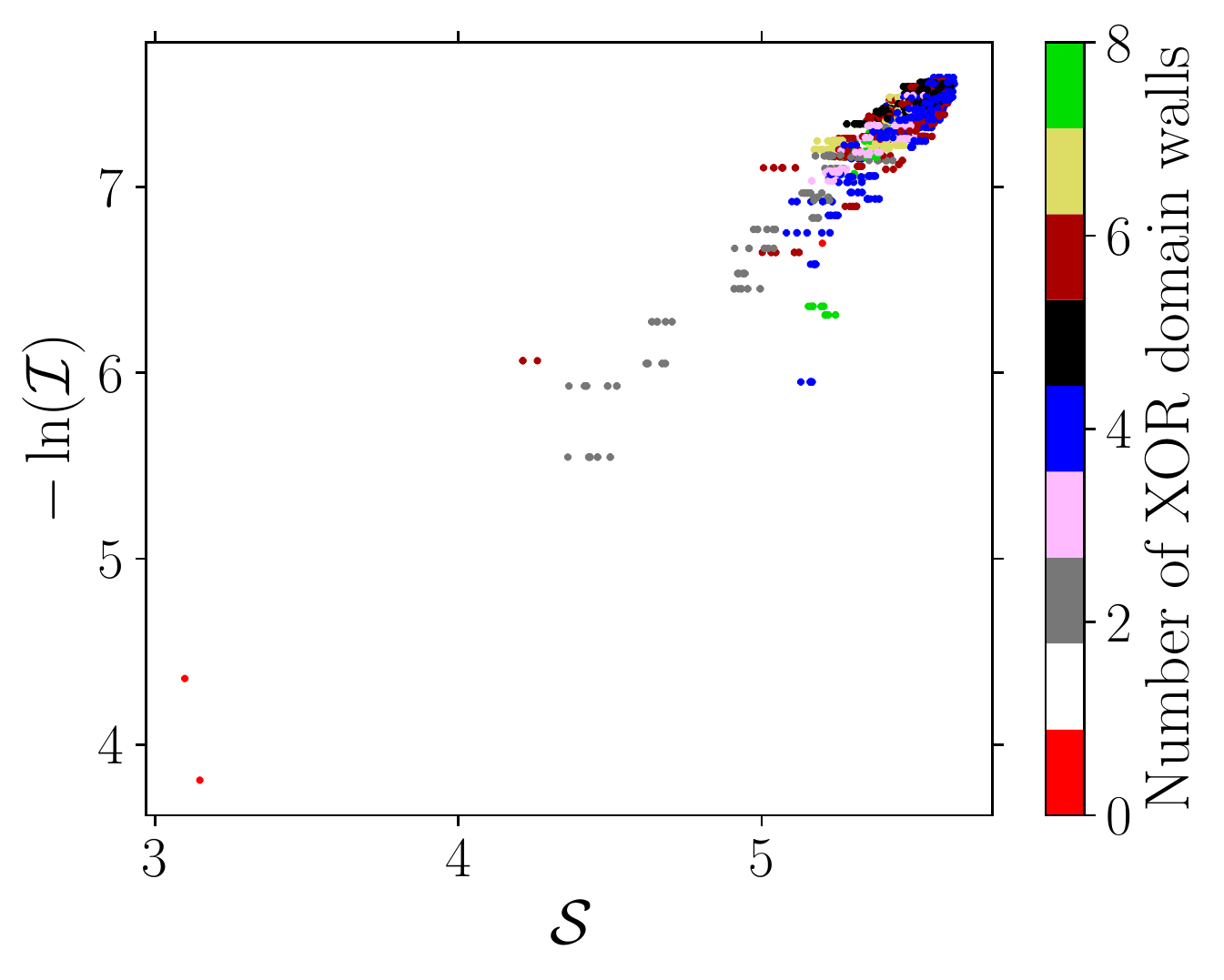}
  \caption{Logarithm of the inverse participation ratio as a function of the half-chain entanglement for $N=18$, $\epsilon=1$ and $t=900$.}
\label{f:log1}
\end{figure}

For $\epsilon=1$ and $t=900$ the situation changes qualitatively. As before, we plot the half-chain entanglement (Fig.~\ref{f:hcelog1}a) and the logarithm of the inverse participation ratio (Fig.~\ref{f:hcelog1}b) for all evolved states. However, we can see that there is no more dynamical fragmentation of the Hilbert space as the points get mixed and no clear group is classified by the number of domain walls of the initial states (see Fig.~\ref{f:log1}). These results are also in accordance with our hypothesis that the breaking of ergodicity comes from the presence of weakly interacting quasiparticles near the automaton limit: for such large value of $\epsilon$ the linear dependency of the quasienergies on $\epsilon$ is no more valid.

The results shown in Figs.~\ref{f:log01} and \ref{f:log1} suggest the existence of a transition between the near automaton regime nonergodic phase, and a more conventional thermal phase around values close to $\epsilon=0.1$.

We saw in Fig.~\ref{f:ekeps} that repulsion of eigenenergies was already present near $0.01$, We argue that well above this value, the quasiparticles eigenstates get mixed so much with other types of quasiparticles that their quasi-conservation breaks down, and the dynamical generation of different ergodic sectors becomes impossible. The investigation of this transition is left to future work.


\begin{thebibliography}{69}%
\makeatletter
\providecommand \@ifxundefined [1]{%
 \@ifx{#1\undefined}
}%
\providecommand \@ifnum [1]{%
 \ifnum #1\expandafter \@firstoftwo
 \else \expandafter \@secondoftwo
 \fi
}%
\providecommand \@ifx [1]{%
 \ifx #1\expandafter \@firstoftwo
 \else \expandafter \@secondoftwo
 \fi
}%
\providecommand \natexlab [1]{#1}%
\providecommand \enquote  [1]{``#1''}%
\providecommand \bibnamefont  [1]{#1}%
\providecommand \bibfnamefont [1]{#1}%
\providecommand \citenamefont [1]{#1}%
\providecommand \href@noop [0]{\@secondoftwo}%
\providecommand \href [0]{\begingroup \@sanitize@url \@href}%
\providecommand \@href[1]{\@@startlink{#1}\@@href}%
\providecommand \@@href[1]{\endgroup#1\@@endlink}%
\providecommand \@sanitize@url [0]{\catcode `\\12\catcode `\$12\catcode
  `\&12\catcode `\#12\catcode `\^12\catcode `\_12\catcode `\%12\relax}%
\providecommand \@@startlink[1]{}%
\providecommand \@@endlink[0]{}%
\providecommand \url  [0]{\begingroup\@sanitize@url \@url }%
\providecommand \@url [1]{\endgroup\@href {#1}{\urlprefix }}%
\providecommand \urlprefix  [0]{URL }%
\providecommand \Eprint [0]{\href }%
\providecommand \doibase [0]{https://doi.org/}%
\providecommand \selectlanguage [0]{\@gobble}%
\providecommand \bibinfo  [0]{\@secondoftwo}%
\providecommand \bibfield  [0]{\@secondoftwo}%
\providecommand \translation [1]{[#1]}%
\providecommand \BibitemOpen [0]{}%
\providecommand \bibitemStop [0]{}%
\providecommand \bibitemNoStop [0]{.\EOS\space}%
\providecommand \EOS [0]{\spacefactor3000\relax}%
\providecommand \BibitemShut  [1]{\csname bibitem#1\endcsname}%
\let\auto@bib@innerbib\@empty
\bibitem [{\citenamefont {Arrighi}(2019)}]{Arrighi-2019a}%
  \BibitemOpen
  \bibfield  {author} {\bibinfo {author} {\bibfnamefont {P.}~\bibnamefont
  {Arrighi}},\ }\bibfield  {title} {\bibinfo {title} {An overview of quantum
  cellular automata},\ }\href {https://doi.org/10.1007/s11047-019-09762-6}
  {\bibfield  {journal} {\bibinfo  {journal} {Nat Comput}\ }\textbf {\bibinfo
  {volume} {18}},\ \bibinfo {pages} {885} (\bibinfo {year} {2019})}\BibitemShut
  {NoStop}%
\bibitem [{\citenamefont {Farrelly}(2020)}]{Farrelly-2020}%
  \BibitemOpen
  \bibfield  {author} {\bibinfo {author} {\bibfnamefont {T.}~\bibnamefont
  {Farrelly}},\ }\bibfield  {title} {\bibinfo {title} {A review of {{Quantum
  Cellular Automata}}},\ }\href {https://doi.org/10.22331/q-2020-11-30-368}
  {\bibfield  {journal} {\bibinfo  {journal} {Quantum}\ }\textbf {\bibinfo
  {volume} {4}},\ \bibinfo {pages} {368} (\bibinfo {year} {2020})}\BibitemShut
  {NoStop}%
\bibitem [{\citenamefont {Deutsch}(1985)}]{Deutsch-1985hl}%
  \BibitemOpen
  \bibfield  {author} {\bibinfo {author} {\bibfnamefont {D.}~\bibnamefont
  {Deutsch}},\ }\bibfield  {title} {\bibinfo {title} {Quantum {{Theory}}, the
  {{Church-Turing Principle}} and the {{Universal Quantum Computer}}},\ }\href
  {https://doi.org/10.1098/rspa.1985.0070} {\bibfield  {journal} {\bibinfo
  {journal} {Proc. R. Soc. Lond. Math. Phys. Eng. Sci.}\ }\textbf {\bibinfo
  {volume} {400}},\ \bibinfo {pages} {97} (\bibinfo {year} {1985})}\BibitemShut
  {NoStop}%
\bibitem [{\citenamefont {Schumacher}\ and\ \citenamefont
  {Werner}(2004)}]{Schumacher-2004}%
  \BibitemOpen
  \bibfield  {author} {\bibinfo {author} {\bibfnamefont {B.}~\bibnamefont
  {Schumacher}}\ and\ \bibinfo {author} {\bibfnamefont {R.~F.}\ \bibnamefont
  {Werner}},\ }\bibfield  {title} {\bibinfo {title} {Reversible quantum
  cellular automata},\ }\bibfield  {journal} {\bibinfo  {journal} {arXiv}\
  }\href {https://doi.org/10.48550/arXiv.quant-ph/0405174}
  {10.48550/arXiv.quant-ph/0405174} (\bibinfo {year} {2004}),\ \Eprint
  {https://arxiv.org/abs/quant-ph/0405174} {arXiv:quant-ph/0405174}
  \BibitemShut {NoStop}%
\bibitem [{\citenamefont {{Pérez-Delgado}}\ and\ \citenamefont
  {Cheung}(2007)}]{Perez-Delgado-2007}%
  \BibitemOpen
  \bibfield  {author} {\bibinfo {author} {\bibfnamefont {C.~A.}\ \bibnamefont
  {{Pérez-Delgado}}}\ and\ \bibinfo {author} {\bibfnamefont {D.}~\bibnamefont
  {Cheung}},\ }\bibfield  {title} {\bibinfo {title} {Local unitary quantum
  cellular automata},\ }\href {https://doi.org/10.1103/PhysRevA.76.032320}
  {\bibfield  {journal} {\bibinfo  {journal} {Phys. Rev. A}\ }\textbf {\bibinfo
  {volume} {76}},\ \bibinfo {pages} {032320} (\bibinfo {year}
  {2007})}\BibitemShut {NoStop}%
\bibitem [{\citenamefont {Watrous}(1995)}]{Watrous-1995}%
  \BibitemOpen
  \bibfield  {author} {\bibinfo {author} {\bibfnamefont {J.}~\bibnamefont
  {Watrous}},\ }\bibfield  {title} {\bibinfo {title} {On one-dimensional
  quantum cellular automata},\ }in\ \href
  {https://doi.org/10.1109/SFCS.1995.492583} {\emph {\bibinfo {booktitle}
  {Proceedings of {{IEEE}} 36th {{Annual Foundations}} of {{Computer
  Science}}}}}\ (\bibinfo {year} {1995})\ pp.\ \bibinfo {pages}
  {528--537}\BibitemShut {NoStop}%
\bibitem [{\citenamefont {Arrighi}\ and\ \citenamefont
  {Grattage}(2012)}]{Arrighi-2012}%
  \BibitemOpen
  \bibfield  {author} {\bibinfo {author} {\bibfnamefont {P.}~\bibnamefont
  {Arrighi}}\ and\ \bibinfo {author} {\bibfnamefont {J.}~\bibnamefont
  {Grattage}},\ }\bibfield  {title} {\bibinfo {title} {Partitioned quantum
  cellular automata are intrinsically universal},\ }\href
  {https://doi.org/10.1007/s11047-011-9277-6} {\bibfield  {journal} {\bibinfo
  {journal} {Nat Comput}\ }\textbf {\bibinfo {volume} {11}},\ \bibinfo {pages}
  {13} (\bibinfo {year} {2012})}\BibitemShut {NoStop}%
\bibitem [{\citenamefont {Wintermantel}\ \emph {et~al.}(2020)\citenamefont
  {Wintermantel}, \citenamefont {Wang}, \citenamefont {Lochead}, \citenamefont
  {Shevate}, \citenamefont {Brennen},\ and\ \citenamefont
  {Whitlock}}]{Wintermantel-2020}%
  \BibitemOpen
  \bibfield  {author} {\bibinfo {author} {\bibfnamefont {T.~M.}\ \bibnamefont
  {Wintermantel}}, \bibinfo {author} {\bibfnamefont {Y.}~\bibnamefont {Wang}},
  \bibinfo {author} {\bibfnamefont {G.}~\bibnamefont {Lochead}}, \bibinfo
  {author} {\bibfnamefont {S.}~\bibnamefont {Shevate}}, \bibinfo {author}
  {\bibfnamefont {G.~K.}\ \bibnamefont {Brennen}},\ and\ \bibinfo {author}
  {\bibfnamefont {S.}~\bibnamefont {Whitlock}},\ }\bibfield  {title} {\bibinfo
  {title} {Unitary and {{Nonunitary Quantum Cellular Automata}} with {{Rydberg
  Arrays}}},\ }\href {https://doi.org/10.1103/PhysRevLett.124.070503}
  {\bibfield  {journal} {\bibinfo  {journal} {Phys. Rev. Lett.}\ }\textbf
  {\bibinfo {volume} {124}},\ \bibinfo {pages} {070503} (\bibinfo {year}
  {2020})}\BibitemShut {NoStop}%
\bibitem [{\citenamefont {Hillberry}\ \emph {et~al.}(2021)\citenamefont
  {Hillberry}, \citenamefont {Jones}, \citenamefont {Vargas}, \citenamefont
  {Rall}, \citenamefont {Halpern}, \citenamefont {Bao}, \citenamefont
  {Notarnicola}, \citenamefont {Montangero},\ and\ \citenamefont
  {Carr}}]{Hillberry-2021a}%
  \BibitemOpen
  \bibfield  {author} {\bibinfo {author} {\bibfnamefont {L.~E.}\ \bibnamefont
  {Hillberry}}, \bibinfo {author} {\bibfnamefont {M.~T.}\ \bibnamefont
  {Jones}}, \bibinfo {author} {\bibfnamefont {D.~L.}\ \bibnamefont {Vargas}},
  \bibinfo {author} {\bibfnamefont {P.}~\bibnamefont {Rall}}, \bibinfo {author}
  {\bibfnamefont {N.~Y.}\ \bibnamefont {Halpern}}, \bibinfo {author}
  {\bibfnamefont {N.}~\bibnamefont {Bao}}, \bibinfo {author} {\bibfnamefont
  {S.}~\bibnamefont {Notarnicola}}, \bibinfo {author} {\bibfnamefont
  {S.}~\bibnamefont {Montangero}},\ and\ \bibinfo {author} {\bibfnamefont
  {L.~D.}\ \bibnamefont {Carr}},\ }\bibfield  {title} {\bibinfo {title}
  {Entangled quantum cellular automata, physical complexity, and {{Goldilocks}}
  rules},\ }\href {https://doi.org/10.1088/2058-9565/ac1c41} {\bibfield
  {journal} {\bibinfo  {journal} {Quantum Sci. Technol.}\ }\textbf {\bibinfo
  {volume} {6}},\ \bibinfo {pages} {045017} (\bibinfo {year}
  {2021})}\BibitemShut {NoStop}%
\bibitem [{\citenamefont {Bernien}\ \emph {et~al.}(2017)\citenamefont
  {Bernien}, \citenamefont {Schwartz}, \citenamefont {Keesling}, \citenamefont
  {Levine}, \citenamefont {Omran}, \citenamefont {Pichler}, \citenamefont
  {Choi}, \citenamefont {Zibrov}, \citenamefont {Endres}, \citenamefont
  {Greiner}, \citenamefont {Vuletić},\ and\ \citenamefont
  {Lukin}}]{Bernien-2017}%
  \BibitemOpen
  \bibfield  {author} {\bibinfo {author} {\bibfnamefont {H.}~\bibnamefont
  {Bernien}}, \bibinfo {author} {\bibfnamefont {S.}~\bibnamefont {Schwartz}},
  \bibinfo {author} {\bibfnamefont {A.}~\bibnamefont {Keesling}}, \bibinfo
  {author} {\bibfnamefont {H.}~\bibnamefont {Levine}}, \bibinfo {author}
  {\bibfnamefont {A.}~\bibnamefont {Omran}}, \bibinfo {author} {\bibfnamefont
  {H.}~\bibnamefont {Pichler}}, \bibinfo {author} {\bibfnamefont
  {S.}~\bibnamefont {Choi}}, \bibinfo {author} {\bibfnamefont {A.~S.}\
  \bibnamefont {Zibrov}}, \bibinfo {author} {\bibfnamefont {M.}~\bibnamefont
  {Endres}}, \bibinfo {author} {\bibfnamefont {M.}~\bibnamefont {Greiner}},
  \bibinfo {author} {\bibfnamefont {V.}~\bibnamefont {Vuletić}},\ and\
  \bibinfo {author} {\bibfnamefont {M.~D.}\ \bibnamefont {Lukin}},\ }\bibfield
  {title} {\bibinfo {title} {Probing many-body dynamics on a 51-atom quantum
  simulator},\ }\href {https://doi.org/10.1038/nature24622} {\bibfield
  {journal} {\bibinfo  {journal} {Nature}\ }\textbf {\bibinfo {volume} {551}},\
  \bibinfo {pages} {579} (\bibinfo {year} {2017})}\BibitemShut {NoStop}%
\bibitem [{\citenamefont {Turner}\ \emph
  {et~al.}(2018{\natexlab{a}})\citenamefont {Turner}, \citenamefont
  {Michailidis}, \citenamefont {Abanin}, \citenamefont {Serbyn},\ and\
  \citenamefont {Papić}}]{Turner-2018}%
  \BibitemOpen
  \bibfield  {author} {\bibinfo {author} {\bibfnamefont {C.~J.}\ \bibnamefont
  {Turner}}, \bibinfo {author} {\bibfnamefont {A.~A.}\ \bibnamefont
  {Michailidis}}, \bibinfo {author} {\bibfnamefont {D.~A.}\ \bibnamefont
  {Abanin}}, \bibinfo {author} {\bibfnamefont {M.}~\bibnamefont {Serbyn}},\
  and\ \bibinfo {author} {\bibfnamefont {Z.}~\bibnamefont {Papić}},\
  }\bibfield  {title} {\bibinfo {title} {Weak ergodicity breaking from quantum
  many-body scars},\ }\href {https://doi.org/10.1038/s41567-018-0137-5}
  {\bibfield  {journal} {\bibinfo  {journal} {Nature Phys}\ }\textbf {\bibinfo
  {volume} {14}},\ \bibinfo {pages} {745} (\bibinfo {year}
  {2018}{\natexlab{a}})}\BibitemShut {NoStop}%
\bibitem [{\citenamefont {Iadecola}\ and\ \citenamefont
  {Vijay}(2020)}]{Iadecola-2020}%
  \BibitemOpen
  \bibfield  {author} {\bibinfo {author} {\bibfnamefont {T.}~\bibnamefont
  {Iadecola}}\ and\ \bibinfo {author} {\bibfnamefont {S.}~\bibnamefont
  {Vijay}},\ }\bibfield  {title} {\bibinfo {title} {Nonergodic quantum dynamics
  from deformations of classical cellular automata},\ }\href
  {https://doi.org/10.1103/PhysRevB.102.180302} {\bibfield  {journal} {\bibinfo
   {journal} {Phys. Rev. B}\ }\textbf {\bibinfo {volume} {102}},\ \bibinfo
  {pages} {180302} (\bibinfo {year} {2020})}\BibitemShut {NoStop}%
\bibitem [{\citenamefont {Yoshida}(2013)}]{Yoshida-2013}%
  \BibitemOpen
  \bibfield  {author} {\bibinfo {author} {\bibfnamefont {B.}~\bibnamefont
  {Yoshida}},\ }\bibfield  {title} {\bibinfo {title} {Exotic topological order
  in fractal spin liquids},\ }\href
  {https://doi.org/10.1103/PhysRevB.88.125122} {\bibfield  {journal} {\bibinfo
  {journal} {Phys. Rev. B}\ }\textbf {\bibinfo {volume} {88}},\ \bibinfo
  {pages} {125122} (\bibinfo {year} {2013})}\BibitemShut {NoStop}%
\bibitem [{\citenamefont {Gopalakrishnan}\ and\ \citenamefont
  {Zakirov}(2018)}]{Gopalakrishnan-2018}%
  \BibitemOpen
  \bibfield  {author} {\bibinfo {author} {\bibfnamefont {S.}~\bibnamefont
  {Gopalakrishnan}}\ and\ \bibinfo {author} {\bibfnamefont {B.}~\bibnamefont
  {Zakirov}},\ }\bibfield  {title} {\bibinfo {title} {Facilitated quantum
  cellular automata as simple models with non-thermal eigenstates and
  dynamics},\ }\href {https://doi.org/10.1088/2058-9565/aad759} {\bibfield
  {journal} {\bibinfo  {journal} {Quantum Sci. Technol.}\ }\textbf {\bibinfo
  {volume} {3}},\ \bibinfo {pages} {044004} (\bibinfo {year}
  {2018})}\BibitemShut {NoStop}%
\bibitem [{\citenamefont {Bobenko}\ \emph {et~al.}(1993)\citenamefont
  {Bobenko}, \citenamefont {Bordemann}, \citenamefont {Gunn},\ and\
  \citenamefont {Pinkall}}]{Bobenko-1993}%
  \BibitemOpen
  \bibfield  {author} {\bibinfo {author} {\bibfnamefont {A.}~\bibnamefont
  {Bobenko}}, \bibinfo {author} {\bibfnamefont {M.}~\bibnamefont {Bordemann}},
  \bibinfo {author} {\bibfnamefont {C.}~\bibnamefont {Gunn}},\ and\ \bibinfo
  {author} {\bibfnamefont {U.}~\bibnamefont {Pinkall}},\ }\bibfield  {title}
  {\bibinfo {title} {On two integrable cellular automata},\ }\href
  {https://doi.org/10.1007/BF02097234} {\bibfield  {journal} {\bibinfo
  {journal} {Commun.Math. Phys.}\ }\textbf {\bibinfo {volume} {158}},\ \bibinfo
  {pages} {127} (\bibinfo {year} {1993})}\BibitemShut {NoStop}%
\bibitem [{\citenamefont {Bremner}\ \emph {et~al.}(2009)\citenamefont
  {Bremner}, \citenamefont {Mora},\ and\ \citenamefont
  {Winter}}]{Bremner-2009}%
  \BibitemOpen
  \bibfield  {author} {\bibinfo {author} {\bibfnamefont {M.~J.}\ \bibnamefont
  {Bremner}}, \bibinfo {author} {\bibfnamefont {C.}~\bibnamefont {Mora}},\ and\
  \bibinfo {author} {\bibfnamefont {A.}~\bibnamefont {Winter}},\ }\bibfield
  {title} {\bibinfo {title} {Are {{Random Pure States Useful}} for {{Quantum
  Computation}}?},\ }\href {https://doi.org/10.1103/PhysRevLett.102.190502}
  {\bibfield  {journal} {\bibinfo  {journal} {Phys. Rev. Lett.}\ }\textbf
  {\bibinfo {volume} {102}},\ \bibinfo {pages} {190502} (\bibinfo {year}
  {2009})}\BibitemShut {NoStop}%
\bibitem [{\citenamefont {Gross}\ \emph {et~al.}(2009)\citenamefont {Gross},
  \citenamefont {Flammia},\ and\ \citenamefont {Eisert}}]{Gross-2009uq}%
  \BibitemOpen
  \bibfield  {author} {\bibinfo {author} {\bibfnamefont {D.}~\bibnamefont
  {Gross}}, \bibinfo {author} {\bibfnamefont {S.~T.}\ \bibnamefont {Flammia}},\
  and\ \bibinfo {author} {\bibfnamefont {J.}~\bibnamefont {Eisert}},\
  }\bibfield  {title} {\bibinfo {title} {Most {{Quantum States Are Too
  Entangled To Be Useful As Computational Resources}}},\ }\href
  {https://doi.org/10.1103/PhysRevLett.102.190501} {\bibfield  {journal}
  {\bibinfo  {journal} {Phys. Rev. Lett.}\ }\textbf {\bibinfo {volume} {102}},\
  \bibinfo {pages} {190501} (\bibinfo {year} {2009})}\BibitemShut {NoStop}%
\bibitem [{\citenamefont {Page}(1993)}]{Page-1993nr}%
  \BibitemOpen
  \bibfield  {author} {\bibinfo {author} {\bibfnamefont {D.~N.}\ \bibnamefont
  {Page}},\ }\bibfield  {title} {\bibinfo {title} {Average entropy of a
  subsystem},\ }\href {https://doi.org/10.1103/PhysRevLett.71.1291} {\bibfield
  {journal} {\bibinfo  {journal} {Phys. Rev. Lett.}\ }\textbf {\bibinfo
  {volume} {71}},\ \bibinfo {pages} {1291} (\bibinfo {year}
  {1993})}\BibitemShut {NoStop}%
\bibitem [{\citenamefont {Hopfield}(1988)}]{Hopfield-1988}%
  \BibitemOpen
  \bibfield  {author} {\bibinfo {author} {\bibfnamefont {J.}~\bibnamefont
  {Hopfield}},\ }\bibfield  {title} {\bibinfo {title} {Artificial neural
  networks},\ }\href {https://doi.org/10.1109/101.8118} {\bibfield  {journal}
  {\bibinfo  {journal} {IEEE Circuits Devices Mag.}\ }\textbf {\bibinfo
  {volume} {4}},\ \bibinfo {pages} {3} (\bibinfo {year} {1988})}\BibitemShut
  {NoStop}%
\bibitem [{\citenamefont {Chamon}(2005)}]{Chamon-2005}%
  \BibitemOpen
  \bibfield  {author} {\bibinfo {author} {\bibfnamefont {C.}~\bibnamefont
  {Chamon}},\ }\bibfield  {title} {\bibinfo {title} {Quantum {{Glassiness}} in
  {{Strongly Correlated Clean Systems}}: {{An Example}} of {{Topological
  Overprotection}}},\ }\href {https://doi.org/10.1103/PhysRevLett.94.040402}
  {\bibfield  {journal} {\bibinfo  {journal} {Phys. Rev. Lett.}\ }\textbf
  {\bibinfo {volume} {94}},\ \bibinfo {pages} {040402} (\bibinfo {year}
  {2005})}\BibitemShut {NoStop}%
\bibitem [{\citenamefont {D’Alessio}\ and\ \citenamefont
  {Rigol}(2014)}]{DAlessio-2014}%
  \BibitemOpen
  \bibfield  {author} {\bibinfo {author} {\bibfnamefont {L.}~\bibnamefont
  {D’Alessio}}\ and\ \bibinfo {author} {\bibfnamefont {M.}~\bibnamefont
  {Rigol}},\ }\bibfield  {title} {\bibinfo {title} {Long-time {{Behavior}} of
  {{Isolated Periodically Driven Interacting Lattice Systems}}},\ }\href
  {https://doi.org/10.1103/PhysRevX.4.041048} {\bibfield  {journal} {\bibinfo
  {journal} {Phys. Rev. X}\ }\textbf {\bibinfo {volume} {4}},\ \bibinfo {pages}
  {041048} (\bibinfo {year} {2014})}\BibitemShut {NoStop}%
\bibitem [{\citenamefont {Mizuta}\ \emph {et~al.}(2020)\citenamefont {Mizuta},
  \citenamefont {Takasan},\ and\ \citenamefont {Kawakami}}]{Mizuta-2020}%
  \BibitemOpen
  \bibfield  {author} {\bibinfo {author} {\bibfnamefont {K.}~\bibnamefont
  {Mizuta}}, \bibinfo {author} {\bibfnamefont {K.}~\bibnamefont {Takasan}},\
  and\ \bibinfo {author} {\bibfnamefont {N.}~\bibnamefont {Kawakami}},\
  }\bibfield  {title} {\bibinfo {title} {Exact {{Floquet}} quantum many-body
  scars under {{Rydberg}} blockade},\ }\href
  {https://doi.org/10.1103/PhysRevResearch.2.033284} {\bibfield  {journal}
  {\bibinfo  {journal} {Phys. Rev. Research}\ }\textbf {\bibinfo {volume}
  {2}},\ \bibinfo {pages} {033284} (\bibinfo {year} {2020})}\BibitemShut
  {NoStop}%
\bibitem [{\citenamefont {Mukherjee}\ \emph {et~al.}(2020)\citenamefont
  {Mukherjee}, \citenamefont {Nandy}, \citenamefont {Sen}, \citenamefont
  {Sen},\ and\ \citenamefont {Sengupta}}]{Mukherjee-2020}%
  \BibitemOpen
  \bibfield  {author} {\bibinfo {author} {\bibfnamefont {B.}~\bibnamefont
  {Mukherjee}}, \bibinfo {author} {\bibfnamefont {S.}~\bibnamefont {Nandy}},
  \bibinfo {author} {\bibfnamefont {A.}~\bibnamefont {Sen}}, \bibinfo {author}
  {\bibfnamefont {D.}~\bibnamefont {Sen}},\ and\ \bibinfo {author}
  {\bibfnamefont {K.}~\bibnamefont {Sengupta}},\ }\bibfield  {title} {\bibinfo
  {title} {Collapse and revival of quantum many-body scars via {{Floquet}}
  engineering},\ }\href {https://doi.org/10.1103/PhysRevB.101.245107}
  {\bibfield  {journal} {\bibinfo  {journal} {Phys. Rev. B}\ }\textbf {\bibinfo
  {volume} {101}},\ \bibinfo {pages} {245107} (\bibinfo {year}
  {2020})}\BibitemShut {NoStop}%
\bibitem [{\citenamefont {Wilkinson}\ \emph {et~al.}(2020)\citenamefont
  {Wilkinson}, \citenamefont {Klobas}, \citenamefont {Prosen},\ and\
  \citenamefont {Garrahan}}]{Wilkinson-2020}%
  \BibitemOpen
  \bibfield  {author} {\bibinfo {author} {\bibfnamefont {J.~W.~P.}\
  \bibnamefont {Wilkinson}}, \bibinfo {author} {\bibfnamefont {K.}~\bibnamefont
  {Klobas}}, \bibinfo {author} {\bibfnamefont {T.}~\bibnamefont {Prosen}},\
  and\ \bibinfo {author} {\bibfnamefont {J.~P.}\ \bibnamefont {Garrahan}},\
  }\bibfield  {title} {\bibinfo {title} {Exact solution of the {{Floquet-PXP}}
  cellular automaton},\ }\href {https://doi.org/10.1103/PhysRevE.102.062107}
  {\bibfield  {journal} {\bibinfo  {journal} {Phys. Rev. E}\ }\textbf {\bibinfo
  {volume} {102}},\ \bibinfo {pages} {062107} (\bibinfo {year}
  {2020})}\BibitemShut {NoStop}%
\bibitem [{\citenamefont {Gopalakrishnan}\ \emph {et~al.}(2018)\citenamefont
  {Gopalakrishnan}, \citenamefont {Huse}, \citenamefont {Khemani},\ and\
  \citenamefont {Vasseur}}]{Gopalakrishnan-2018b}%
  \BibitemOpen
  \bibfield  {author} {\bibinfo {author} {\bibfnamefont {S.}~\bibnamefont
  {Gopalakrishnan}}, \bibinfo {author} {\bibfnamefont {D.~A.}\ \bibnamefont
  {Huse}}, \bibinfo {author} {\bibfnamefont {V.}~\bibnamefont {Khemani}},\ and\
  \bibinfo {author} {\bibfnamefont {R.}~\bibnamefont {Vasseur}},\ }\bibfield
  {title} {\bibinfo {title} {Hydrodynamics of operator spreading and
  quasiparticle diffusion in interacting integrable systems},\ }\href
  {https://doi.org/10.1103/PhysRevB.98.220303} {\bibfield  {journal} {\bibinfo
  {journal} {Phys. Rev. B}\ }\textbf {\bibinfo {volume} {98}},\ \bibinfo
  {pages} {220303} (\bibinfo {year} {2018})}\BibitemShut {NoStop}%
\bibitem [{\citenamefont {Friedman}\ \emph {et~al.}(2019)\citenamefont
  {Friedman}, \citenamefont {Gopalakrishnan},\ and\ \citenamefont
  {Vasseur}}]{Friedman-2019}%
  \BibitemOpen
  \bibfield  {author} {\bibinfo {author} {\bibfnamefont {A.~J.}\ \bibnamefont
  {Friedman}}, \bibinfo {author} {\bibfnamefont {S.}~\bibnamefont
  {Gopalakrishnan}},\ and\ \bibinfo {author} {\bibfnamefont {R.}~\bibnamefont
  {Vasseur}},\ }\bibfield  {title} {\bibinfo {title} {Integrable {{Many-Body
  Quantum Floquet-Thouless Pumps}}},\ }\href
  {https://doi.org/10.1103/PhysRevLett.123.170603} {\bibfield  {journal}
  {\bibinfo  {journal} {Phys. Rev. Lett.}\ }\textbf {\bibinfo {volume} {123}},\
  \bibinfo {pages} {170603} (\bibinfo {year} {2019})}\BibitemShut {NoStop}%
\bibitem [{\citenamefont {Turner}\ \emph
  {et~al.}(2018{\natexlab{b}})\citenamefont {Turner}, \citenamefont
  {Michailidis}, \citenamefont {Abanin}, \citenamefont {Serbyn},\ and\
  \citenamefont {Papić}}]{Turner-2018a}%
  \BibitemOpen
  \bibfield  {author} {\bibinfo {author} {\bibfnamefont {C.~J.}\ \bibnamefont
  {Turner}}, \bibinfo {author} {\bibfnamefont {A.~A.}\ \bibnamefont
  {Michailidis}}, \bibinfo {author} {\bibfnamefont {D.~A.}\ \bibnamefont
  {Abanin}}, \bibinfo {author} {\bibfnamefont {M.}~\bibnamefont {Serbyn}},\
  and\ \bibinfo {author} {\bibfnamefont {Z.}~\bibnamefont {Papić}},\
  }\bibfield  {title} {\bibinfo {title} {Quantum scarred eigenstates in a
  {{Rydberg}} atom chain: {{Entanglement}}, breakdown of thermalization, and
  stability to perturbations},\ }\href
  {https://doi.org/10.1103/PhysRevB.98.155134} {\bibfield  {journal} {\bibinfo
  {journal} {Phys. Rev. B}\ }\textbf {\bibinfo {volume} {98}},\ \bibinfo
  {pages} {155134} (\bibinfo {year} {2018}{\natexlab{b}})}\BibitemShut
  {NoStop}%
\bibitem [{\citenamefont {Shiraishi}(2018)}]{Shiraishi-2018}%
  \BibitemOpen
  \bibfield  {author} {\bibinfo {author} {\bibfnamefont {N.}~\bibnamefont
  {Shiraishi}},\ }\bibfield  {title} {\bibinfo {title} {Analytic model of
  thermalization: {{Quantum}} emulation of classical cellular automata},\
  }\href {https://doi.org/10.1103/PhysRevE.97.062144} {\bibfield  {journal}
  {\bibinfo  {journal} {Phys. Rev. E}\ }\textbf {\bibinfo {volume} {97}},\
  \bibinfo {pages} {062144} (\bibinfo {year} {2018})}\BibitemShut {NoStop}%
\bibitem [{\citenamefont {Papić}(2021)}]{Papic-2021}%
  \BibitemOpen
  \bibfield  {author} {\bibinfo {author} {\bibfnamefont {Z.}~\bibnamefont
  {Papić}},\ }\bibfield  {title} {\bibinfo {title} {Weak ergodicity breaking
  through the lens of quantum entanglement},\ }\bibfield  {journal} {\bibinfo
  {journal} {ArXiv210803460 Cond-Mat}\ }\href
  {https://doi.org/10.48550/arXiv.2108.03460} {10.48550/arXiv.2108.03460}
  (\bibinfo {year} {2021}),\ \Eprint {https://arxiv.org/abs/2108.03460}
  {arXiv:2108.03460 [cond-mat]} \BibitemShut {NoStop}%
\bibitem [{\citenamefont {Duranthon}\ and\ \citenamefont
  {Di~Molfetta}(2021)}]{Duranthon-2021}%
  \BibitemOpen
  \bibfield  {author} {\bibinfo {author} {\bibfnamefont {O.}~\bibnamefont
  {Duranthon}}\ and\ \bibinfo {author} {\bibfnamefont {G.}~\bibnamefont
  {Di~Molfetta}},\ }\bibfield  {title} {\bibinfo {title} {Coarse-grained
  quantum cellular automata},\ }\href
  {https://doi.org/10.1103/PhysRevA.103.032224} {\bibfield  {journal} {\bibinfo
   {journal} {Phys. Rev. A}\ }\textbf {\bibinfo {volume} {103}},\ \bibinfo
  {pages} {032224} (\bibinfo {year} {2021})}\BibitemShut {NoStop}%
\bibitem [{\citenamefont {Lindner}\ \emph {et~al.}(2017)\citenamefont
  {Lindner}, \citenamefont {Berg},\ and\ \citenamefont
  {Rudner}}]{Lindner-2017}%
  \BibitemOpen
  \bibfield  {author} {\bibinfo {author} {\bibfnamefont {N.~H.}\ \bibnamefont
  {Lindner}}, \bibinfo {author} {\bibfnamefont {E.}~\bibnamefont {Berg}},\ and\
  \bibinfo {author} {\bibfnamefont {M.~S.}\ \bibnamefont {Rudner}},\ }\bibfield
   {title} {\bibinfo {title} {Universal {{Chiral Quasisteady States}} in
  {{Periodically Driven Many-Body Systems}}},\ }\href
  {https://doi.org/10.1103/PhysRevX.7.011018} {\bibfield  {journal} {\bibinfo
  {journal} {Phys. Rev. X}\ }\textbf {\bibinfo {volume} {7}},\ \bibinfo {pages}
  {011018} (\bibinfo {year} {2017})}\BibitemShut {NoStop}%
\bibitem [{\citenamefont {Friedman}(2019)}]{Friedman-2019a}%
  \BibitemOpen
  \bibfield  {author} {\bibinfo {author} {\bibfnamefont {A.~J.}\ \bibnamefont
  {Friedman}},\ }\emph {\bibinfo {title} {Isolated {{Quantum Systems}}:
  {{Dynamics}} and {{Phase Structure Far From Equilibrium}}}},\ \href
  {https://escholarship.org/uc/item/0xt8x61v} {Ph.D. thesis},\ \bibinfo
  {school} {UC Irvine} (\bibinfo {year} {2019})\BibitemShut {NoStop}%
\bibitem [{\citenamefont {Serbyn}\ \emph {et~al.}(2021)\citenamefont {Serbyn},
  \citenamefont {Abanin},\ and\ \citenamefont {Papić}}]{Serbyn-2021}%
  \BibitemOpen
  \bibfield  {author} {\bibinfo {author} {\bibfnamefont {M.}~\bibnamefont
  {Serbyn}}, \bibinfo {author} {\bibfnamefont {D.~A.}\ \bibnamefont {Abanin}},\
  and\ \bibinfo {author} {\bibfnamefont {Z.}~\bibnamefont {Papić}},\
  }\bibfield  {title} {\bibinfo {title} {Quantum many-body scars and weak
  breaking of ergodicity},\ }\href {https://doi.org/10.1038/s41567-021-01230-2}
  {\bibfield  {journal} {\bibinfo  {journal} {Nat. Phys.}\ }\textbf {\bibinfo
  {volume} {17}},\ \bibinfo {pages} {675} (\bibinfo {year} {2021})}\BibitemShut
  {NoStop}%
\bibitem [{\citenamefont {Peres}(1984)}]{Peres-1984ai}%
  \BibitemOpen
  \bibfield  {author} {\bibinfo {author} {\bibfnamefont {A.}~\bibnamefont
  {Peres}},\ }\bibfield  {title} {\bibinfo {title} {Stability of quantum motion
  in chaotic and regular systems},\ }\href
  {https://doi.org/10.1103/PhysRevA.30.1610} {\bibfield  {journal} {\bibinfo
  {journal} {Phys. Rev. A}\ }\textbf {\bibinfo {volume} {30}},\ \bibinfo
  {pages} {1610} (\bibinfo {year} {1984})}\BibitemShut {NoStop}%
\bibitem [{\citenamefont {Gorin}\ \emph {et~al.}(2006)\citenamefont {Gorin},
  \citenamefont {Prosen}, \citenamefont {Seligman},\ and\ \citenamefont
  {Žnidarič}}]{Gorin-2006}%
  \BibitemOpen
  \bibfield  {author} {\bibinfo {author} {\bibfnamefont {T.}~\bibnamefont
  {Gorin}}, \bibinfo {author} {\bibfnamefont {T.}~\bibnamefont {Prosen}},
  \bibinfo {author} {\bibfnamefont {T.~H.}\ \bibnamefont {Seligman}},\ and\
  \bibinfo {author} {\bibfnamefont {M.}~\bibnamefont {Žnidarič}},\ }\bibfield
   {title} {\bibinfo {title} {Dynamics of {{Loschmidt}} echoes and fidelity
  decay},\ }\href {https://doi.org/10.1016/j.physrep.2006.09.003} {\bibfield
  {journal} {\bibinfo  {journal} {Physics Reports}\ }\textbf {\bibinfo {volume}
  {435}},\ \bibinfo {pages} {33} (\bibinfo {year} {2006})}\BibitemShut
  {NoStop}%
\bibitem [{\citenamefont {Coffman}\ \emph {et~al.}(2000)\citenamefont
  {Coffman}, \citenamefont {Kundu},\ and\ \citenamefont
  {Wootters}}]{Coffman-2000}%
  \BibitemOpen
  \bibfield  {author} {\bibinfo {author} {\bibfnamefont {V.}~\bibnamefont
  {Coffman}}, \bibinfo {author} {\bibfnamefont {J.}~\bibnamefont {Kundu}},\
  and\ \bibinfo {author} {\bibfnamefont {W.~K.}\ \bibnamefont {Wootters}},\
  }\bibfield  {title} {\bibinfo {title} {Distributed entanglement},\ }\href
  {https://doi.org/10.1103/PhysRevA.61.052306} {\bibfield  {journal} {\bibinfo
  {journal} {Phys. Rev. A}\ }\textbf {\bibinfo {volume} {61}},\ \bibinfo
  {pages} {052306} (\bibinfo {year} {2000})}\BibitemShut {NoStop}%
\bibitem [{\citenamefont {Meyer}\ and\ \citenamefont
  {Wallach}(2002)}]{Meyer-2002}%
  \BibitemOpen
  \bibfield  {author} {\bibinfo {author} {\bibfnamefont {D.~A.}\ \bibnamefont
  {Meyer}}\ and\ \bibinfo {author} {\bibfnamefont {N.~R.}\ \bibnamefont
  {Wallach}},\ }\bibfield  {title} {\bibinfo {title} {Global entanglement in
  multiparticle systems},\ }\href {https://doi.org/10.1063/1.1497700}
  {\bibfield  {journal} {\bibinfo  {journal} {J. Math. Phys.}\ }\textbf
  {\bibinfo {volume} {43}},\ \bibinfo {pages} {4273} (\bibinfo {year}
  {2002})}\BibitemShut {NoStop}%
\bibitem [{\citenamefont {Brennen}(2003)}]{Brennen-2003a}%
  \BibitemOpen
  \bibfield  {author} {\bibinfo {author} {\bibfnamefont {G.~K.}\ \bibnamefont
  {Brennen}},\ }\bibfield  {title} {\bibinfo {title} {An observable measure of
  entanglement for pure states of multi-qubit systems},\ }\href
  {https://doi.org/10.26421/QIC3.6-5} {\bibfield  {journal} {\bibinfo
  {journal} {Quantum Info. Comput.}\ }\textbf {\bibinfo {volume} {3}},\
  \bibinfo {pages} {619} (\bibinfo {year} {2003})}\BibitemShut {NoStop}%
\bibitem [{\citenamefont {Lakshminarayan}\ and\ \citenamefont
  {Subrahmanyam}(2005)}]{Lakshminarayan-2005}%
  \BibitemOpen
  \bibfield  {author} {\bibinfo {author} {\bibfnamefont {A.}~\bibnamefont
  {Lakshminarayan}}\ and\ \bibinfo {author} {\bibfnamefont {V.}~\bibnamefont
  {Subrahmanyam}},\ }\bibfield  {title} {\bibinfo {title} {Multipartite
  entanglement in a one-dimensional time-dependent {{Ising}} model},\ }\href
  {https://doi.org/10.1103/PhysRevA.71.062334} {\bibfield  {journal} {\bibinfo
  {journal} {Phys. Rev. A}\ }\textbf {\bibinfo {volume} {71}},\ \bibinfo
  {pages} {062334} (\bibinfo {year} {2005})}\BibitemShut {NoStop}%
\bibitem [{\citenamefont {Bluvstein}\ \emph {et~al.}(2021)\citenamefont
  {Bluvstein}, \citenamefont {Omran}, \citenamefont {Levine}, \citenamefont
  {Keesling}, \citenamefont {Semeghini}, \citenamefont {Ebadi}, \citenamefont
  {Wang}, \citenamefont {Michailidis}, \citenamefont {Maskara}, \citenamefont
  {Ho}, \citenamefont {Choi}, \citenamefont {Serbyn}, \citenamefont {Greiner},
  \citenamefont {Vuletić},\ and\ \citenamefont {Lukin}}]{Bluvstein-2021}%
  \BibitemOpen
  \bibfield  {author} {\bibinfo {author} {\bibfnamefont {D.}~\bibnamefont
  {Bluvstein}}, \bibinfo {author} {\bibfnamefont {A.}~\bibnamefont {Omran}},
  \bibinfo {author} {\bibfnamefont {H.}~\bibnamefont {Levine}}, \bibinfo
  {author} {\bibfnamefont {A.}~\bibnamefont {Keesling}}, \bibinfo {author}
  {\bibfnamefont {G.}~\bibnamefont {Semeghini}}, \bibinfo {author}
  {\bibfnamefont {S.}~\bibnamefont {Ebadi}}, \bibinfo {author} {\bibfnamefont
  {T.~T.}\ \bibnamefont {Wang}}, \bibinfo {author} {\bibfnamefont {A.~A.}\
  \bibnamefont {Michailidis}}, \bibinfo {author} {\bibfnamefont
  {N.}~\bibnamefont {Maskara}}, \bibinfo {author} {\bibfnamefont {W.~W.}\
  \bibnamefont {Ho}}, \bibinfo {author} {\bibfnamefont {S.}~\bibnamefont
  {Choi}}, \bibinfo {author} {\bibfnamefont {M.}~\bibnamefont {Serbyn}},
  \bibinfo {author} {\bibfnamefont {M.}~\bibnamefont {Greiner}}, \bibinfo
  {author} {\bibfnamefont {V.}~\bibnamefont {Vuletić}},\ and\ \bibinfo
  {author} {\bibfnamefont {M.~D.}\ \bibnamefont {Lukin}},\ }\bibfield  {title}
  {\bibinfo {title} {Controlling quantum many-body dynamics in driven
  {{Rydberg}} atom arrays},\ }\href {https://doi.org/10.1126/science.abg2530}
  {\bibfield  {journal} {\bibinfo  {journal} {Science}\ }\textbf {\bibinfo
  {volume} {371}},\ \bibinfo {pages} {1355} (\bibinfo {year}
  {2021})}\BibitemShut {NoStop}%
\bibitem [{\citenamefont {Sugiura}\ \emph {et~al.}(2021)\citenamefont
  {Sugiura}, \citenamefont {Kuwahara},\ and\ \citenamefont
  {Saito}}]{Sugiura-2021}%
  \BibitemOpen
  \bibfield  {author} {\bibinfo {author} {\bibfnamefont {S.}~\bibnamefont
  {Sugiura}}, \bibinfo {author} {\bibfnamefont {T.}~\bibnamefont {Kuwahara}},\
  and\ \bibinfo {author} {\bibfnamefont {K.}~\bibnamefont {Saito}},\ }\bibfield
   {title} {\bibinfo {title} {Many-body scar state intrinsic to periodically
  driven system},\ }\href {https://doi.org/10.1103/PhysRevResearch.3.L012010}
  {\bibfield  {journal} {\bibinfo  {journal} {Phys. Rev. Research}\ }\textbf
  {\bibinfo {volume} {3}},\ \bibinfo {pages} {L012010} (\bibinfo {year}
  {2021})}\BibitemShut {NoStop}%
\bibitem [{\citenamefont {Jurcevic}\ \emph {et~al.}(2014)\citenamefont
  {Jurcevic}, \citenamefont {Lanyon}, \citenamefont {Hauke}, \citenamefont
  {Hempel}, \citenamefont {Zoller}, \citenamefont {Blatt},\ and\ \citenamefont
  {Roos}}]{Jurcevic-2014}%
  \BibitemOpen
  \bibfield  {author} {\bibinfo {author} {\bibfnamefont {P.}~\bibnamefont
  {Jurcevic}}, \bibinfo {author} {\bibfnamefont {B.~P.}\ \bibnamefont
  {Lanyon}}, \bibinfo {author} {\bibfnamefont {P.}~\bibnamefont {Hauke}},
  \bibinfo {author} {\bibfnamefont {C.}~\bibnamefont {Hempel}}, \bibinfo
  {author} {\bibfnamefont {P.}~\bibnamefont {Zoller}}, \bibinfo {author}
  {\bibfnamefont {R.}~\bibnamefont {Blatt}},\ and\ \bibinfo {author}
  {\bibfnamefont {C.~F.}\ \bibnamefont {Roos}},\ }\bibfield  {title} {\bibinfo
  {title} {Quasiparticle engineering and entanglement propagation in a quantum
  many-body system},\ }\href {https://doi.org/10.1038/nature13461} {\bibfield
  {journal} {\bibinfo  {journal} {Nature}\ }\textbf {\bibinfo {volume} {511}},\
  \bibinfo {pages} {202} (\bibinfo {year} {2014})}\BibitemShut {NoStop}%
\bibitem [{\citenamefont {Wootters}(1998)}]{Wootters-1998}%
  \BibitemOpen
  \bibfield  {author} {\bibinfo {author} {\bibfnamefont {W.~K.}\ \bibnamefont
  {Wootters}},\ }\bibfield  {title} {\bibinfo {title} {Entanglement of
  {{Formation}} of an {{Arbitrary State}} of {{Two Qubits}}},\ }\href
  {https://doi.org/10.1103/PhysRevLett.80.2245} {\bibfield  {journal} {\bibinfo
   {journal} {Phys. Rev. Lett.}\ }\textbf {\bibinfo {volume} {80}},\ \bibinfo
  {pages} {2245} (\bibinfo {year} {1998})}\BibitemShut {NoStop}%
\bibitem [{\citenamefont {Plenio}(2005)}]{Plenio-2005fj}%
  \BibitemOpen
  \bibfield  {author} {\bibinfo {author} {\bibfnamefont {M.~B.}\ \bibnamefont
  {Plenio}},\ }\bibfield  {title} {\bibinfo {title} {Logarithmic
  {{Negativity}}: {{A Full Entanglement Monotone That}} is not {{Convex}}},\
  }\href {https://doi.org/10.1103/PhysRevLett.95.090503} {\bibfield  {journal}
  {\bibinfo  {journal} {Phys. Rev. Lett.}\ }\textbf {\bibinfo {volume} {95}},\
  \bibinfo {pages} {090503} (\bibinfo {year} {2005})}\BibitemShut {NoStop}%
\bibitem [{\citenamefont {Li}\ and\ \citenamefont {Haldane}(2008)}]{Li-2008fk}%
  \BibitemOpen
  \bibfield  {author} {\bibinfo {author} {\bibfnamefont {H.}~\bibnamefont
  {Li}}\ and\ \bibinfo {author} {\bibfnamefont {F.~D.~M.}\ \bibnamefont
  {Haldane}},\ }\bibfield  {title} {\bibinfo {title} {Entanglement {{Spectrum}}
  as a {{Generalization}} of {{Entanglement Entropy}}: {{Identification}} of
  {{Topological Order}} in {{Non-Abelian Fractional Quantum Hall Effect
  States}}},\ }\href {https://doi.org/10.1103/PhysRevLett.101.010504}
  {\bibfield  {journal} {\bibinfo  {journal} {Phys. Rev. Lett.}\ }\textbf
  {\bibinfo {volume} {101}},\ \bibinfo {pages} {010504} (\bibinfo {year}
  {2008})}\BibitemShut {NoStop}%
\bibitem [{\citenamefont {Iaconis}(2021)}]{Iaconis-2021}%
  \BibitemOpen
  \bibfield  {author} {\bibinfo {author} {\bibfnamefont {J.}~\bibnamefont
  {Iaconis}},\ }\bibfield  {title} {\bibinfo {title} {Quantum {{State
  Complexity}} in {{Computationally Tractable Quantum Circuits}}},\ }\href
  {https://doi.org/10.1103/PRXQuantum.2.010329} {\bibfield  {journal} {\bibinfo
   {journal} {PRX Quantum}\ }\textbf {\bibinfo {volume} {2}},\ \bibinfo {pages}
  {010329} (\bibinfo {year} {2021})}\BibitemShut {NoStop}%
\bibitem [{\citenamefont {Oganesyan}\ and\ \citenamefont
  {Huse}(2007)}]{Oganesyan-2007zr}%
  \BibitemOpen
  \bibfield  {author} {\bibinfo {author} {\bibfnamefont {V.}~\bibnamefont
  {Oganesyan}}\ and\ \bibinfo {author} {\bibfnamefont {D.~A.}\ \bibnamefont
  {Huse}},\ }\bibfield  {title} {\bibinfo {title} {Localization of interacting
  fermions at high temperature},\ }\href
  {https://doi.org/10.1103/PhysRevB.75.155111} {\bibfield  {journal} {\bibinfo
  {journal} {Phys. Rev. B}\ }\textbf {\bibinfo {volume} {75}},\ \bibinfo
  {pages} {155111} (\bibinfo {year} {2007})}\BibitemShut {NoStop}%
\bibitem [{\citenamefont {Atas}\ \emph {et~al.}(2013)\citenamefont {Atas},
  \citenamefont {Bogomolny}, \citenamefont {Giraud},\ and\ \citenamefont
  {Roux}}]{Atas-2013}%
  \BibitemOpen
  \bibfield  {author} {\bibinfo {author} {\bibfnamefont {Y.~Y.}\ \bibnamefont
  {Atas}}, \bibinfo {author} {\bibfnamefont {E.}~\bibnamefont {Bogomolny}},
  \bibinfo {author} {\bibfnamefont {O.}~\bibnamefont {Giraud}},\ and\ \bibinfo
  {author} {\bibfnamefont {G.}~\bibnamefont {Roux}},\ }\bibfield  {title}
  {\bibinfo {title} {Distribution of the {{Ratio}} of {{Consecutive Level
  Spacings}} in {{Random Matrix Ensembles}}},\ }\href
  {https://doi.org/10.1103/PhysRevLett.110.084101} {\bibfield  {journal}
  {\bibinfo  {journal} {Phys. Rev. Lett.}\ }\textbf {\bibinfo {volume} {110}},\
  \bibinfo {pages} {084101} (\bibinfo {year} {2013})}\BibitemShut {NoStop}%
\bibitem [{\citenamefont {Zhang}\ \emph {et~al.}(2015)\citenamefont {Zhang},
  \citenamefont {Kim},\ and\ \citenamefont {Huse}}]{Zhang-2015}%
  \BibitemOpen
  \bibfield  {author} {\bibinfo {author} {\bibfnamefont {L.}~\bibnamefont
  {Zhang}}, \bibinfo {author} {\bibfnamefont {H.}~\bibnamefont {Kim}},\ and\
  \bibinfo {author} {\bibfnamefont {D.~A.}\ \bibnamefont {Huse}},\ }\bibfield
  {title} {\bibinfo {title} {Thermalization of entanglement},\ }\href
  {https://doi.org/10.1103/PhysRevE.91.062128} {\bibfield  {journal} {\bibinfo
  {journal} {Phys. Rev. E}\ }\textbf {\bibinfo {volume} {91}},\ \bibinfo
  {pages} {062128} (\bibinfo {year} {2015})}\BibitemShut {NoStop}%
\bibitem [{\citenamefont {Giraud}\ \emph {et~al.}(2007)\citenamefont {Giraud},
  \citenamefont {Martin},\ and\ \citenamefont {Georgeot}}]{Giraud-2007}%
  \BibitemOpen
  \bibfield  {author} {\bibinfo {author} {\bibfnamefont {O.}~\bibnamefont
  {Giraud}}, \bibinfo {author} {\bibfnamefont {J.}~\bibnamefont {Martin}},\
  and\ \bibinfo {author} {\bibfnamefont {B.}~\bibnamefont {Georgeot}},\
  }\bibfield  {title} {\bibinfo {title} {Entanglement of localized states},\
  }\href {https://doi.org/10.1103/PhysRevA.76.042333} {\bibfield  {journal}
  {\bibinfo  {journal} {Phys. Rev. A}\ }\textbf {\bibinfo {volume} {76}},\
  \bibinfo {pages} {042333} (\bibinfo {year} {2007})}\BibitemShut {NoStop}%
\bibitem [{\citenamefont {Atas}\ and\ \citenamefont
  {Bogomolny}(2012)}]{Atas-2012}%
  \BibitemOpen
  \bibfield  {author} {\bibinfo {author} {\bibfnamefont {Y.~Y.}\ \bibnamefont
  {Atas}}\ and\ \bibinfo {author} {\bibfnamefont {E.}~\bibnamefont
  {Bogomolny}},\ }\bibfield  {title} {\bibinfo {title} {Multifractality of
  eigenfunctions in spin chains},\ }\href
  {https://doi.org/10.1103/PhysRevE.86.021104} {\bibfield  {journal} {\bibinfo
  {journal} {Phys. Rev. E}\ }\textbf {\bibinfo {volume} {86}},\ \bibinfo
  {pages} {021104} (\bibinfo {year} {2012})}\BibitemShut {NoStop}%
\bibitem [{\citenamefont {De~Tomasi}\ and\ \citenamefont
  {Khaymovich}(2020)}]{DeTomasi-2020}%
  \BibitemOpen
  \bibfield  {author} {\bibinfo {author} {\bibfnamefont {G.}~\bibnamefont
  {De~Tomasi}}\ and\ \bibinfo {author} {\bibfnamefont {I.~M.}\ \bibnamefont
  {Khaymovich}},\ }\bibfield  {title} {\bibinfo {title} {Multifractality
  {{Meets Entanglement}}: {{Relation}} for {{Nonergodic Extended States}}},\
  }\href {https://doi.org/10.1103/PhysRevLett.124.200602} {\bibfield  {journal}
  {\bibinfo  {journal} {Phys. Rev. Lett.}\ }\textbf {\bibinfo {volume} {124}},\
  \bibinfo {pages} {200602} (\bibinfo {year} {2020})}\BibitemShut {NoStop}%
\bibitem [{\citenamefont {Luitz}\ \emph {et~al.}(2014)\citenamefont {Luitz},
  \citenamefont {Alet},\ and\ \citenamefont {Laflorencie}}]{Luitz-2014}%
  \BibitemOpen
  \bibfield  {author} {\bibinfo {author} {\bibfnamefont {D.~J.}\ \bibnamefont
  {Luitz}}, \bibinfo {author} {\bibfnamefont {F.}~\bibnamefont {Alet}},\ and\
  \bibinfo {author} {\bibfnamefont {N.}~\bibnamefont {Laflorencie}},\
  }\bibfield  {title} {\bibinfo {title} {Universal {{Behavior}} beyond
  {{Multifractality}} in {{Quantum Many-Body Systems}}},\ }\href
  {https://doi.org/10.1103/PhysRevLett.112.057203} {\bibfield  {journal}
  {\bibinfo  {journal} {Phys. Rev. Lett.}\ }\textbf {\bibinfo {volume} {112}},\
  \bibinfo {pages} {057203} (\bibinfo {year} {2014})}\BibitemShut {NoStop}%
\bibitem [{\citenamefont {Lee}\ and\ \citenamefont {Vidal}(2013)}]{Lee-2013}%
  \BibitemOpen
  \bibfield  {author} {\bibinfo {author} {\bibfnamefont {Y.~A.}\ \bibnamefont
  {Lee}}\ and\ \bibinfo {author} {\bibfnamefont {G.}~\bibnamefont {Vidal}},\
  }\bibfield  {title} {\bibinfo {title} {Entanglement negativity and
  topological order},\ }\href {https://doi.org/10.1103/PhysRevA.88.042318}
  {\bibfield  {journal} {\bibinfo  {journal} {Phys. Rev. A}\ }\textbf {\bibinfo
  {volume} {88}},\ \bibinfo {pages} {042318} (\bibinfo {year}
  {2013})}\BibitemShut {NoStop}%
\bibitem [{\citenamefont {Vidal}\ and\ \citenamefont
  {Werner}(2002)}]{Vidal-2002zr}%
  \BibitemOpen
  \bibfield  {author} {\bibinfo {author} {\bibfnamefont {G.}~\bibnamefont
  {Vidal}}\ and\ \bibinfo {author} {\bibfnamefont {R.~F.}\ \bibnamefont
  {Werner}},\ }\bibfield  {title} {\bibinfo {title} {Computable measure of
  entanglement},\ }\href {https://doi.org/10.1103/PhysRevA.65.032314}
  {\bibfield  {journal} {\bibinfo  {journal} {Phys. Rev. A}\ }\textbf {\bibinfo
  {volume} {65}},\ \bibinfo {pages} {032314} (\bibinfo {year}
  {2002})}\BibitemShut {NoStop}%
\bibitem [{\citenamefont {Peres}(1996)}]{Peres-1996}%
  \BibitemOpen
  \bibfield  {author} {\bibinfo {author} {\bibfnamefont {A.}~\bibnamefont
  {Peres}},\ }\bibfield  {title} {\bibinfo {title} {Separability {{Criterion}}
  for {{Density Matrices}}},\ }\href
  {https://doi.org/10.1103/PhysRevLett.77.1413} {\bibfield  {journal} {\bibinfo
   {journal} {Phys. Rev. Lett.}\ }\textbf {\bibinfo {volume} {77}},\ \bibinfo
  {pages} {1413} (\bibinfo {year} {1996})}\BibitemShut {NoStop}%
\bibitem [{\citenamefont {Choi}\ \emph {et~al.}(2020)\citenamefont {Choi},
  \citenamefont {Bao}, \citenamefont {Qi},\ and\ \citenamefont
  {Altman}}]{Choi-2020}%
  \BibitemOpen
  \bibfield  {author} {\bibinfo {author} {\bibfnamefont {S.}~\bibnamefont
  {Choi}}, \bibinfo {author} {\bibfnamefont {Y.}~\bibnamefont {Bao}}, \bibinfo
  {author} {\bibfnamefont {X.-L.}\ \bibnamefont {Qi}},\ and\ \bibinfo {author}
  {\bibfnamefont {E.}~\bibnamefont {Altman}},\ }\bibfield  {title} {\bibinfo
  {title} {Quantum {{Error Correction}} in {{Scrambling Dynamics}} and
  {{Measurement-Induced Phase Transition}}},\ }\href
  {https://doi.org/10.1103/PhysRevLett.125.030505} {\bibfield  {journal}
  {\bibinfo  {journal} {Phys. Rev. Lett.}\ }\textbf {\bibinfo {volume} {125}},\
  \bibinfo {pages} {030505} (\bibinfo {year} {2020})}\BibitemShut {NoStop}%
\bibitem [{\citenamefont {Wildeboer}\ \emph {et~al.}(2022)\citenamefont
  {Wildeboer}, \citenamefont {Iadecola},\ and\ \citenamefont
  {Williamson}}]{Wildeboer-2022}%
  \BibitemOpen
  \bibfield  {author} {\bibinfo {author} {\bibfnamefont {J.}~\bibnamefont
  {Wildeboer}}, \bibinfo {author} {\bibfnamefont {T.}~\bibnamefont
  {Iadecola}},\ and\ \bibinfo {author} {\bibfnamefont {D.~J.}\ \bibnamefont
  {Williamson}},\ }\bibfield  {title} {\bibinfo {title} {Symmetry-{{Protected
  Infinite-Temperature Quantum Memory}} from {{Subsystem Codes}}},\ }\href
  {https://doi.org/10.1103/PRXQuantum.3.020330} {\bibfield  {journal} {\bibinfo
   {journal} {PRX Quantum}\ }\textbf {\bibinfo {volume} {3}},\ \bibinfo {pages}
  {020330} (\bibinfo {year} {2022})}\BibitemShut {NoStop}%
\bibitem [{\citenamefont {Scherg}\ \emph {et~al.}(2021)\citenamefont {Scherg},
  \citenamefont {Kohlert}, \citenamefont {Sala}, \citenamefont {Pollmann},
  \citenamefont {Hebbe~Madhusudhana}, \citenamefont {Bloch},\ and\
  \citenamefont {Aidelsburger}}]{Scherg-2021a}%
  \BibitemOpen
  \bibfield  {author} {\bibinfo {author} {\bibfnamefont {S.}~\bibnamefont
  {Scherg}}, \bibinfo {author} {\bibfnamefont {T.}~\bibnamefont {Kohlert}},
  \bibinfo {author} {\bibfnamefont {P.}~\bibnamefont {Sala}}, \bibinfo {author}
  {\bibfnamefont {F.}~\bibnamefont {Pollmann}}, \bibinfo {author}
  {\bibfnamefont {B.}~\bibnamefont {Hebbe~Madhusudhana}}, \bibinfo {author}
  {\bibfnamefont {I.}~\bibnamefont {Bloch}},\ and\ \bibinfo {author}
  {\bibfnamefont {M.}~\bibnamefont {Aidelsburger}},\ }\bibfield  {title}
  {\bibinfo {title} {Observing non-ergodicity due to kinetic constraints in
  tilted {{Fermi-Hubbard}} chains},\ }\href
  {https://doi.org/10.1038/s41467-021-24726-0} {\bibfield  {journal} {\bibinfo
  {journal} {Nat Commun}\ }\textbf {\bibinfo {volume} {12}},\ \bibinfo {pages}
  {4490} (\bibinfo {year} {2021})}\BibitemShut {NoStop}%
\bibitem [{\citenamefont {Verga}\ and\ \citenamefont
  {Elías}(2019)}]{Verga-2019b}%
  \BibitemOpen
  \bibfield  {author} {\bibinfo {author} {\bibfnamefont {A.~D.}\ \bibnamefont
  {Verga}}\ and\ \bibinfo {author} {\bibfnamefont {R.~G.}\ \bibnamefont
  {Elías}},\ }\bibfield  {title} {\bibinfo {title} {Thermal state entanglement
  entropy on a quantum graph},\ }\href
  {https://doi.org/10.1103/PhysRevE.100.062137} {\bibfield  {journal} {\bibinfo
   {journal} {Phys. Rev. E}\ }\textbf {\bibinfo {volume} {100}},\ \bibinfo
  {pages} {062137} (\bibinfo {year} {2019})}\BibitemShut {NoStop}%
\bibitem [{\citenamefont {Stephen}\ \emph {et~al.}(2019)\citenamefont
  {Stephen}, \citenamefont {Nautrup}, \citenamefont {{Bermejo-Vega}},
  \citenamefont {Eisert},\ and\ \citenamefont {Raussendorf}}]{Stephen-2019}%
  \BibitemOpen
  \bibfield  {author} {\bibinfo {author} {\bibfnamefont {D.~T.}\ \bibnamefont
  {Stephen}}, \bibinfo {author} {\bibfnamefont {H.~P.}\ \bibnamefont
  {Nautrup}}, \bibinfo {author} {\bibfnamefont {J.}~\bibnamefont
  {{Bermejo-Vega}}}, \bibinfo {author} {\bibfnamefont {J.}~\bibnamefont
  {Eisert}},\ and\ \bibinfo {author} {\bibfnamefont {R.}~\bibnamefont
  {Raussendorf}},\ }\bibfield  {title} {\bibinfo {title} {Subsystem symmetries,
  quantum cellular automata, and computational phases of quantum matter},\
  }\href {https://doi.org/10.22331/q-2019-05-20-142} {\bibfield  {journal}
  {\bibinfo  {journal} {Quantum}\ }\textbf {\bibinfo {volume} {3}},\ \bibinfo
  {pages} {142} (\bibinfo {year} {2019})}\BibitemShut {NoStop}%
\bibitem [{\citenamefont {Sellapillay}\ \emph {et~al.}(2022)\citenamefont
  {Sellapillay}, \citenamefont {Arrighi},\ and\ \citenamefont
  {Di~Molfetta}}]{Sellapillay-2022}%
  \BibitemOpen
  \bibfield  {author} {\bibinfo {author} {\bibfnamefont {K.}~\bibnamefont
  {Sellapillay}}, \bibinfo {author} {\bibfnamefont {P.}~\bibnamefont
  {Arrighi}},\ and\ \bibinfo {author} {\bibfnamefont {G.}~\bibnamefont
  {Di~Molfetta}},\ }\bibfield  {title} {\bibinfo {title} {A discrete
  relativistic spacetime formalism for 1 + 1-{{QED}} with continuum limits},\
  }\href {https://doi.org/10.1038/s41598-022-06241-4} {\bibfield  {journal}
  {\bibinfo  {journal} {Sci Rep}\ }\textbf {\bibinfo {volume} {12}},\ \bibinfo
  {pages} {2198} (\bibinfo {year} {2022})}\BibitemShut {NoStop}%
\bibitem [{\citenamefont {Kitaev}(2003)}]{Kitaev-2003fk}%
  \BibitemOpen
  \bibfield  {author} {\bibinfo {author} {\bibfnamefont {A.~Y.}\ \bibnamefont
  {Kitaev}},\ }\bibfield  {title} {\bibinfo {title} {Fault-tolerant quantum
  computation by anyons},\ }\href
  {https://doi.org/10.1016/S0003-4916(02)00018-0} {\bibfield  {journal}
  {\bibinfo  {journal} {Ann. Phys.}\ }\textbf {\bibinfo {volume} {303}},\
  \bibinfo {pages} {2} (\bibinfo {year} {2003})}\BibitemShut {NoStop}%
\bibitem [{\citenamefont {Satzinger}\ \emph {et~al.}(2021)\citenamefont
  {Satzinger}, \citenamefont {Liu}, \citenamefont {Smith}, \citenamefont
  {Knapp}, \citenamefont {Newman}, \citenamefont {Jones}, \citenamefont {Chen},
  \citenamefont {Quintana}, \citenamefont {Mi},\ and\ \citenamefont
  {Dunsworth}}]{Satzinger-2021a}%
  \BibitemOpen
  \bibfield  {author} {\bibinfo {author} {\bibfnamefont {K.~J.}\ \bibnamefont
  {Satzinger}}, \bibinfo {author} {\bibfnamefont {Y.-J.}\ \bibnamefont {Liu}},
  \bibinfo {author} {\bibfnamefont {A.}~\bibnamefont {Smith}}, \bibinfo
  {author} {\bibfnamefont {C.}~\bibnamefont {Knapp}}, \bibinfo {author}
  {\bibfnamefont {M.}~\bibnamefont {Newman}}, \bibinfo {author} {\bibfnamefont
  {C.}~\bibnamefont {Jones}}, \bibinfo {author} {\bibfnamefont
  {Z.}~\bibnamefont {Chen}}, \bibinfo {author} {\bibfnamefont {C.}~\bibnamefont
  {Quintana}}, \bibinfo {author} {\bibfnamefont {X.}~\bibnamefont {Mi}},\ and\
  \bibinfo {author} {\bibfnamefont {A.}~\bibnamefont {Dunsworth}},\ }\bibfield
  {title} {\bibinfo {title} {Realizing topologically ordered states on a
  quantum processor},\ }\href@noop {} {\bibfield  {journal} {\bibinfo
  {journal} {Science}\ }\textbf {\bibinfo {volume} {374}},\ \bibinfo {pages}
  {1237} (\bibinfo {year} {2021})}\BibitemShut {NoStop}%
\bibitem [{\citenamefont {Sala}\ \emph {et~al.}(2020)\citenamefont {Sala},
  \citenamefont {Rakovszky}, \citenamefont {Verresen}, \citenamefont {Knap},\
  and\ \citenamefont {Pollmann}}]{Sala-2020}%
  \BibitemOpen
  \bibfield  {author} {\bibinfo {author} {\bibfnamefont {P.}~\bibnamefont
  {Sala}}, \bibinfo {author} {\bibfnamefont {T.}~\bibnamefont {Rakovszky}},
  \bibinfo {author} {\bibfnamefont {R.}~\bibnamefont {Verresen}}, \bibinfo
  {author} {\bibfnamefont {M.}~\bibnamefont {Knap}},\ and\ \bibinfo {author}
  {\bibfnamefont {F.}~\bibnamefont {Pollmann}},\ }\bibfield  {title} {\bibinfo
  {title} {Ergodicity {{Breaking Arising}} from {{Hilbert Space Fragmentation}}
  in {{Dipole-Conserving Hamiltonians}}},\ }\href
  {https://doi.org/10.1103/PhysRevX.10.011047} {\bibfield  {journal} {\bibinfo
  {journal} {Phys. Rev. X}\ }\textbf {\bibinfo {volume} {10}},\ \bibinfo
  {pages} {011047} (\bibinfo {year} {2020})}\BibitemShut {NoStop}%
\bibitem [{\citenamefont {Khemani}\ \emph {et~al.}(2020)\citenamefont
  {Khemani}, \citenamefont {Hermele},\ and\ \citenamefont
  {Nandkishore}}]{Khemani-2020}%
  \BibitemOpen
  \bibfield  {author} {\bibinfo {author} {\bibfnamefont {V.}~\bibnamefont
  {Khemani}}, \bibinfo {author} {\bibfnamefont {M.}~\bibnamefont {Hermele}},\
  and\ \bibinfo {author} {\bibfnamefont {R.}~\bibnamefont {Nandkishore}},\
  }\bibfield  {title} {\bibinfo {title} {Localization from {{Hilbert}} space
  shattering: {{From}} theory to physical realizations},\ }\href
  {https://doi.org/10.1103/PhysRevB.101.174204} {\bibfield  {journal} {\bibinfo
   {journal} {Phys. Rev. B}\ }\textbf {\bibinfo {volume} {101}},\ \bibinfo
  {pages} {174204} (\bibinfo {year} {2020})}\BibitemShut {NoStop}%
\bibitem [{\citenamefont {Li}\ and\ \citenamefont {Fisher}(2021)}]{Li-2021a}%
  \BibitemOpen
  \bibfield  {author} {\bibinfo {author} {\bibfnamefont {Y.}~\bibnamefont
  {Li}}\ and\ \bibinfo {author} {\bibfnamefont {M.~P.~A.}\ \bibnamefont
  {Fisher}},\ }\bibfield  {title} {\bibinfo {title} {Statistical mechanics of
  quantum error correcting codes},\ }\href
  {https://doi.org/10.1103/PhysRevB.103.104306} {\bibfield  {journal} {\bibinfo
   {journal} {Phys. Rev. B}\ }\textbf {\bibinfo {volume} {103}},\ \bibinfo
  {pages} {104306} (\bibinfo {year} {2021})}\BibitemShut {NoStop}%
\bibitem [{\citenamefont {Quan}\ \emph {et~al.}(2006)\citenamefont {Quan},
  \citenamefont {Song}, \citenamefont {Liu}, \citenamefont {Zanardi},\ and\
  \citenamefont {Sun}}]{Quan-2006}%
  \BibitemOpen
  \bibfield  {author} {\bibinfo {author} {\bibfnamefont {H.~T.}\ \bibnamefont
  {Quan}}, \bibinfo {author} {\bibfnamefont {Z.}~\bibnamefont {Song}}, \bibinfo
  {author} {\bibfnamefont {X.~F.}\ \bibnamefont {Liu}}, \bibinfo {author}
  {\bibfnamefont {P.}~\bibnamefont {Zanardi}},\ and\ \bibinfo {author}
  {\bibfnamefont {C.~P.}\ \bibnamefont {Sun}},\ }\bibfield  {title} {\bibinfo
  {title} {Decay of {{Loschmidt Echo Enhanced}} by {{Quantum Criticality}}},\
  }\href {https://doi.org/10.1103/PhysRevLett.96.140604} {\bibfield  {journal}
  {\bibinfo  {journal} {Phys. Rev. Lett.}\ }\textbf {\bibinfo {volume} {96}},\
  \bibinfo {pages} {140604} (\bibinfo {year} {2006})}\BibitemShut {NoStop}%
\bibitem [{\citenamefont {Heyl}(2018)}]{Heyl-2018}%
  \BibitemOpen
  \bibfield  {author} {\bibinfo {author} {\bibfnamefont {M.}~\bibnamefont
  {Heyl}},\ }\bibfield  {title} {\bibinfo {title} {Dynamical quantum phase
  transitions: A review},\ }\href {https://doi.org/10.1088/1361-6633/aaaf9a}
  {\bibfield  {journal} {\bibinfo  {journal} {Rep. Prog. Phys.}\ }\textbf
  {\bibinfo {volume} {81}},\ \bibinfo {pages} {054001} (\bibinfo {year}
  {2018})}\BibitemShut {NoStop}%
\end{thebibliography}
\end{document}